# What Rate Theories Compute

*Chemical kinetics from Arrhenius to machine-learned potentials*

Rodney S. Ruoff
Center for Multidimensional Carbon Materials, Institute for Basic Science (IBS),
Ulsan 44919, Republic of Korea
Department of Chemistry; Department of Materials Science and Engineering;
and School of Energy and Chemical Engineering,
Ulsan National Institute of Science and Technology (UNIST),
Ulsan 44919, Republic of Korea
Corresponding author. Email: ruofflab@gmail.com

**Abstract**

Chemical rate theories have advanced by changing how a reaction is represented and which molecular quantities are calculated. Several advances also supplied controlled corrections to earlier theories. I connect this history, from Arrhenius to machine-learned potentials, to tests of the assumptions behind a calculated rate. Original sources clarify Marcelin's treatment of incomplete transmission and reactive-population depletion, and Christiansen's descriptions of orientation and collisional deactivation. Five numerical comparisons distinguish tests against experiment from tests using the same molecular model and from calibration. For hydrogen desorption from platinum, a published barrier adjustment improves absolute rate agreement but increases the spread of calculated-to-experimental ratios across temperatures; the source supplies no rate uncertainties. For triosephosphate isomerase, changing the prior on microscopic rates shifts the median inferred intermediate energy from 7.3 to 17.2 kJ mol$^{-1}$ with the same inputs, likelihood, and bounds. This sensitivity calculation does not determine a preferred physical energy. I also examine how later molecular information can change the interpretation of older measurements. Hydrogen–oxygen chain branching provides a published reconciliation involving revised thermochemistry and dynamics. For pyrophosphate hydrolysis, conditional speciation identifies a potentially important complex containing two magnesium ions whose hydrolysis rate is not supplied by the recent calculations. Proposed tests use independent species populations, a later optical calibration of Criegee intermediates, and controlled temperature histories and local recession measurements in diamond dissolution.

## 1 Introduction

An accurate fit of rate constants measured at different temperatures does not uniquely identify a reaction mechanism. Nor does an accurate energy barrier, by itself, give an accurate rate. The calculation must also describe how reactants reach the barrier and what happens after they cross it. Molecules may return to the reactants, for example, or escape may deplete a reactive population faster than collisions or conformational changes can replenish it.

The aim of rate theory has remained the calculation of a rate for a specified process. Its major advances changed how that calculation was made. Arrhenius related the temperature dependence to a population of reactive molecules. Collision theories counted molecular encounters. Marcelin calculated flux through a surface in phase space, and Eyring evaluated the population and forward motion of an activated complex from molecular partition functions. Kramers treated reaction as escape driven by energy exchange with the surroundings. Later theories resolved populations by energy, included quantum dynamics, or sampled trajectories that connect reactants and products.

These developments also clarified when earlier theories remain useful. Kramers's description, for example, approaches the transition-state estimate when the reactive population is replenished rapidly enough and few crossing trajectories return. I argue that the history is best understood through both kinds of advance: changes in what is calculated and quantitative corrections to earlier estimates.

The original papers often state the restrictions with considerable precision. Marcelin examined whether reaction depleted the population near his critical surface and allowed only a fraction of crossings to produce transformation. Pelzer and Wigner distinguished trajectories that proceed to products from those that return to reactants. Kramers calculated how the rate changes when energy supply or motion through the surrounding medium limits escape. Later major reviews also discussed these restrictions. I use the original accounts to identify the authors' contributions and to examine how their assumptions can be tested in present-day calculations and experiments.

Several reviews provide the theoretical and historical foundation. Hänggi, Talkner, and Borkovec surveyed dynamical rate theory after Kramers; Pollak and Talkner traced transition-state theory and its extensions; and Laidler and King examined its early development. [1, 2, 3] Tuñón and Williams connected the meanings of transition states, activated complexes, dividing surfaces, and reaction coordinates to modern calculations. [4] Peters linked reaction-coordinate identification and dynamical regime to the choice of rate method. He also emphasized that agreement with one measured rate can result from compensating errors and that kinetic trends provide further tests. [5]

Recent reviews treat several of these problems in detail. Kim and co-workers discuss model architectures and training data for reactive machine-learning interatomic potentials. [6] Weare and Dinner's September 2026 review preprint uses an operator formulation to describe learning committors and other kinetic quantities from molecular dynamics. [7] McLeod and Thorne examine activation parameters, temperature dependence, and enthalpy–entropy compensation in enzymes. [8] Bello, Yang, and Heyden assess uncertainty in heterogeneous catalysis, from atomistic calculations through microkinetics to reactor models. [9]

I add a detailed analysis of the original formulations and a quantitative reassessment of selected rate calculations. The historical analysis distinguishes the introduction of a physical idea from its evaluation as a rate, including Marcelin's transmission fraction and Christiansen's orientation and deactivation arguments. For five modern studies, I reconstruct the numerical comparisons and identify their reference systems, fitted quantities, and variation with sampling or numerical settings. The platinum example gives a specific result: the published barrier adjustment improves absolute agreement while increasing the variation of calculated-to-experimental ratios with temperature. The chignolin example separates agreement in one run from reproducibility across independent runs. The source passages, rate data, and calculations supporting these assessments are given in the Supplementary Materials.

I also examine what later molecular information and tests of parameter inference can add to the interpretation of existing rate measurements. Section 9 distinguishes a published reconciliation from calculations performed here and proposed tests. The examples show which additional measurements, calibrations, or calculations could constrain a rate model, and when the information already available is insufficient.

Sections 3 and 4 follow the main theoretical developments, grouped by their conceptual connections rather than strict chronology. Sections 5 and 6 describe tests of their assumptions and examine the effects of nonideal thermodynamics, environmental dynamics, and the relation between elementary reactions and measured kinetics. Section 7 discusses examples from solution, surfaces, and enzymes. Section 8 considers learned potentials, sampling, and rate estimation. Section 9 examines how independent molecular information and tests of parameter inference can be used to revisit older rate measurements. Nonadiabatic, electrochemical, and microkinetic theories are discussed only where they affect these comparisons. The Supplementary Materials provide the historical quotations and translations, derivations, and numerical details.

## 2 What measured rates and activation parameters establish

### 2.1 Activation energy from the temperature dependence of a rate

A rate constant is inferred from a measurement using a kinetic model. The measurement may follow concentration with time, product flux, the time to first reach a product state, jump frequency, or enzyme turnover. The chosen rate law determines the units of the constant. An apparent constant obtained for a multistep process need not describe any one elementary reaction, even if its temperature dependence fits the Arrhenius equation.

For constant parameters over a specified interval, the Arrhenius relation is

$$k(T) = A \exp\left(-\frac{E_a}{RT}\right). \tag{1}$$

More generally, the slope of a plot of the logarithm of the rate constant against reciprocal absolute temperature defines a local activation energy:

$$E_a(T) = -R\frac{d\ln[k(T)/k_{ref}]}{d(1/T)} = RT^2\frac{d\ln[k(T)/k_{ref}]}{dT}, \tag{2}$$

where the temperature-independent reference $k_{ref}$ has the same units as $k$. A straight Arrhenius plot indicates an approximately constant activation energy over the measured interval. The same linear behavior can arise from different mechanisms, including ones with tunneling; it does not identify a potential-energy barrier. The prefactor $A$ has the units of $k$ and cannot generally be interpreted as a molecular vibration frequency. Section 3 develops the statistical interpretation of the Arrhenius slope.

The Arrhenius exponential is often interpreted as the fraction of molecules with enough energy to react. That fraction must include the number of states available at each energy, and only particular energy distributions give a pure exponential. Section S3.1 gives an example for quadratic degrees of freedom. A rate calculation must also account for the probability that an energetically accessible encounter produces products.

**2.2 Free energy barriers and standard states**

A potential-energy saddle is a stationary point on the molecular energy surface. A free-energy profile, or potential of mean force, includes the statistical contributions of configurations sampled over the other degrees of freedom. Its barrier depends on the chosen coordinate and on how configurations are counted. An apparent activation energy, in contrast, is obtained from the temperature dependence of a rate. Figure 1 illustrates these different definitions. [4, 5]

For an elementary reaction of molecularity $n$, a concentration-based rate constant can be written in an Eyring form as

$$k=\kappa(T)\frac{k_{\mathrm{B}}T}{h}(c^{\circ})^{1-n}\exp\left(-\frac{\Delta G_c^{\ddagger}}{RT}\right). \tag{3}$$

Here $c^{\circ}$ is the fixed standard concentration, and $\Delta G_c^{\ddagger}$ is the activation free energy for that standard state. Section 3 describes how the equilibrium population of the activated complex and its forward motion give this expression. A concentration rate law for a nonideal system also requires activity factors (Section 6). $R$ is the gas constant, $k_{\mathrm{B}}$ Boltzmann's constant, $h$ Planck's constant, and $T$ absolute temperature. Molar energies use $R$ and molecular energies use $k_{\mathrm{B}}$, with $R=N_{\mathrm{A}}k_{\mathrm{B}}$. I use $\hbar=h/(2\pi)$ and $\beta=1/(k_{\mathrm{B}}T)$ below. A superscript circle denotes a standard state, and ‡ denotes an activated-complex or transition-state quantity.

The parameters inferred from an Eyring plot depend on the standard state and on what is held fixed as temperature changes. To distinguish them from thermodynamic enthalpy and entropy, define $\mathscr{H}_c^{\ddagger}=-T^2\,\mathrm{d}(\Delta G_c^{\ddagger}/T)/\mathrm{d}T$ and $\mathcal{S}_c^{\ddagger}=-\mathrm{d}\Delta G_c^{\ddagger}/\mathrm{d}T$ along the specified temperature path, with $c^{\circ}$ fixed. They satisfy $\Delta G_c^{\ddagger}=\mathscr{H}_c^{\ddagger}-T\,\mathcal{S}_c^{\ddagger}$. These parameters describe the local slope and intercept of the Eyring plot. Whether they equal physical activation enthalpy and entropy depends on the thermodynamic convention, as shown in Section 2.3.

With $\kappa=1$, a dimensionless Eyring plot follows from

$$\ln\left[\frac{hk(c^{\circ})^{n-1}}{k_{\mathrm{B}}T}\right]=\frac{\mathcal{S}_c^{\ddagger}}{R}-\frac{\mathscr{H}_c^{\ddagger}}{RT}. \tag{4}$$

Its local slope against $1/T$ is $-\mathscr{H}_c^{\ddagger}/R$, and its tangent intercept is $\mathcal{S}_c^{\ddagger}/R$. Fitting a straight line treats these parameters as constant over the fitted interval. Variation may remain unresolved within the experimental uncertainty. Observable curvature can result from activation heat capacity, changing populations or pathways, environmental dynamics, tunneling, or changes in the contributions of elementary steps to the measured rate. [8]

Changing the standard concentration changes the inferred activation free energy and entropy, although a consistent conversion leaves the predicted rate unchanged. For bimolecular association, conversion from a 1 bar ideal-gas standard to a 1 mol L$^{-1}$ concentration standard lowers $\Delta G^{\ddagger}$ by $RT\ln(c^{\circ}RT/p^{\circ})$: about 7.95 kJ mol$^{-1}$ at 298 K. This change comes entirely from the standard-state conversion and contains no solvation contribution. A negative activation entropy can therefore reflect the standard-state convention, molecular restrictions, solvent organization, or a combination of these contributions.

### 2.3 Effects of dynamics and standard states on fitted activation parameters

Differentiating Equation 3 at fixed $c^\circ$ gives

$$E_\mathrm{a} = \mathscr{H}_c^{\ddagger} + RT + RT^2 \frac{\mathrm{d}\ln\kappa}{\mathrm{d}T}. \qquad (5)$$

For an ideal-gas elementary reaction, $\mathscr{H}_c^{\ddagger} = \Delta H_p^{\ddagger} + (n-1)RT = \Delta U^{\ddagger}$, where $\Delta H_p^{\ddagger}$ is the physical activation enthalpy at the pressure standard state. Thus $E_\mathrm{a} = \Delta H_p^{\ddagger} + nRT + RT^2\,\mathrm{d}\ln\kappa/\mathrm{d}T$. Holding the standard concentration fixed as temperature varies differs from holding the standard pressure fixed; the derivative therefore gives a different parameter. At a given temperature, the physical ideal-gas enthalpy is independent of the chosen pressure standard. The equality to $\Delta U^{\ddagger}$ applies to an ideal gas and must not be carried over to solution reactions. Here molecularity counts the reactants in an elementary event, rather than the empirical order of a multistep reaction. Section S3.2 derives these relations and the additional terms from temperature-dependent activity factors.

If a measured rate contains $\kappa(T)$ but its Eyring fit assumes unity, the apparent slope parameter $\mathscr{H}_{\mathrm{app}}^{\ddagger}$ includes $RT^2\,\mathrm{d}\ln\kappa/\mathrm{d}T$. The apparent intercept parameter $\mathcal{S}_{\mathrm{app}}^{\ddagger}$ includes both $R\ln\kappa$ and $RT\,\mathrm{d}\ln\kappa/\mathrm{d}T$. A constant $\kappa$ shifts only the intercept. The fitted parameters can therefore contain contributions from dynamics as well as equilibrium populations. This interpretation requires a rate for which Equation 3 is applicable; an overall multistep rate may not admit one physically meaningful transmission coefficient.

For classical dynamics, an equilibrium transition-state calculation counts every forward crossing of a surface separating reactants and products. If some trajectories return to reactants, this count exceeds the number of successful reactions. With equilibrated reactants and recrossing as the only correction, the rate calculated for the same molecular model is bounded by the transition-state estimate, and $0 \le \kappa \le 1$. A factor that also accounts for quantum tunneling need not satisfy this bound. [1, 10]

Activation enthalpy and entropy estimated from the same temperature series have correlated errors. Krug, Hunter, and Grieger and Exner showed that these errors can produce an apparent enthalpy–entropy compensation relation. [11, 12, 13] Proposed physical explanations should therefore be tested against the rate data and their uncertainties, preferably at temperatures where their predictions differ. Isotope substitution can provide a further test, as the lipoxygenase example in Section 7 shows. The temperature interval, rate law, standard state, and treatment of $\kappa$ are needed to interpret the reported activation parameters.

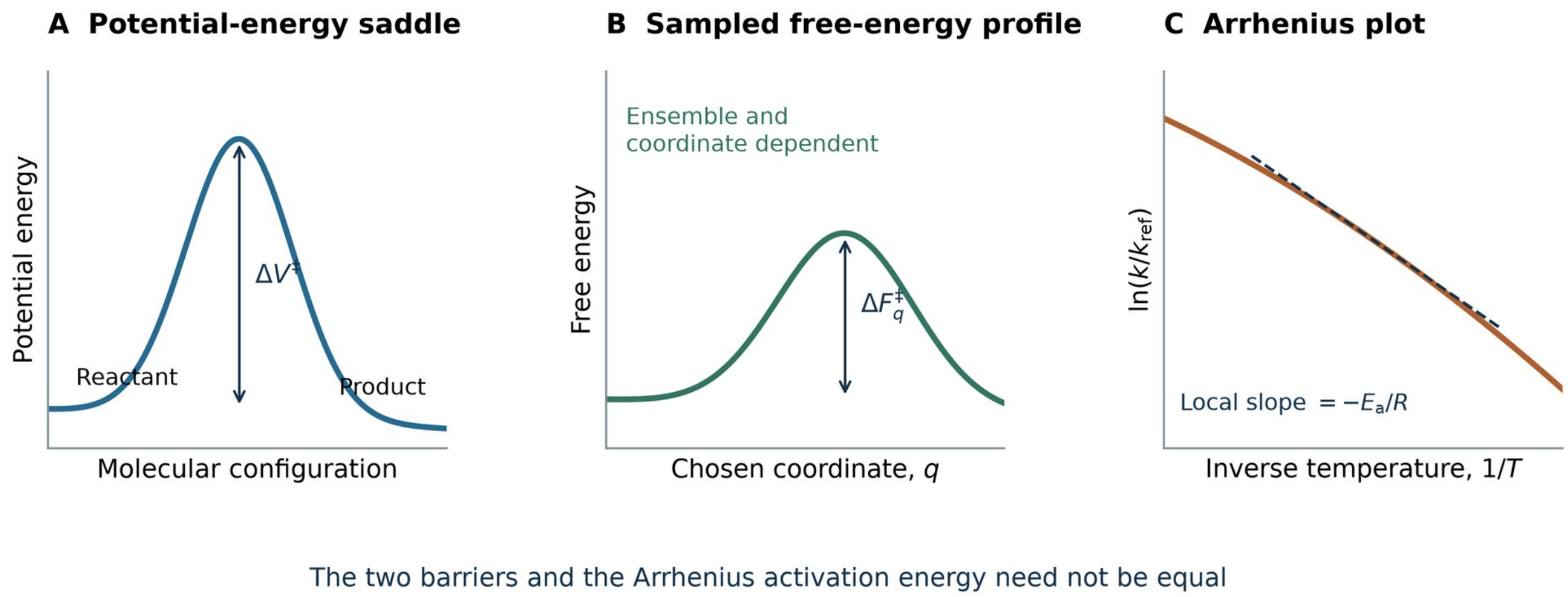

Figure 1. Potential-energy barriers, free-energy barriers, and apparent activation energies. (A) A potential-energy saddle describes a molecular configuration. (B) A free-energy profile includes the statistical contributions of other configurations and depends on the chosen coordinate and counting measure. (C) The local slope of the logarithmic rate constant against reciprocal temperature defines an apparent activation energy. The curves are schematics, not measurements of the same system. Converting the free-energy profile into a rate also requires a description of the dynamics. [4, 5]

## 3 From reactive populations to statistical flux

### 3.1 Temperature dependence and reactive encounters

Van ’t Hoff obtained the differential form of the temperature dependence in 1884 by relating forward and reverse rates to the equilibrium isochore. He treated constant coefficients as an approximation and allowed them to vary with temperature. [14] Arrhenius used a temperature-independent activation energy as a first approximation in his 1889 study of acid-catalyzed cane-sugar inversion. [15] He proposed that a reactive form of the sugar, produced by atomic rearrangement or addition of water, was in equilibrium with ordinary sugar. This proposed chemical species was distinct from the transition state later defined on a potential-energy surface. The interpretation in terms of molecules in the high-energy part of a thermal distribution came later. Section S2.1 gives the original statements and translations.

Trautz and W. C. McC. Lewis attempted to calculate absolute rates from molecular collisions, velocities, sizes, and activation requirements. [16, 17] They also had to specify how many collisions would lead to reaction. Trautz recognized that observed rates could be much smaller than collision counts, whereas Lewis assumed that every collision between active molecules was successful. Neither account supplied the later orientation or steric factor. Christiansen identified molecular orientation as a restriction on nitric-oxide collision complexes in 1921. [18] Pelzer and Wigner subsequently estimated how the energy cost of an off-axis approach reduced its contribution to reaction. [19]

Lewis sought molecular quantities that could be determined independently of reaction-rate measurements. However, the collision diameter was poorly known: he adopted an approximate common value and fitted it in two cases. His hydrogen-iodide calculation also used a critical energy increment obtained from the measured temperature dependence. His calculated rates therefore depended in part on kinetic measurements, despite the aim of predicting rates from independently known molecular properties.

Tolman gave the Arrhenius activation energy a statistical interpretation in 1920. [20] For a canonical ensemble, with a microscopic energy function and reaction weighting that have no explicit temperature dependence, his result can be written in modern molar notation as

$$E_{\mathrm{a}} = \langle E \rangle_{\mathrm{reacting}} - \langle E \rangle_{\mathrm{reactants}}. \tag{6}$$

The first average weights each reactant state by its contribution to reaction; the second averages over the equilibrium reactant population. The energies include all reactants involved in an event and are expressed per mole of events. If the reaction weighting depends explicitly on temperature, its logarithmic derivative adds another term. Tolman's original thermal discussion also included radiation modes, which are omitted here. Section S2.3 gives the context.

The activation energy is therefore generally a difference of average energies, rather than the minimum energy required for reaction. It can vary with temperature and can be negative. A rate that increases as temperature falls is consequently compatible with this statistical description. The Arrhenius slope alone cannot identify a collision mechanism or the height of a potential-energy saddle.

### 3.2 Marcelin and reaction flux through phase space

Marcelin sought a rate law for irreversible transformation that could be expressed in thermodynamic quantities. His doctoral thesis was published in Paris by Gauthier-Villars et Cie in 1914; a digitized copy is available in the Internet Archive's Leiden University collection. [21] The work was republished posthumously in *Annales de Physique* in 1915. [22] A footnote on its first page records that he was killed in action in September 1914.

Marcelin defined a critical surface in molecular phase space and calculated the flux through it from the Gibbs distribution. He divided this flux by the equilibrium reactant population to obtain a rate. For a unimolecular reaction, the resulting logarithmic temperature derivative contained the difference between the critical energy, weighted by contributions to the flux, and the mean reactant energy. This has the statistical form later given by Tolman. Marcelin assumed a temperature-independent critical surface. He also defined a critical affinity through a surface integral, connecting his statistical description to thermodynamics without evaluating a molecular prefactor.

Marcelin allowed only a fraction of crossings to produce transformation by including a transmission fraction inside the surface integral. If this fraction is constant over the surface and independent of temperature, it cancels from the logarithmic derivative. Variation over the surface changes the weighting, whereas explicit temperature dependence adds a derivative term.

He also examined whether reaction would disturb the equilibrium population near the critical surface. In his analogy of a swarm of bees, bees that crossed outward were removed. He argued that removal would scarcely change the active fraction if the surface lay far outside the bulk of the swarm, but that this argument failed when the surface cut through it. The analogy addressed depletion and replenishment of the reactive population. It did not calculate the return trajectories treated in later recrossing theories. Section S2.4 gives the passages and explains why his critical energy is not simply the height of a geometric barrier.

Marcelin thus formulated reaction as a statistical flux and considered both incomplete transmission and depletion of reactants. His phase-space integrals remained unevaluated, however. He did not obtain a universal prefactor from molecular partition functions or calculate a dynamical correction from trajectories.

### 3.3 Potential-energy surfaces and the activated complex

Eyring and Polanyi represented reaction as motion on a multidimensional potential-energy surface in 1931. [23] They constructed an approximate surface from diatomic energy curves using an adiabatic treatment. The surface identified reactant and product regions and a pass between them, allowing the reaction path and zero-point effects to be examined in terms of molecular geometry. The accuracy of a rate calculated from this surface remained limited by the approximate interaction energies.

Pelzer and Wigner calculated an absolute rate for an exchange reaction in 1932 and examined the adiabatic approximation. [19] They distinguished crossings of the ridge between reactants and products from completed reactions and considered trajectories that returned. Their low-temperature estimate of a factor of one-half applied to the system examined; it was not a universal lower bound on transmission. They warned that agreement with measurement could be partly accidental because of uncertainty in the potential, return crossings, and the classical treatment of hydrogen.

Wigner later stated the transition-state assumptions as motion on one Born–Oppenheimer surface, classical nuclear dynamics, and a dividing surface crossed only once. [24] Locating a saddle on an energy surface does not establish the last condition, which concerns the subsequent motion of the nuclei. Wigner also separated three calculations: determining the molecular energy surface and dynamics, calculating elementary rates statistically, and combining elementary reactions to obtain the observed kinetics. [24] Sections 5–8 examine tests at each of these stages.

Eyring's 1935 formulation calculated the constrained population of the activated complex from molecular partition functions and combined it with forward motion along the reaction coordinate. [25, 26] The calculation includes translational, rotational, vibrational, electronic, symmetry, isotope, and zero-point contributions. The expression in Section 2 contains $k_B T / h$, an activation free energy, the standard-state concentration factor required by molecularity, and a transmission coefficient $\kappa$. The activated complex represents an ensemble constrained to the dividing surface, so its free energy includes more than the electronic energy of one saddle configuration.

At 300 K, $k_B T / h$ is approximately $6.25 \times 10^{12} \mathrm{s}^{-1}$. Although it has units of frequency, it is not the frequency of a molecular vibration or the inverse lifetime of the activated complex. Eyring obtained it by combining the partition function for the reaction coordinate with the mean forward velocity. He treated motion near a sufficiently flat barrier top as translational and neglected tunneling there. In a mode representation, the classical factor $k_B T / (h \nu_{rc})$ combines with the crossing frequency, and the assigned $\nu_{\mathrm{rc}}$ cancels. The full classical transition-state expression also has no residual dependence on *h*. Section S2.7 gives the original derivation.

Writing $\Delta G^{\ddagger} = \Delta H^{\ddagger} - T \Delta S^{\ddagger}$ connects the partition-function calculation to thermodynamic activation parameters. Their values depend on the standard state and on the translational, rotational, vibrational, and environmental contributions included in those functions. Eyring retained a transmission coefficient and justified setting it to unity for particular applications; the approximation was explicit in the original theory.

Hirschfelder and Wigner subsequently derived a correction from reflection probabilities. [27] They assumed that those probabilities did not depend on which side the system came from or how many times it had already crossed, and stated that this assumption was not always justified. Glasstone, Laidler, and Eyring reproduced the result with attribution. [28] The correction therefore treated a specified class of trajectory statistics; applying it to a molecular system required those statistics to be appropriate for that system.

Evans and Polanyi independently developed the transition-state treatment of solution reactions in 1935 and related pressure dependence to activation volume. [29] Their result described the pressure derivative of the quasi-equilibrium constant for forming the transition state. A negative activation volume means that increasing pressure raises its population. The rate then increases if transmission and the other pressure-dependent factors remain fixed. Interpreting measured pressure dependence therefore also requires the concentration or activity convention and any pressure dependence of transmission. Wynne-Jones and Eyring treated condensed phases, while Glasstone, Laidler, and Eyring consolidated applications to reaction, diffusion, viscosity, and electrode processes. [28, 30]

### 3.4 Unimolecular reactions and energy resolved populations

Unimolecular reactions required a description of both collisional activation and the subsequent fate of the energized molecule. In the 1922 Faraday discussion, Lindemann proposed collisional activation followed by delayed reaction. He predicted that pressure-independent first-order kinetics would fail when the interval between collisions became comparable to the activated molecule’s lifetime. [31] Christiansen had already noted in his 1921 thesis that another collision during the delay could deactivate the molecule. [18] His statement challenged a radiation-only interpretation but did not give a steady-state kinetic scheme. Lindemann likewise did not write the later equations for competition between reaction and collisional deactivation.

Hinshelwood, Rice and Ramsperger, and Kassel treated internal modes as an energy reservoir and considered energy redistribution between collisions. [32, 33, 34] Rice and Ramsperger explicitly balanced reaction against collisional deactivation. They examined the assumption that every collision deactivates an energized molecule and calculated the consequences when it fails. Their treatment and Kassel’s became RRK theory. Section S2.5 documents the separate contributions.

Marcus's 1952 development, now called RRKM theory, combined an energy-resolved transition-state count with the reactant density of states and treated rotational restrictions approximately. [35] A modern formulation that resolves both internal energy and angular momentum is

$$k(E,J)=\frac{\sigma N^{\ddagger}(E,J)}{h\rho(E,J)}. \tag{7}$$

$N^{\ddagger}$ counts accessible transition-state states for one path, excluding the reaction-coordinate mode; $\rho$ is the reactant density of states per unit energy; and $\sigma$ counts equivalent paths not already included. Both counts use the same energy zero, and $E$ is energy per molecule. The expression assumes sufficiently rapid intramolecular redistribution and no recrossing. In the 1952 treatment, Marcus distinguished active, adiabatic, and inactive modes and averaged the approximately conserved rotational quantum numbers into partition-function factors. He also identified anharmonic energy transfer as an unresolved part of a more fundamental description.

A canonical rate follows by thermal averaging of microcanonical rates when the reactant population has a thermal energy distribution. At lower pressure, or after chemical activation, molecules may react before energy transfer establishes that distribution. Energy-resolved rate coefficients must then be combined with equations for the evolving populations, commonly a master equation. Dependence on pressure and bath gas can test this treatment of energy transfer and reaction. Recent studies of gas-phase ions and radiative dissociation illustrate such tests. [36, 37] The microcanonical theory calculates rates at specified energy; thermal averaging connects those results to the canonical theory.

## 4 Dynamics and reaction rates

### 4.1 Kramers and energy exchange during escape

Kramers recast reaction as stochastic escape, with energy exchange between a reactive coordinate and its surroundings entering the rate calculation. [38] His 1940 treatment used a classical one-dimensional potential, a barrier large compared with $k_{\mathrm{B}}T$, and position-independent friction. At weak friction, the surroundings supply energy too slowly to replenish the escaping population; increasing friction therefore raises the rate. At strong friction, motion across the barrier is slow and diffusive, and increasing friction lowers the rate. The rate passes through a maximum between these regimes. Kramers derived results for the limiting regimes and identified this turnover, but did not obtain one expression valid across the entire range.

The transition-state estimate reappears as an approximate limiting case when the reactant population is replenished sufficiently rapidly and few crossings return to reactants. Kramers identified this intervening regime as well as the failures at weak and strong friction. His treatment both recast reaction as stochastic escape and supplied a dynamical correction to transition-state theory. Its moderate-to-strong-friction expression cannot be extended into the weak-friction regime simply by taking friction toward zero.

For a molecular reaction in solution, the free-energy profile along a selected coordinate can include the statistical contribution of solvent configurations. This profile is a potential of mean force. Eliminating the solvent coordinates from the explicit dynamics can also leave memory: the force at a given time depends on earlier motion. Grote and Hynes showed how frequency-dependent friction changes the rate associated with a given barrier. [39] Bulk viscosity alone need not describe the solvent motions that couple to the reaction. Hänggi, Talkner, and Borkovec review the development from Kramers's treatment to memory-dependent friction and reactive-flux calculations. [1]

### 4.2 Activation through solvent reorganization and lattice vibrations

Marcus's electron-transfer theory made solvent polarization and nuclear reorganization central to activation. [40] For harmonic diabatic free-energy surfaces of equal curvature, the classical crossing barrier is

$$\Delta G^{\ddagger} = \frac{(\lambda + \Delta G^{\circ})^2}{4\lambda}. \tag{8}$$

$\Delta G^{\circ}$ is the reaction Gibbs energy and $\lambda$ is the energy required to reorganize nuclei and solvent from the reactant equilibrium configuration to the product configuration without transferring the electron. The barrier vanishes at $-\Delta G^{\circ} = \lambda$ and increases again in the classical inverted region. Converting this barrier to a rate requires the electronic coupling and the rates of competing steps. A maximum in the electron-transfer rate follows only when the other rate factors are held fixed.

Marcus's 1956 model assumed slight electronic overlap and treated the solvent outside rigid, dielectrically saturated cavities as a continuum. His simple expression for the overall rate assumed that electron transfer occurred about as rapidly as, or more rapidly than, the reorganized pair returned to its initial state. When electron transfer was slower, the competing steps had to be treated explicitly. Quantum nuclear modes modify the classical inverted-region prediction; Siders and Marcus showed that including them did not remove every discrepancy with experiment. [41] Modern electron-transfer models extend the descriptions of electronic coupling and solvent response. [42] The activation barrier here describes collective nuclear reorganization that allows an electronic transition; it need not correspond to a bond-breaking saddle.

Vineyard expressed the classical harmonic transition-state prefactor in terms of the normal-mode frequencies of the entire crystal. [43] In this approximation,

$$\nu^{*} = \frac{\prod_{j=1}^{N} \nu_j}{\prod_{j=1}^{N-1} \nu_j'}. \tag{9}$$

The numerator contains the stable frequencies in the initial well. The denominator contains the real saddle frequencies and omits the unstable mode. Zero-frequency translations and rotations are removed by imposing consistent constraints on both spectra. Because these frequencies describe the surrounding lattice as well as the jumping atom, the prefactor cannot be assigned to one local hopping frequency. Vineyard neglected quantum effects, assumed nearly harmonic motion, and equated the crossing rate with the jump rate. He did not calculate the return of trajectories after crossing. Solvent reorganization and the lattice vibration spectrum can both contribute to activation entropy.

### 4.3 Optimizing a surface and calculating recrossing

For equilibrated reactants evolving under a specified classical dynamical model, counting every forward crossing overestimates the reaction rate if some trajectories return. Changing the dividing surface can reduce this overcount. Canonical variational transition-state theory minimizes the one-way equilibrium flux over a chosen family of surfaces,

$$k_{\mathrm{CVT}}(T)=\min_{s} k_{\mathrm{TST}}(s,T). \tag{10}$$

The parameter $s$ labels the trial surfaces. The minimum gives the closest upper bound within this family, but recrossing can remain. The bound requires sufficiently rapid replenishment of reactant states and consistent definitions of reactants and products. It also requires the same potential and classical dynamics in the transition-state calculation and the rate used for comparison. [10]

Wigner stated the variational principle in 1937, and Horiuti varied a dividing surface in 1938. [44, 45] Keck applied the principle to multidimensional recombination; Pechukas and McLafferty derived conditions under which it gives the exact rate in collinear collisions; and Garrett and Truhlar developed canonical variational calculations for molecular reactions. [46, 47, 48] In Horiuti's two-atom potential model, minimizing the flux reduced the conventional transition-state estimate by a factor of 1.1–1.5 at 25 °C. Surface optimization is particularly useful when fragments form a loose transition region or associate without a sharp saddle. The surface with the lowest flux need not pass through a stationary point of the potential. [49]

This upper bound applies to the rate for the chosen molecular model, not necessarily to the experimental rate. Wigner's estimates for iodine recombination were below the measured rates. For the atomic bath gases, he identified a conflict if his molecular model was accurate and discussed uncertainties in its inputs. [44] An inaccurate potential-energy surface or omitted electronic states can produce an incorrect rate even when the mathematical bound holds. This remains true when a learned potential supplies the energies and forces.

Reactive-flux theory calculates how much of the initial crossing flux survives as reaction. Following Yamamoto's work on quantum statistical rates, [50] Chandler showed that the transition-state estimate is the short-time limit of a flux correlation treatment of liquid-state isomerization. [51] The correlation must reach a plateau after rapid return crossings have decayed and before slow population relaxation becomes important. That plateau gives the rate over the intervening time interval; the initial flux includes crossings that subsequently return.

Figure 2 shows how averaging over energy, varying a dividing surface, and following trajectories connect the rate calculations.

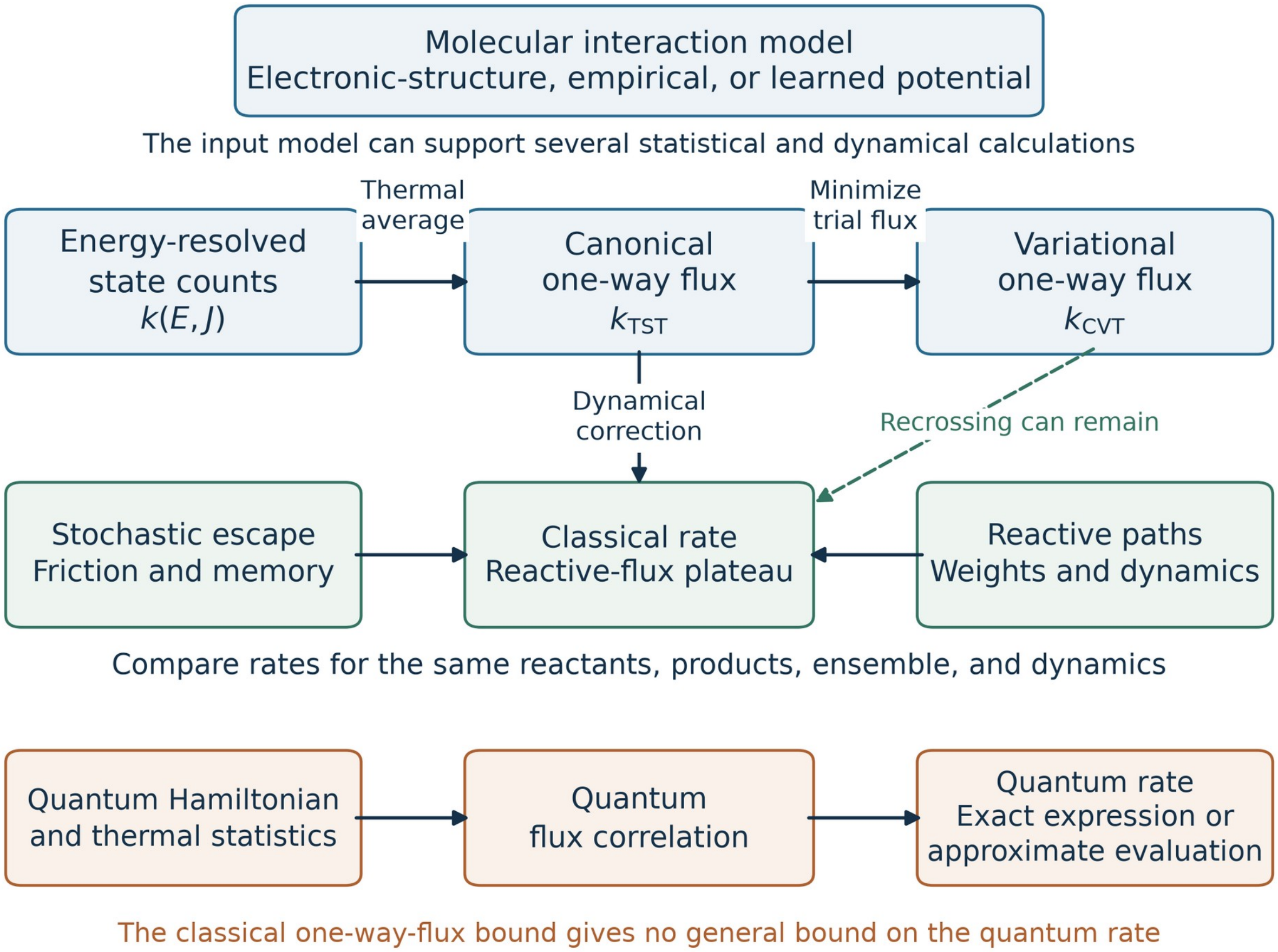

Figure 2. How rate calculations are related. Thermal averaging of microcanonical rate coefficients gives the corresponding canonical transition-state rate when the state counts and ensemble are consistent. Varying the dividing surface minimizes the one-way flux within a chosen family; trajectories can still recross the optimized surface. Stochastic escape calculations and properly weighted reactive trajectories include the dynamics needed to calculate a classical rate. Comparing these rates requires the same state definitions and dynamics. The lower row distinguishes an exact quantum correlation expression from its approximate numerical evaluation. A learned potential can supply energies and forces to several of these calculations; each still requires a choice of ensemble, dynamics, and rate estimator. [10, 35, 38, 51, 52, 53, 54, 55]

### 4.4 Quantum corrections and correlation functions

Eckart analyzed penetration through an analytic barrier, while Wigner and Bell applied quantum treatments to chemical kinetics. [56, 57, 58] Bell's later compact correction for a parabolic barrier is

$$Q=\frac{u/2}{\sin(u/2)},u=\frac{h\nu^{\ddagger}}{k_{\mathrm{B}}T}. \tag{11}$$

Here $\nu^{\ddagger}$ is the magnitude of the imaginary barrier frequency in cycles per unit time, and $Q$ multiplies the classical rate. [59] This expression requires $u<2\pi$, or $T>T_c=h\nu^{\ddagger}/(2\pi k_{\mathrm{B}})$, together with Bell's approximation conditions given in Section S2.8. It diverges at the crossover temperature and cannot describe deep tunneling. Bell provided fuller expressions for other regimes. Later reviews explicitly describe the divergence and the limitations of a one-dimensional tunneling model. [10, 49]

Yamamoto expressed reaction rates through quantum statistical correlations. [50] Miller derived a reactive-flux expression from quantum scattering, and Miller, Schwartz, and Tromp developed equivalent correlation forms suited to direct numerical evaluation. [52, 60] For reactive scattering with asymptotic reactant and product channels, the thermal flux–flux expression is

$$k = \frac{1}{Q_r} \int_0^{\infty} C_{ff}(t)\,\mathrm{d}t. \tag{12}$$

$Q_r$ is the reactant partition function, with volume normalization appropriate to molecularity, and $C_{ff}$ is a thermally weighted correlation of quantum flux operators. The Hamiltonian determines both the quantum statistical weights and the time evolution. The rate is independent of the dividing surface if that surface separates the same reactant and product channels, although its location affects numerical convergence. A closed, equilibrating system requires a different interpretation. As the populations relax, the flux-side correlation decays again, so the infinite-time flux–flux integral does not give the metastable rate. A plateau resolved before population relaxation gives the metastable rate, as discussed in Section 4.3. A finite discrete representation can also introduce numerical recurrences that restrict the usable time interval.

Voth, Chandler, and Miller related quantum statistics, imaginary-time paths, centroids, and dynamical corrections. [61] Craig and Manolopoulos used ring-polymer molecular dynamics to approximate quantum rate correlations; Richardson and Althorpe connected its deep-tunneling limit with semiclassical instanton theory. [53, 62] Increasing the number of path discretization points can converge quantum statistics while leaving the real-time dynamics approximate. Ring-polymer dynamics is exact for certain cases, including a parabolic barrier above crossover. Its accuracy for other barriers must be assessed separately. The classical one-way-flux bound does not generally apply to quantum rates; tunneling can raise the quantum rate above the classical transition-state estimate.

### 4.5 Reactive trajectories and reaction coordinates

Transition path sampling generates reactive trajectories without requiring a reaction coordinate or transition state in advance. [54, 63] It requires definitions of reactants and products and a specified dynamical model. Transition-path theory describes the trajectories through committors and reactive currents. [55] The forward committor is the probability of reaching product before reactant from a specified state. For inertial dynamics, the state generally includes momenta; a configuration committor averages over a specified momentum distribution. A committor of one-half identifies equal probabilities of reaching the two states. It does not imply that every crossing of that surface reaches product or that the surface minimizes the transition-state estimate.

String methods seek pathways in configuration space or selected collective variables. [64, 65] A string can converge even if the selected variables omit motions needed to distinguish reactive from nonreactive configurations. Committor and recrossing tests examine this possibility (Sections 5 and 7). A pathway or commitment probability alone supplies no reaction timescale; calculating a rate still requires dynamical weights and time normalization.

Sections S2.1–S2.10 document the stated restrictions, unresolved problems, and later treatments of these theories. [1, 28, 49]

## 5 Testing assumptions before choosing a rate method

Choosing a rate method begins with defining the event and the quantity to be compared with experiment. A first-passage time between specified states, an elementary rate coefficient, and a catalytic turnover number describe different measurements. The historical theories distinguish the quantities needed to calculate them: state counting gives a population, crossing a dividing surface gives a flux, and dynamics determine which crossings lead to reaction. Table 1 lists tests of the assumptions used in these calculations.

### 5.1 Reactant equilibration and escape

Equilibrium rate calculations require sufficiently rapid replenishment of the reactant configurations from which reaction occurs. In a simulation, prepare several distributions within the reactant basin and follow the relaxation of energy, conformation, and relevant solvent variables. Compare these relaxation times with the time to escape. If the subsequent escape statistics agree after a short preparation-dependent transient, the preparations are consistent with a common reacting ensemble. The relaxed populations should also agree with the equilibrium distribution assumed in the rate calculation; rapid relaxation of one variable can miss slow changes in others. For pressure-dependent unimolecular reactions, competition between energy transfer and reaction requires an energy-resolved population calculation. [33, 35, 51]

Let $S_A(t)$ be the probability that a trajectory prepared in reactant state $A$ has not yet reached product state $B$ by time $t$. Its time-dependent escape rate is

$$\lambda_A(t) = -\frac{\mathrm{d}\ln S_A(t)}{\mathrm{d}t}. \tag{13}$$

If $S_A(t)=\exp(-kt)$, then $\lambda_A(t)=k$ and the mean first-passage time is $1/k$. A varying $\lambda_A(t)$ means that one constant does not describe the first-passage distribution over the examined interval. Possible causes include incomplete preparation, unresolved intermediates, slowly exchanging conformations, and a changing environment. Trajectories that have not reacted by the end of observation are censored observations and must be included in the analysis; discarding them preferentially removes long waiting times. Section S5 distinguishes first-passage from population-relaxation measurements.

A barrier height expressed in multiples of $k_\mathrm{B}T$ does not by itself establish whether one rate constant is adequate. The relevant comparison is between relaxation within a state and transfer between states. Both depend on the dynamics and the state definitions. In Chandler's flux formulation, their separation appears as a plateau after molecular transients and before slow population relaxation. [51] An unresolved plateau may reflect insufficient statistics or simulation duration rather than the absence of this separation. Longer observations and different initial preparations can help distinguish these possibilities.

### 5.2 Coordinates, recrossing, and competing pathways

To test a proposed reaction coordinate, take configurations on its putative transition surface and launch trajectories with a specified momentum distribution. The fraction reaching $B$ before $A$ estimates the committor. A broad or bimodal committor distribution shows that the surface groups configurations with substantially different probabilities of reacting. Additional solvent or molecular coordinates may distinguish them. The uncertainty caused by using a finite number of trajectories for each configuration must be separated from the variation among configurations. [65, 66]

A distribution concentrated near one-half supports the identification of equal commitment, but does not establish negligible recrossing. For inertial dynamics, the configuration committor averages over momenta; it does not measure the flux that survives after return crossings. The flux correlation and its transmission coefficient must be calculated separately. In the ion-pair example in Section 7.1, a narrower committor distribution was accompanied by a smaller transmission coefficient. [66]

Several reaction pathways can give a single exponential survival curve. If parallel channels have constant rates and draw from the same equilibrated reactant population, their rates add to give the total escape rate. A nonexponential curve, conversely, does not establish multiple chemical pathways. Path ensembles, intermediate populations, isotope-dependent branching, and perturbations that affect different channels can help distinguish the pathways. Transition path sampling provides trajectories for this analysis; rate calculations also require their statistical weights and the corresponding time normalization. [55, 63]

| Assumption being used | What to examine | If the assumption fails |
|---|---|---|
| Reactants form a common ensemble before escape | Relaxation of relevant basin variables; escape statistics after different preparations | Resolve slow populations with a kinetic network or energy-resolved master equation. |
| One constant describes escape | Survival distribution; time dependence of the escape rate; a flux plateau | Check sampling and preparation, then retain time dependence or resolve additional states. |
| The chosen coordinate identifies the transition region | Committor distribution on the proposed surface, including estimation uncertainty | Add physically relevant coordinates or sample pathways without that coordinate constraint. |
| Forward crossings represent completed reactions | Reactive-flux decay and transmission coefficient | Calculate the dynamical correction or use a trajectory-based rate method. Optimizing the coordinate may not remove recrossing. |
| The calculated elementary rate represents the measurement | Dependence on concentration, coverage, isotope, transport, and intermediate populations | Connect elementary steps through a kinetic model and include transport where needed. |
| The comparison tests an independent prediction | Identify data used to fit inputs; state whether the reference is a calculation or an experiment | Test observables excluded from fitting. Compare methods with the same potential and dynamics before assessing agreement with experiment. |

*Table 1. Measurements and simulation tests for choosing and assessing a rate calculation. Their interpretation depends on the ensemble, dynamics, and measured quantity, as discussed in Sections 6–8 and S5.*

## 6 Activities, environmental dynamics, and measured rates

Composition can change the reacting populations through chemical activities, while the environment can change the dynamics of crossing and return. Binding, product release, and transport can then alter the measured rate even if the elementary reaction coefficient remains unchanged.

### 6.1 Thermodynamic populations and activities

For a species $i$, the chemical potential is $\mu_i = \mu_i^\circ + RT \ln a_i$, where $a_i$ is its dimensionless activity. In a concentration convention, $a_i = \gamma_i c_i / c^\circ$. Concentration alone therefore does not determine the chemical potential used to calculate equilibrium populations in a nonideal solution. Gas fugacity, surface occupation, and interactions between adsorbed species can likewise enter the thermodynamic calculation. Eyring's solution treatment recovered the activity-coefficient relation for rates associated with Brønsted. [26, 30]

For an elementary bimolecular activated-complex model at fixed temperature and consistent concentration standard states, the concentration-based rate coefficient contains the factor $\gamma_A \gamma_B / \gamma^\ddagger$. The numerator describes the reactants and the denominator the activated ensemble. Replacing reactant concentrations by activities still leaves the activity coefficient of the activated ensemble, as well as any composition dependence of the dynamical correction. A useful test varies composition while monitoring speciation and, where possible, independently determining activities. A change in apparent reaction order alone does not identify thermodynamic nonideality.

Reaction order can change even when adsorbed species do not interact. Consider identical independent sites that reversibly bind $A$. Adsorbed $A$ converts to product and frees the site, with conversion slow enough for adsorption to remain equilibrated. For the initial rate, assume negligible reverse conversion and no product inhibition. For a dimensionless adsorption constant $K_A$,

$$\theta_A = \frac{K_A a_A}{1 + K_A a_A}, r_{\mathrm{site}} = k_s \theta_A. \tag{14}$$

Here $\theta_A$ is the occupied-site fraction and $r_{\mathrm{site}}$ is product formation per site per unit time. The apparent order changes from approximately one at low occupation to approximately zero near saturation, although the sites are ideal and the elementary conversion coefficient $k_s$ is unchanged. Measuring occupation alongside rate tests whether site saturation explains the change. This idealized calculation is separate from the Pt(111) measurements discussed below.

### 6.2 Environmental dynamics

A solvent can change both the equilibrium free-energy profile and the dynamics of crossing. Viscosity alone does not separate these effects: changing solvent composition or temperature can also change solvation, activities, and reactive conformations. Kramers and Grote–Hynes treatments distinguish energy exchange, friction, and memory within their approximations. [38, 39] An atomistic calculation can compare the free-energy profile, the decay of solvent-force correlations, and reactive-flux decay. Solvent-force correlations that persist during crossing indicate a possible failure of the approximation of instantaneous friction. [66]

A learned potential can supply the energies and forces for these calculations, provided its accuracy has been checked for the compositions and configurations sampled. Even with accurate forces, a simulation may end before slow solvent rearrangements or reactant conformations have equilibrated. Comparing initial preparations and relaxation times tests whether those motions have been sampled, while reactive-flux calculations test how many crossings return (Sections 5 and 7).

### 6.3 From elementary events to the measured signal

An enzyme turnover number or catalyst production rate can depend on binding, reaction, product release, competing pathways, and transport. Comparing a calculated elementary rate with either measurement requires a kinetic model of those steps under the experimental conditions. Varying substrate and product concentrations tests changes in populations and competition among steps; measuring transient intermediates constrains their formation and consumption. Dependence on flow rate, stirring rate, or particle dimensions can reveal transport contributions when the other chemical conditions are controlled.

In the lipoxygenase study, assigning turnover measurements to hydrogen transfer required evidence that this chemical step limited the measured rate. [67] For hydrogen desorption from Pt(111), a zero-coverage calculation left finite-coverage occupation and interactions untested. [68] Sections 7.2 and 7.3 examine these comparisons.

## 7 Three worked examples

### 7.1 Ion pair dissociation and recrossing

Mullen, Shea, and Peters examined one NaCl pair in TIP3P water at 300 K and 1 bar. [66] They first used the distance between the ions as the reaction coordinate. At the maximum of the corresponding free-energy profile, the committor distribution was bimodal: configurations at the same ion separation predominantly returned to contact or proceeded to solvent-separated states. Solvent configurations distinguished these outcomes. Ion separation alone therefore assigned many configurations to the transition state even though their probabilities of reaching the two states were far from equal.

The authors tested additional solvent coordinates using transition-path data. Choosing a coordinate to predict the configuration committor more accurately narrowed its distribution near one-half, but the transmission coefficient fell from $0.29 \pm 0.02$ for ion separation to approximately 0.07. Forward and time-reversed trajectories explained the result: a configuration could have equal probabilities of reaching either state while trajectories recrossed the surface in both directions. Improving the committor estimate therefore did not improve the rate estimated by counting forward crossings alone.

When the optimization also included dynamical information, it selected ion separation, water density between the ions, and the number of waters jointly coordinated to them. The transmission coefficient rose to $0.43 \pm 0.02$. Even trajectories initiated on the surface with a configuration committor of one-half gave $\kappa = 0.38 \pm 0.09$, estimated from 100 trajectories. The authors attributed the remaining recrossing to anharmonicity, shallow intermediates, and multiple channels in the free-energy description that included solvent coordinates (their Figs. 5–10). They examined surfaces in configuration space and allowed that a surface defined in full phase space might remove recrossing.

For this system, the committor tests identified solvent variables missing from the initial coordinate, while the trajectory calculations measured the remaining recrossing. In a converged calculation with the same Hamiltonian, states, ensemble, and target dynamics, changing the dividing surface should change the transition-state estimate and transmission coefficient by compensating amounts, leaving the corrected rate unchanged. This agreement would test the rate calculation within the chosen molecular model; it would not establish the experimental accuracy of the ion and water force fields.

### 7.2 Hydrogen desorption from platinum and a fitted barrier

Zhang and co-workers calculated the zero-coverage second-order coefficient for recombinative hydrogen desorption, $2\,H^* \rightarrow H_2$, using ring-polymer molecular dynamics. [68] A neural network represented 1457 density-functional energies, with the metal atoms held fixed. The adsorbed hydrogen atoms were approximated as freely translating in two dimensions. The authors calculated both the free energy and the dynamical correction for recrossing.

At 653–953 K, the transmission coefficients on the original potential-energy surface were 0.36–0.37. With identical symmetry and electronic-spin factors in both estimates, recrossing reduced the centroid transition-state rate by factors of 2.7–2.8. Even after this correction, the calculated coefficients exceeded experiment by factors of 3.9–6.2.

The authors report a computed Arrhenius activation energy of 63.7 kJ $mol^{-1}$ and compare it with an experimentally estimated zero-coverage barrier of 72.4 kJ $mol^{-1}$. They used that difference to increase the desorption barrier by 8.7 kJ $mol^{-1}$. On the adjusted potential, the recalculated coefficients were 0.95–2.12 times the measured values at the four tabulated temperatures. Relative branching trends among hydrogen isotopologues were also consistent with experiment. Section S4.2 distinguishes the reported activation energy from a regression reconstructed from the published rates.

I compared rate magnitude and temperature dependence separately using the four entries in the authors' Table 2. The barrier adjustment brings every calculated-to-experimental ratio closer to unity. However, the largest ratio divided by the smallest increases from 1.59 before adjustment to 2.24 afterward. Multiplying all calculated rates by a temperature-independent factor cannot change this measure. Thus, the improvement in absolute magnitude does not also improve this measure of temperature dependence over 653–953 K. The source table gives no rate uncertainties, so this comparison of central values does not establish statistical significance.

The authors identified the fixed metal atoms, the treatment of hydrogen translation, the density functional, and approximate quantum dynamics as possible sources of error, but did not isolate their individual contributions. One could test the first and third explanations by separately allowing lattice motion or changing the electronic-structure description, keeping the other ingredients fixed and checking convergence. In the reported study, comparison of classical and ring-polymer calculations indicated that zero-point energy, rather than strong tunneling, was the main nuclear quantum contribution over the experimental temperature range.

Further tests could use conditions not included in the barrier adjustment. Extension to finite coverage would require an explicit description of adsorbate occupation and interactions, followed by comparison with measurements at those coverages. Section S4 gives the numerical results before and after the barrier adjustment separately.

### 7.3 Isotope substitution in lipoxygenase

Knapp, Rickert, and Klinman studied hydrogen transfer catalyzed by soybean lipoxygenase-1 in pH 9 borate buffer over 5–50 °C. [67] They measured the turnover number $k_{cat}$. Agreement between steady-state and single-turnover properties supported hydrogen abstraction as the limiting step in both measurements. The authors also discussed earlier tests of alternative explanations for the isotope effect and corrected the wild-type and Ile553→Ala measurements for incomplete oxygen saturation.

For the wild-type enzyme, the protium turnover number at 30 °C was $297 \pm 12\ s^{-1}$ and its activation energy was $8.8 \pm 0.8$ kJ $mol^{-1}$. Ile553→Ala gave similar values: $280 \pm 10\ s^{-1}$ and $7.9 \pm 0.8$ kJ $mol^{-1}$. Measurements with protium alone could therefore suggest that the mutation had little effect on the reaction dynamics or their temperature dependence. The authors' Table 1 credits the wild-type data to the earlier Rickert–Klinman study.

Deuterium measurements revealed a substantial difference. The difference $E_{a,D} - E_{a,H}$ increased from $3.8 \pm 0.8$ to $16.7 \pm 1.3$ kJ mol$^{-1}$, while the Arrhenius prefactor ratio $A_H / A_D$ decreased from $18 \pm 5$ to $0.12 \pm 0.06$. The wild-type isotope effect at 30 °C was $81 \pm 5$. A model that reproduced the protium rates and activation energies alone would miss this response to the mutation.

The authors modeled hydrogen tunneling coupled to environmental reorganization and fluctuations in the donor–acceptor distance. Their treatment reproduced important isotope trends, but the gated model gave an activation energy of about 33.9 kJ mol$^{-1}$ for Ile553→Ala, compared with the measured $7.9 \pm 0.8$ kJ mol$^{-1}$ (their Tables 1–2). They explicitly discussed this discrepancy. The isotope measurements support tunneling and environmental coupling, but do not establish the full gating model or all its parameters.

A small Arrhenius prefactor alone would not rule out transition-state theory: activation entropy and the transmission coefficient also contribute to that prefactor.

## 8 Learned potentials and tests of calculated rates

### 8.1 What a learned potential supplies

A machine-learned potential approximates molecular energies and forces. Reactive potentials can describe bond breaking and formation. Machine learning can also be used to choose a reaction coordinate or estimate a committor, but these tasks provide different information: a coordinate describes progress between states, and a committor gives the probability of reaching product before reactant. [6, 7, 69]

For a learned potential, energies and forces should be checked against electronic-structure calculations for configurations not used in training. The tests should include reactive configurations, competing pathways, and the environments and thermodynamic conditions of interest. A small average error for a test set dominated by stable configurations can hide substantial errors near the transition state. More sampled trajectories cannot correct a systematic error in the potential. An accurate potential also cannot compensate for a slow coordinate omitted from the kinetic description or an incorrect conversion of biased simulation time to physical time. [9, 70, 71]

At 300 K, changing an activation free energy by 4.184 kJ mol$^{-1}$ changes a rate by a factor of approximately 5.4 if the prefactor and other contributions remain fixed. Errors in an actual rate calculation can also involve entropy, recrossing, nuclear quantum effects, and the populations of competing pathways; they need not be equivalent to a single barrier shift.

Park and co-workers distinguish energy and force accuracy, dynamical stability, comparison with ab initio trajectories, and agreement with measured transport properties. Their validation hierarchy and method-selection guide address different physical regimes, including reactive interfaces and rare events. [71] Figure 3 applies these established distinctions to the rate comparisons considered here. Independent electronic-structure calculations test the learned energies and forces. A converged rate calculation using the same Hamiltonian, state definitions, ensemble, and target dynamics tests the sampling and rate algorithm. Comparison with experiment tests the combined molecular model and kinetic calculation under the experimental conditions.

## 8.2 What the numerical comparisons show

Section S4 gives the numerical details for five studies, including the Pt(111) calculation discussed in Section 7.2. [68] These studies compare different quantities and do not provide a ranking of methods.

For OH + $CH_4$, a learned potential and ring-polymer molecular dynamics gave calculated-to-experimental rate ratios of about 0.79–1.26 over 200–1500 K. [72, 73] The comparison tests the potential and rate calculation together. The low-temperature nuclear dynamics remain approximate. Because numerical treatments also differed among calculations using earlier potentials, the improvement cannot be attributed entirely to the new potential. Section S4.1 gives the extracted source data, experimental references, and convergence limitations.

For radical recombination, a neural interaction model agreed within about 10–20% with CASPT2 reference calculations using the same variable-reaction-coordinate transition-state treatment. [74] The comparison tests how accurately the neural model represents the reference interactions within that treatment. It leaves approximations shared by both calculations untested and does not establish the same accuracy relative to experiment.

For chignolin folding, three neural-committor estimates were 2.49, 13.4, and 10.1 $\mu s^{-1}$, compared with 2.5 ± 0.5 $\mu s^{-1}$ from a long equilibrium simulation using the same Hamiltonian. [75] One run agreed closely, but the 5.4-fold spread among independent runs exposed a sampling problem. These results test kinetics for the simulated protein model; they do not test that model against an experimental protein.

For alanine dipeptide, conventional infrequent metadynamics using a poor collective variable could give a rate more than three orders of magnitude from the reference value. [76] Fitting the short-time behavior reduced the discrepancy to about an order of magnitude in the reported comparison. The reference used the same Hamiltonian. The initial preparation and treatment of trajectories that had not completed a transition affected the comparison. Section S4.5 gives those details, the reference fit, and the uncertainties; the reported improvement applies to the tested preparation and fitting procedure.

## 8.3 Information needed to assess a calculated rate

I recommend reporting the reactant and product definitions, rate law, units, and standard state, together with the molecular model, its electronic-structure or training data, and the sampling and rate methods. Each comparison should identify whether the reference is experiment, an independent calculation with the same Hamiltonian, or data used to fit the model. Temperature, pressure, composition, coverage, and other relevant conditions should be given alongside the rates. Convergence tests, uncertainties, treatment of unfinished trajectories, and variation among independent runs should accompany the numerical results. The discussion should identify which sources of error were examined separately and which could not be separated.

Predictions can be tested at a temperature, isotope composition, solvent condition, or coverage not used to fit the model, provided the model applies under the new conditions. Conditions for which competing models predict different rates are particularly useful.

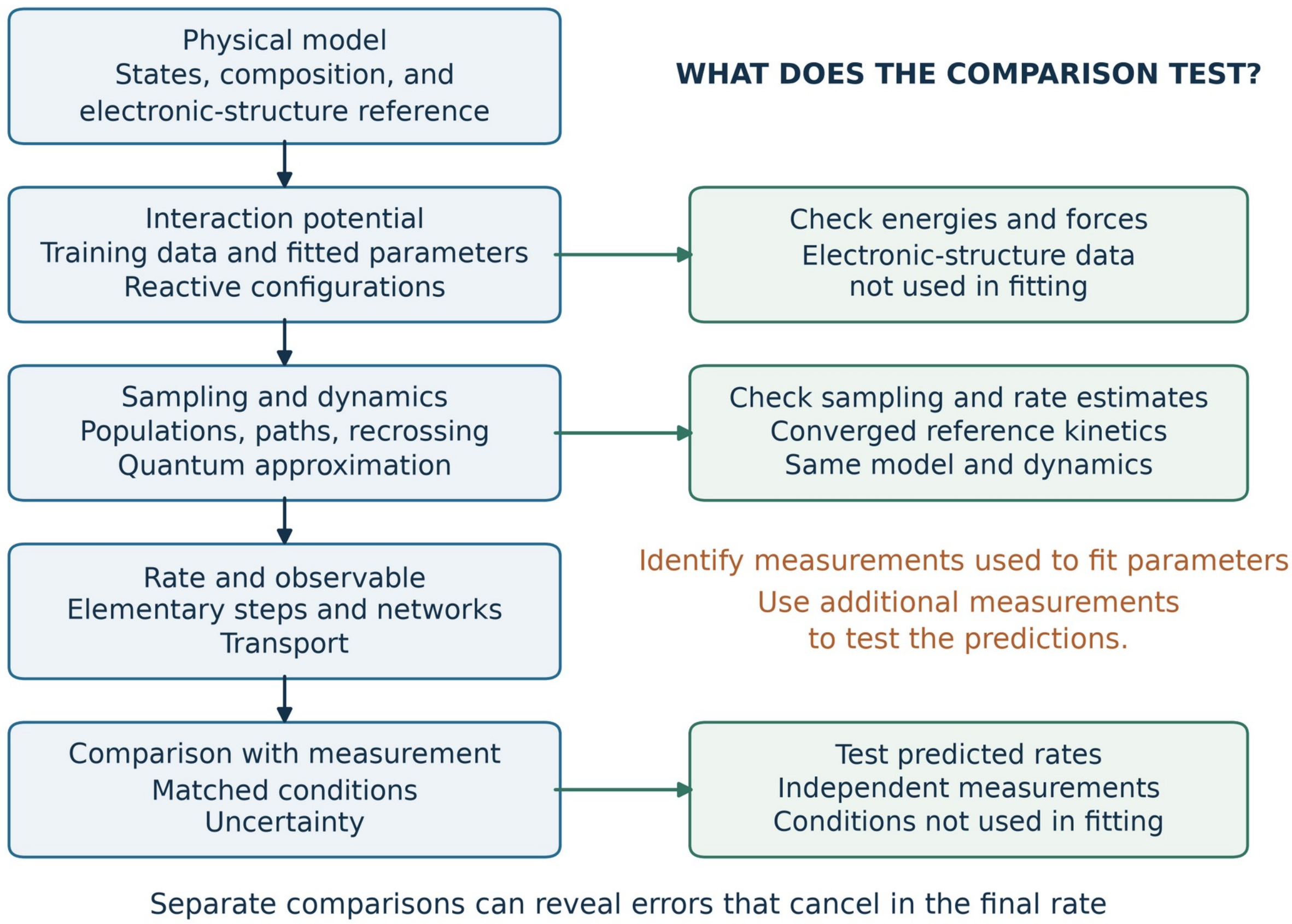

Figure 3. Tests of a rate calculation. Independent electronic-structure energies and forces test the learned potential. Reference kinetics obtained with the same Hamiltonian, state definitions, ensemble, and target dynamics test sampling and rate estimation. Experiment tests the combined molecular model, kinetic calculation, and relation to the measured quantity. Data used to fit a barrier or another parameter do not independently test that fitted quantity; measurements under additional conditions can test its predictions. Comparing the intermediate energies and kinetics can expose compensating errors in the final rate. Sections 7, 8, and S4 give numerical examples. [9, 68, 70, 71, 72, 73, 74, 75, 76]

## 9 Revisiting older rate measurements

Understanding what different rate theories compute can help identify which information is needed to reconsider an industrial reaction or a test of reaction theory. A disagreement between a calculated and measured rate may arise from the assumed populations of reactive species, uncertain thermochemistry, omitted reaction pathways, or the treatment of molecular dynamics. Each possibility calls for different evidence. Later measurements of equilibrium properties or improved calculations may therefore change the interpretation of existing rate data without requiring those rates to be measured again.

The following examples distinguish a published reconciliation, calculations performed here, and proposed tests. Hydrogen–oxygen chain branching shows how independent thermochemistry and reaction dynamics changed the interpretation of older measurements. Pyrophosphate hydrolysis requires a test of solution populations and their reactivities. Triosephosphate isomerase allows a calculation of how an inferred intermediate energy depends on assumptions about microscopic rates. The Criegee example asks what a later optical calibration changes in the inference of rate coefficients. Diamond dissolution motivates experiments that separate temperature history, local surface recession, and changes at a buried interface.

### 9.1 Hydrogen–oxygen chain branching and independent thermochemistry

The chain-branching reaction $\mathrm{H+O_2 \rightleftharpoons OH+O}$ is a key step in combustion. Its forward direction converts a hydrogen atom into an OH radical and an oxygen atom, which can propagate further reactions. A published reconciliation of its calculated and measured rates shows why thermochemistry and reaction dynamics must be checked separately. [77, 78, 79] Harding and co-workers' statistical treatment of energized $HO_2$ appeared consistent with raising the assumed OH formation enthalpy by about 2 kJ $mol^{-1}$ to improve agreement with one group of rate determinations. Independent photoionization measurements and electronic-structure calculations by Ruscic and co-workers instead supported lowering that enthalpy. [80]

Holding the published forward-rate fit fixed isolates the consequence of this thermochemical revision. The equilibrium-constant expressions used in the 2000 and 2001 studies lower the inferred reverse rate by $\exp(-260\,\mathrm{K}/T)$, or 22.9% at 1000 K and 12.2% at 2000 K. Troe and Ushakov also replaced statistical reactive branching with complete classical trajectories on an analytical ab initio potential. In their model, additional nonstatistical return of energized $HO_2$ to O + OH lowered the calculated reverse rate. [78] Thermochemistry therefore changed the reverse rate inferred from experiment, while the trajectory treatment changed the calculated probability of reaction. At 2000 K, the reverse rate inferred from experiment is $1.54 \times 10^{-17}$ $m^3$ $molecule^{-1}$ $s^{-1}$, compared with $1.47 \times 10^{-17}$ $m^3$ $molecule^{-1}$ $s^{-1}$ from the trajectory calculation. Troe subsequently described this reconciliation and its remaining limitations. [79] Section S6.1 reproduces the published expressions and separates the two contributions.

### 9.2 Pyrophosphate hydrolysis and reactive populations

Pyrophosphate contains two phosphate groups joined by a bridging oxygen. Hydrolysis cleaves this P–O–P linkage to produce two orthophosphate ions. Its slow reaction in water provides a benchmark for the acceleration achieved by inorganic pyrophosphatases. Stockbridge and Wolfenden measured hydrolysis in magnesium-containing solutions and attributed the accelerated reaction to the magnesium–pyrophosphate complex. [81] García-Martínez and co-workers later combined a machine-learned potential trained on density-functional calculations with reactive-trajectory sampling and free-energy calculations. They proposed that the more reactive protonated complex, $MgHP_2O_7^-$, contributes substantially to the apparent magnesium acceleration. [82]

Their calculated barriers for $MgP_2O_7^{2-}$ and $MgHP_2O_7^-$ are approximately 138.1 and 118.8 kJ $mol^{-1}$ at 298.15 K. Treating these potential-of-mean-force heights as Eyring activation free energies, with a common prefactor and unit transmission coefficients, gives a conditional rate ratio of approximately 2350. In a mixture restricted to these two complexes, fitting the extrapolated experimental rate requires only about 3% protonated complex. That fitted fraction illustrates the possible importance of a minor species; it does not establish its abundance in the experimental solutions. The authors had already used published equilibrium information to argue that both complexes could be present.

I propose that this interpretation could be tested by predicting the rate for each experimental solution from independently constrained species populations and hydrolysis rates. The calculation must include complexes containing two magnesium ions when they are appreciably populated. Using the 25 °C constants compiled by Baykov and co-workers, the conditional calculation in Section S6 places much of the dissolved pyrophosphate in $Mg_2P_2O_7$ when the total magnesium and pyrophosphate concentrations match one of the reported recipes. [83] The 2026 study supplies no hydrolysis barrier for that complex. Its direct reaction and its effect on the populations of other species therefore need to be calculated or bounded.

Baykov and co-workers had already shown that controlling ionic strength and using equilibrium constants appropriate to the assay medium changed the interpretation of enzyme-catalyzed pyrophosphate hydrolysis. [84] The six constants used here were measured by Shafranskii and co-workers at 25 °C in a tetramethylammonium chloride medium, at a nominal ionic strength of 0.1 M. [85]

Earlier titrations measured the temperature dependence of pyrophosphate protonation, but precipitation prevented measurement of two magnesium-binding constants at 65 °C. [86] Later calorimetry measured apparent magnesium–pyrophosphate binding at 37 °C in HEPES buffer; it did not separate the enthalpy changes for individual protonation and magnesium-binding steps. [87]

The historical measurements were made at elevated temperatures and pooled across conditions that differed in pH, buffer, magnesium and pyrophosphate concentrations, and assay method. A quantitative test requires the individual measurements, solution composition and pH at each incubation temperature, and species-specific rates at those temperatures. The existing room-temperature potentials of mean force and equilibrium constants cannot by themselves supply those predictions. Section S6 sets out the conditional calculations and the additional inputs needed.

### 9.3 Triosephosphate isomerase and uncertainty in an inferred energy profile

Triosephosphate isomerase interconverts dihydroxyacetone phosphate and glyceraldehyde 3-phosphate through an enzyme-bound enediol(ate) intermediate. Toney combined earlier turnover, isotope-effect, viscosity, and intermediate-partition measurements with NMR limits on bound species. His analysis revised parts of the inferred free-energy profile but left the intermediate energy weakly constrained, because rapid decomposition of the intermediate contributed little to the measured turnover. Barr subsequently estimated distributions of the microscopic rates by Bayesian inference. [88, 89]

I examined how strongly the inferred intermediate energy depends on the rate prior, using the implementation associated with the 2022 version of Barr's preprint. The quantity of interest is $\Delta G_{EZ-ES} = G_{EZ} - G_{ES} = -RT\ln(k_2/k_{-2})$, the free energy of the enediol(ate) intermediate relative to the enzyme-bound substrate, evaluated at 298 K. Here $k_2$ and $k_{-2}$ describe forward and reverse conversion between these two bound states. This difference is not an activation barrier. Replacing the shared-scale prior on six microscopic rates by independent bounded log-uniform priors, which give equal weight to each decade of each rate within the stated bounds, shifted the median from 7.3 to 17.2 kJ $mol^{-1}$. The central 90% posterior intervals changed from 6.0–9.1 to 7.8–38.2 kJ $mol^{-1}$. Both calculations used the same input values, likelihood, and parameter bounds. This change measures sensitivity of the inference and does not select a revised physical energy; the resulting interval is conditional on the imposed rate ceiling of $10^{12}$ $s^{-1}$.

Fits without a rate prior, with energy differences fixed at 10 and 25 kJ $mol^{-1}$, placed every model prediction within one supplied standard deviation of its input value. In contrast, the inferred equilibrium ratio of bound product to bound substrate changed little between priors. Later infrared measurements estimated an upper bound on aldehydic glyceraldehyde 3-phosphate relative to keto-dihydroxyacetone phosphate in enzyme/substrate mixtures. [90] Comparing that bound with the modeled bound-product/substrate ratio requires accounting for the chemical forms detected, free and enzyme-bound contributions, and differences in solvent, pH, and temperature. Section S6.4 gives the calculation, including the treatment of an experimental upper limit as a Gaussian input in the archived code. These results concern that 2022 implementation and do not reproduce the posterior reported in the 2023 revision.

### 9.4 Criegee intermediates and optical calibration

When ozone reacts with an alkene, cleavage of its carbon–carbon double bond can produce a carbonyl oxide, known as a Criegee intermediate. These reactive species contribute to atmospheric oxidation. Acetaldehyde oxide, $CH_3CHOO$, has two conformers, *syn* and *anti*, which differ in the orientation of the terminal oxygen relative to the methyl group. They do not readily interconvert at room temperature. Here I consider their self-reactions, involving two molecules of the same conformer, and their cross-reaction, involving one of each. Determining these three rate coefficients requires distinguishing the two conformer concentrations.

Kao and co-workers generated both conformers by photolyzing $CH_3CHI_2$ in oxygen. Ultraviolet absorption at 335 nm detected both, while infrared absorption near 11 μm selectively followed the *syn* conformer; a third probe measured precursor consumption. Analysis of these records gave an *anti* self-reaction coefficient approximately four times the *syn* value. The model also included the cross-reaction between the two conformers. The authors tested different cross-reaction coefficients and assessed uncertainties in concentrations and calibration. These measurements concern reactant loss at laboratory concentrations; the self-reactions need not dominate loss in the atmosphere. [92]

Lin and Lin subsequently provided a direct absolute calibration for the *anti* conformer's ultraviolet spectrum by measuring photodepletion at 352 nm. They calibrated the photon fluence with $NO_2$ and used the reported near-unity photodissociation yield to relate photodissociation to absorption. Their spectrum gives a cross section at 335 nm approximately one quarter larger than that used by Kao and co-workers, although the two values agree within the earlier uncertainty. For a fixed, isolated *anti* absorption signal, the larger cross section implies fewer molecules. In the coupled experiment, however, the optical signals overlap and the conformers react with each other. The inferred rate coefficients cannot all be corrected by one scale factor. [93]

Trac and co-workers combined calculated potential-energy surfaces with variational transition-state and RRKM treatments, including collisional stabilization of the initially formed dimers. Their reported total loss coefficients agreed with the experimental values within the quoted uncertainties. Both stabilization and fragmentation remove the Criegee reactants measured in the experiment. However, the paper's narrative and supplementary tables give inconsistent branching between these channels, so their individual contributions require clarification. [94]

I propose using Lin and Lin's calibration to test how tightly the original records constrain the ratio of the two self-reaction coefficients and the cross-reaction coefficient. A joint fit of the ultraviolet, infrared and precursor records should retain the correlations introduced by the shared calibrations and include uncertainty in initial conformer yields. Comparing otherwise identical fits with the earlier and later calibrations would separate that effect from changes in the kinetic model. The result could confirm the reported ordering while changing its magnitude or uncertainty; it could also show that some coefficients remain poorly determined. Trac's reported total loss coefficients would provide a separate comparison at matched temperature, pressure, and bath gas. This test requires the original numerical records and calibration information, which are not supplied in Kao and co-workers' article and SI, but does not require new electronic-structure calculations.

### 9.5 Diamond dissolution and local recession rates

Writing this review led me to revisit our study of diamond dissolution beneath nickel and cobalt films. [95] Carbon released from diamond at the buried interface passes through the solid metal film and is removed at its exposed surface in the presence of water vapor. Our main interest is Regime II, which we interpreted as dissolution controlled by removal of carbon atoms from diamond at the buried metal–diamond interface. The measured recession rates and their temperature dependence do not establish which atomic arrangements at that interface allow carbon atoms to leave diamond.

Our group is extending the measurements to a wider temperature range and achieving finer control of the sample temperature throughout each run, particularly during rapid heating and cooling with our Joule heating systems. We envision that varying the heating and cooling times while keeping the hold temperature and duration fixed will help determine how the full temperature history affects the amount of diamond removed.

We also plan to map the same diamond surface regions before metal deposition and after dissolution and metal removal, using a laser microscope and possibly AFM. Differences in surface height, measured relative to protected regions, will give local recession depths. Matched specimens exposed for different durations will give average local recession rates over those intervals and show how the surface shape changes.

A further experiment would compare dissolution at a chosen temperature before and after a brief excursion to a higher temperature. Runs held at the chosen temperature for the same total time would show changes occurring without the excursion. Matched specimens stopped at successive stages would let us estimate the rate during each interval, separating recession after return from material removed during the excursion. We would examine whether any persistent rate change accompanies changes in the metal film or buried interface.

The assignment to Regime II also needs testing over the wider temperature range. Comparisons at different water supplies and film thicknesses, together with measurements of film continuity and oxidation, would test whether carbon removal at the exposed surface or transport through the film affects the measured rates. The rates and structural observations could then be compared with models of carbon removal at the buried interface. Section S6.6 gives the proposed comparisons and the quantities they could establish.

## 10 Conclusions and proposed tests

The history of rate theory is not mainly a story of one equation being polished. Its major advances changed the physical description of reaction and the quantities used to calculate a rate. Kramers, for example, made energy exchange with the surroundings part of a stochastic escape problem and obtained corrections to the transition-state estimate.

The numerical comparisons show why agreement must be tested beyond one rate or barrier. For platinum, a fitted barrier improves absolute rate agreement while increasing the spread of calculated-to-experimental ratios across temperatures. For triosephosphate isomerase, the same inputs, likelihood, and bounds give substantially different intermediate-energy distributions under different rate priors. The inferred bound-product/substrate ratio is more stable under that prior change and provides a separate quantity to compare with independent population measurements.

Revisiting older measurements requires identifying which additional information could change their interpretation. The published hydrogen–oxygen reconciliation separates the effects of revised thermochemistry and dynamics. The proposed pyrophosphate test requires species populations and hydrolysis rates under the historical solution conditions, including the contribution of complexes containing two magnesium ions. The Criegee test requires a joint fit of the original optical records using the later calibration. For diamond dissolution, controlled temperature histories and local recession maps, combined with observations of the metal film and buried interface, could test whether persistent rate changes after a temperature excursion accompany structural changes.

### Acknowledgments

I thank Thomas Schultz for reviewing the German-to-English translations and suggesting clearer wording. I thank Wolfgang Bacsa and Revathi Bacsa for reviewing the French-to-English translations and suggesting revisions. I also thank Wolfgang Bacsa for providing René Marcelin’s 1915 *Annales de Physique* paper, “Contribution à l’étude de la cinétique physico-chimique,” and Peter Bøggild for providing Christiansen’s 1921 thesis and reviewing the two Danish-to-English translations.

### Use of artificial intelligence tools

I used ChatGPT-6 Astra Ultra and Claude Cowork Opus 5.5 Extra to assist with literature searches and organization, source and quotation checks, working translations, mathematical and numerical checks, and drafting, critical review, editing, and preparation of the manuscript and Supplementary Materials. I also used ChatGPT in Codex during revision. I reviewed and edited the resulting material and take full responsibility for the accuracy, integrity, interpretation, and final content of the article.

### Author contributions

R. S. Ruoff conceived the review, researched the primary literature, and wrote and edited the manuscript.

### Funding

This work was supported by the Institute for Basic Science [IBS-R019-D1], Republic of Korea.

### Competing interests

The author declares no conflict of interest regarding publication of this article.

### Data availability

The numerical comparisons use published results identified in the text and Supplementary Materials. Section S4.2 provides the data and definitions for the platinum reanalysis, including the comparison of rate magnitudes and temperature dependence and the reconstruction of the reported Arrhenius fit. Section S6 gives the published hydrogen-reaction expressions, the conditional pyrophosphate calculations, the triosephosphate-isomerase model, inputs, and sensitivity analysis, and the proposed Criegee and diamond tests. Supplementary Data 1 (https://doi.org/10.5281/zenodo.23557458) contains the executable enzyme calculation, saved posterior samples, numerical outputs, and code for Figure S1. No new experimental data were generated. Main-text figures are explanatory schematics; Figure S1 presents numerical results.

## References

[1] P. Hänggi, P. Talkner and M. Borkovec, "Reaction-rate theory: fifty years after Kramers", Rev. Mod. Phys., 1990, 62, 251–341. DOI: 10.1103/RevModPhys.62.251.

[2] E. Pollak and P. Talkner, "Reaction rate theory: What it was, where is it today, and where is it going?", Chaos, 2005, 15, 026116. DOI: 10.1063/1.1858782.

[3] K. J. Laidler and M. C. King, "The Development of Transition-State Theory", J. Phys. Chem., 1983, 87, 2657–2664. DOI: 10.1021/j100238a002.

[4] I. Tuñón and I. H. Williams, "The Transition State and Cognate Concepts", Adv. Phys. Org. Chem., 2019, 53, 29–68. DOI: 10.1016/bs.apoc.2019.09.001.

[5] B. Peters, "Common Features of Extraordinary Rate Theories", J. Phys. Chem. B, 2015, 119, 6349–6356. DOI: 10.1021/acs.jpcb.5b02547.

[6] J. Kim, H. Cho, H. Jeon, J. Jung and S. Han, "Reactive Machine Learning Interatomic Potentials for Chemistry and Materials Science", Chem. Rev., 2026, 126, 4467–4510. DOI: 10.1021/acs.chemrev.5c00728.

[7] J. Weare and A. R. Dinner, "Machine learning kinetics from molecular dynamics data", arXiv:2609.17736v1, September 15, 2026. Review preprint. DOI: 10.48550/arxiv.2609.17736. https://arxiv.org/abs/2609.17736.

[8] M. J. McLeod and R. E. Thorne, "Activation parameters, enthalpy–entropy compensation and the temperature-dependent activity of enzymes", Protein Sci., 2026, 35, e70722. DOI: 10.1002/pro.70722.

[9] M. Bello, W. Yang and A. Heyden, "First-Principles Modeling in Heterogenous Catalysis: Where Are We?", ACS Catal., 2026, 16, 9618–9644. DOI: 10.1021/acscatal.6c01630.

[10] D. G. Truhlar, B. C. Garrett and S. J. Klippenstein, “Current Status of Transition-State Theory”, J. Phys. Chem., 1996, 100, 12771–12800. DOI: 10.1021/jp953748q.

[11] R. R. Krug, W. G. Hunter and R. A. Grieger, “Enthalpy-Entropy Compensation. 1. Some Fundamental Statistical Problems Associated with the Analysis of van’t Hoff and Arrhenius Data”, J. Phys. Chem., 1976, 80, 2335–2341. DOI: 10.1021/j100562a006; “Enthalpy-Entropy Compensation. 2. Separation of the Chemical from the Statistical Effect”, J. Phys. Chem., 1976, 80, 2341–2351. DOI: 10.1021/j100562a007.

[12] O. Exner, “Concerning the Isokinetic Relationship”, Nature, 1964, 201, 488–490. DOI: 10.1038/201488b0.

[13] O. Exner, “The Enthalpy-Entropy Relationship in Organic Reactions”, Collect. Czech. Chem. Commun., 1975, 40, 2762–2780. DOI: 10.1135/cccc19752762.

[14] J. H. van ’t Hoff, Études de dynamique chimique, Frederik Muller & Co., Amsterdam, 1884.

[15] S. Arrhenius, “Über die Reaktionsgeschwindigkeit bei der Inversion von Rohrzucker durch Säuren”, Z. Phys. Chem., 1889, 4, 226–248. DOI: 10.1515/zpch-1889-0416.

[16] M. Trautz, “Das Gesetz der Reaktionsgeschwindigkeit und der Gleichgewichte in Gasen. Bestätigung der Additivität von $C_v$ – 3/2 R. Neue Bestimmung der Integrationskonstanten und der Moleküldurchmesser”, Z. Anorg. Allg. Chem., 1916, 96, 1–28. DOI: 10.1002/zaac.19160960102.

[17] W. C. McC. Lewis, “Studies in Catalysis. Part IX. The Calculation in Absolute Measure of Velocity Constants and Equilibrium Constants in Gaseous Systems”, J. Chem. Soc., Trans., 1918, 113, 471–492. DOI: 10.1039/CT9181300471.

[18] J. A. Christiansen, Reaktionskinetiske Studier: Problemer fra de kemiske Reaktioners Kinetik med en eksperimentel Undersøgelse af Fosgenets termiske Spaltning, doctoral (dr. phil.) thesis, University of Copenhagen, Vilhelm Priors Kgl. Hofboghandel, Copenhagen, 1921.

[19] H. Pelzer and E. Wigner, “Über die Geschwindigkeitskonstante von Austauschreaktionen”, Z. Phys. Chem. B, 1932, 15, 445–471. DOI: 10.1515/zpch-1932-1539.

[20] R. C. Tolman, “Statistical Mechanics Applied to Chemical Kinetics”, J. Am. Chem. Soc., 1920, 42, 2506–2528. DOI: 10.1021/ja01457a008.

[21] R. Marcelin, Contribution à l’étude de la cinétique physico-chimique, doctoral thesis, Faculté des sciences de Paris, Gauthier-Villars et Cie, Paris, 1914. Digitized copy, Internet Archive, Leiden University collection: https://archive.org/details/IA41551612_0077.

[22] R. Marcelin, “Contribution à l’étude de la cinétique physico-chimique”, Ann. Phys. (Paris), 1915, série 9, 3, 120–231. DOI: 10.1051/anphys/191509030120.

[23] H. Eyring and M. Polanyi, “Über einfache Gasreaktionen”, Z. Phys. Chem. B, 1931, 12, 279–311; English translation, “On Simple Gas Reactions”, Z. Phys. Chem., 2013, 227, 1221–1245. DOI: 10.1524/zpch.2013.9023.

[24] E. Wigner, “The Transition State Method”, Trans. Faraday Soc., 1938, 34, 29–41. DOI: 10.1039/tf9383400029.

[25] H. Eyring, “The Activated Complex in Chemical Reactions”, J. Chem. Phys., 1935, 3, 107–115. DOI: 10.1063/1.1749604.

[26] H. Eyring, “The Activated Complex and the Absolute Rate of Chemical Reactions”, Chem. Rev., 1935, 17, 65–77. DOI: 10.1021/cr60056a006.

[27] J. O. Hirschfelder and E. Wigner, “Some Quantum-Mechanical Considerations in the Theory of Reactions Involving an Activation Energy”, J. Chem. Phys., 1939, 7, 616–628. DOI: 10.1063/1.1750500.

[28] S. Glasstone, K. J. Laidler and H. Eyring, The Theory of Rate Processes: The Kinetics of Chemical Reactions, Viscosity, Diffusion and Electrochemical Phenomena, McGraw-Hill, New York, 1941.

[29] M. G. Evans and M. Polanyi, “Some applications of the transition state method to the calculation of reaction velocities, especially in solution”, Trans. Faraday Soc., 1935, 31, 875–894. DOI: 10.1039/tf9353100875.

[30] W. F. K. Wynne-Jones and H. Eyring, “The Absolute Rate of Reactions in Condensed Phases”, J. Chem. Phys., 1935, 3, 492–502. DOI: 10.1063/1.1749713; erratum, “Errata: The Absolute Rate of Reactions in Condensed Phases”, J. Chem. Phys., 1936, 4, 740. DOI: 10.1063/1.1749779.

[31] F. A. Lindemann, S. Arrhenius, I. Langmuir, N. R. Dhar, J. Perrin and W. C. McC. Lewis, “Discussion on ‘The Radiation Theory of Chemical Action’”, Trans. Faraday Soc., 1922, 17, 598–606. DOI: 10.1039/TF9221700598.

[32] C. N. Hinshelwood, “On the Theory of Unimolecular Reactions”, Proc. R. Soc. London, Ser. A, 1926, 113, 230–233. DOI: 10.1098/rspa.1926.0149.

[33] O. K. Rice and H. C. Ramsperger, “Theories of Unimolecular Gas Reactions at Low Pressures”, J. Am. Chem. Soc., 1927, 49, 1617–1629. DOI: 10.1021/ja01406a001.

[34] L. S. Kassel, “Studies in Homogeneous Gas Reactions I”, J. Phys. Chem., 1928, 32, 225–242. DOI: 10.1021/j150284a007.

[35] R. A. Marcus, “Unimolecular Dissociations and Free Radical Recombination Reactions”, J. Chem. Phys., 1952, 20, 359–364. DOI: 10.1063/1.1700424.

[36] T. Auth and K. Koszinowski, “From the Modeling of Experiments to the Ab Initio Prediction of Rate Constants: Statistical Rate Theory for a Quantitative Understanding of Gas-Phase Ion Chemistry”, ChemPhysChem, 2026, 27, e70362. DOI: 10.1002/cphc.70362.

[37] M. Salzburger, M. Reimann, J. C. Hartmann, M. Ončák and M. K. Beyer, “A practical guide to RRKM theory, its simplified multi-well version AWATAR and master equation modelling of radiative processes”, Phys. Chem. Chem. Phys., 2026, 28, 12172–12187. DOI: 10.1039/D6CP00705H.

[38] H. A. Kramers, “Brownian motion in a field of force and the diffusion model of chemical reactions”, Physica, 1940, 7, 284–304. DOI: 10.1016/S0031-8914(40)90098-2.

[39] R. F. Grote and J. T. Hynes, “The stable states picture of chemical reactions. II. Rate constants for condensed and gas phase reaction models”, J. Chem. Phys., 1980, 73, 2715–2732. DOI: 10.1063/1.440485.

[40] R. A. Marcus, “On the Theory of Oxidation-Reduction Reactions Involving Electron Transfer. I”, J. Chem. Phys., 1956, 24, 966–978. DOI: 10.1063/1.1742723.

[41] P. Siders and R. A. Marcus, “Quantum Effects for Electron-Transfer Reactions in the ‘Inverted Region’”, J. Am. Chem. Soc., 1981, 103, 748–752. DOI: 10.1021/ja00394a004.

[42] M. Zhang, Y. Chen, M. M. Melander and J. Huang, “Electrochemical Electron Transfer: Key Concepts, Theories, and Parameterization via Atomistic Simulations”, Chem. Rev., 2026, 126, 7407–7464. DOI: 10.1021/acs.chemrev.5c00926.

[43] G. H. Vineyard, “Frequency factors and isotope effects in solid state rate processes”, J. Phys. Chem. Solids, 1957, 3, 121–127. DOI: 10.1016/0022-3697(57)90059-8.

[44] E. Wigner, “Calculation of the Rate of Elementary Association Reactions”, J. Chem. Phys., 1937, 5, 720–725. DOI: 10.1063/1.1750107.

[45] J. Horiuti, “On the Statistical Mechanical Treatment of the Absolute Rate of Chemical Reaction”, Bull. Chem. Soc. Jpn., 1938, 13, 210–216. DOI: 10.1246/bcsj.13.210.

[46] J. C. Keck, “Variational Theory of Chemical Reaction Rates Applied to Three-Body Recombinations”, J. Chem. Phys., 1960, 32, 1035–1050. DOI: 10.1063/1.1730846.

[47] P. Pechukas and F. J. McLafferty, “On transition-state theory and the classical mechanics of collinear collisions”, J. Chem. Phys., 1973, 58, 1622–1625. DOI: 10.1063/1.1679404.

[48] B. C. Garrett and D. G. Truhlar, “Generalized Transition State Theory. Canonical Variational Calculations Using the Bond Energy-Bond Order Method for Bimolecular Reactions of Combustion Products”, J. Am. Chem. Soc., 1979, 101, 5207–5217. DOI: 10.1021/ja00512a016.

[49] J. L. Bao and D. G. Truhlar, “Variational transition state theory: theoretical framework and recent developments”, Chem. Soc. Rev., 2017, 46, 7548–7596. DOI: 10.1039/c7cs00602k.

[50] T. Yamamoto, “Quantum Statistical Mechanical Theory of the Rate of Exchange Chemical Reactions in the Gas Phase”, J. Chem. Phys., 1960, 33, 281–289. DOI: 10.1063/1.1731099.

[51] D. Chandler, “Statistical mechanics of isomerization dynamics in liquids and the transition state approximation”, J. Chem. Phys., 1978, 68, 2959–2970. DOI: 10.1063/1.436049.

[52] W. H. Miller, S. D. Schwartz and J. W. Tromp, “Quantum mechanical rate constants for bimolecular reactions”, J. Chem. Phys., 1983, 79, 4889–4898. DOI: 10.1063/1.445581.

[53] I. R. Craig and D. E. Manolopoulos, “Chemical reaction rates from ring polymer molecular dynamics”, J. Chem. Phys., 2005, 122, 084106. DOI: 10.1063/1.1850093.

[54] P. G. Bolhuis, D. Chandler, C. Dellago and P. L. Geissler, “Transition Path Sampling: Throwing Ropes Over Rough Mountain Passes, in the Dark”, Annu. Rev. Phys. Chem., 2002, 53, 291–318. DOI: 10.1146/annurev.physchem.53.082301.113146.

[55] W. E and E. Vanden-Eijnden, “Towards a Theory of Transition Paths”, J. Stat. Phys., 2006, 123, 503–523. DOI: 10.1007/s10955-005-9003-9.

[56] E. Wigner, “Über das Überschreiten von Potentialschwellen bei chemischen Reaktionen”, Z. Phys. Chem. B, 1932, 19, 203–216. DOI: 10.1515/zpch-1932-1920.

[57] C. Eckart, “The Penetration of a Potential Barrier by Electrons”, Phys. Rev., 1930, 35, 1303–1309. DOI: 10.1103/PhysRev.35.1303.

[58] R. P. Bell, “The Application of Quantum Mechanics to Chemical Kinetics”, Proc. R. Soc. London, Ser. A, 1933, 139, 466–474. DOI: 10.1098/rspa.1933.0031.

[59] R. P. Bell, “The tunnel effect correction for parabolic potential barriers”, Trans. Faraday Soc., 1959, 55, 1–4. DOI: 10.1039/tf9595500001.

[60] W. H. Miller, “Quantum mechanical transition state theory and a new semiclassical model for reaction rate constants”, J. Chem. Phys., 1974, 61, 1823–1834. DOI: 10.1063/1.1682181.

[61] G. A. Voth, D. Chandler and W. H. Miller, “Rigorous formulation of quantum transition state theory and its dynamical corrections”, J. Chem. Phys., 1989, 91, 7749–7760. DOI: 10.1063/1.457242.

[62] J. O. Richardson and S. C. Althorpe, “Ring-polymer molecular dynamics rate-theory in the deep-tunneling regime: Connection with semiclassical instanton theory”, J. Chem. Phys., 2009, 131, 214106. DOI: 10.1063/1.3267318.

[63] C. Dellago, P. G. Bolhuis, F. S. Csajka and D. Chandler, “Transition path sampling and the calculation of rate constants”, J. Chem. Phys., 1998, 108, 1964–1977. DOI: 10.1063/1.475562.

[64] W. E, W. Ren and E. Vanden-Eijnden, “String method for the study of rare events”, Phys. Rev. B, 2002, 66, 052301. DOI: 10.1103/PhysRevB.66.052301.

[65] L. Maragliano, A. Fischer, E. Vanden-Eijnden and G. Ciccotti, “String method in collective variables: Minimum free energy paths and isocommittor surfaces”, J. Chem. Phys., 2006, 125, 024106. DOI: 10.1063/1.2212942.

[66] R. G. Mullen, J.-E. Shea and B. Peters, “Transmission Coefficients, Committors, and Solvent Coordinates in Ion-Pair Dissociation”, J. Chem. Theory Comput., 2014, 10, 659–667. DOI: 10.1021/ct4009798.

[67] M. J. Knapp, K. Rickert and J. P. Klinman, “Temperature-Dependent Isotope Effects in Soybean Lipoxygenase-1: Correlating Hydrogen Tunneling with Protein Dynamics”, J. Am. Chem. Soc., 2002, 124, 3865–3874. DOI: 10.1021/ja012205t.

[68] L. Zhang, F. Nitz, D. Borodin, A. M. Wodtke and H. Guo, “Ring Polymer Molecular Dynamics Rates for Hydrogen Recombinative Desorption on Pt(111)”, Precis. Chem., 2025, 3, 319–325. DOI: 10.1021/prechem.4c00099.

[69] Y. Yang, S. Zhang, K. D. Ranasinghe, O. Isayev and A. E. Roitberg, “Machine Learning of Reactive Potentials”, Annu. Rev. Phys. Chem., 2024, 75, 371–395. DOI: 10.1146/annurev-physchem-062123-024417.

[70] M. Ferrer, J. Munarriz, T. Stuyver and R. Laplaza, “The Future of Foundation Machine Learning Potentials and DFT in Homogeneous Catalysis: Competition or Synergy?”, Chem. Eur. J., 2026, 32, e71022. DOI: 10.1002/chem.71022.

[71] I. K. Park, S. Pourasad, J. Mun, A. C. Rajan, T. H. Park, A. Rohit, M. Zafari, T.-L. Pham, E. Jung, B. Koo, M. Ha, G. B. Sim, H. G. Park, W. W. Choi, K. Jo, S. Y. Willow, D. C. Yang, C. W. Myung, G. Lee and K. S. Kim, “Machine Learning Interatomic Potentials for Energy Materials: Architectures, Training Strategies, and Applications”, Adv. Energy Mater., 2026, e71046. DOI: 10.1002/aenm.71046.

[72] X. Gui, W. Fan, J. Sun and Y. Li, “New Stable and Fast Ring-Polymer Molecular Dynamics for Calculating Bimolecular Rate Coefficients with an Example of OH + $CH_4$”, J. Chem. Theory Comput., 2022, 18, 5203–5212. DOI: 10.1021/acs.jctc.2c00522.

[73] J. Li and H. Guo, “Communication: An accurate full 15 dimensional permutationally invariant potential energy surface for the OH + $CH_4$ → $H_2O$ + $CH_3$ reaction”, J. Chem. Phys., 2015, 143, 221103. DOI: 10.1063/1.4937570.

[74] S. Vari, C. de Falco and C. Cavallotti, “NN-VRCTST: Neural Network Potentials Meet Variable Reaction Coordinate Transition State Theory for the Rate Constant Determination of Barrierless Reactions”, J. Chem. Theory Comput., 2025, 21, 10373–10384. DOI: 10.1021/acs.jctc.5c01288.

[75] G. Lazzeri, H. Jung, P. G. Bolhuis and R. Covino, “Molecular Free Energies, Rates, and Mechanisms from Data-Efficient Path Sampling Simulations”, J. Chem. Theory Comput., 2023, 19, 9060–9076. DOI: 10.1021/acs.jctc.3c00821.

[76] O. Blumer, S. Reuveni and B. Hirshberg, “Short-Time Infrequent Metadynamics for Improved Kinetics Inference”, J. Chem. Theory Comput., 2024, 20, 3484–3491. DOI: 10.1021/acs.jctc.4c00170.

[77] L. B. Harding, A. I. Maergoiz, J. Troe and V. G. Ushakov, “Statistical rate theory for the HO + O $\rightleftharpoons$ $HO_2$ $\rightleftharpoons$ H + $O_2$ reaction system: SACM/CT calculations between 0 and 5000 K”, J. Chem. Phys., 2000, 113, 11019–11034. DOI: 10.1063/1.1314374.

[78] J. Troe and V. G. Ushakov, “Theoretical studies of the HO + O $\rightleftharpoons$ $HO_2$ $\rightleftharpoons$ H + $O_2$ reaction. II. Classical trajectory calculations on an ab initio potential for temperatures between 300 and 5000 K”, J. Chem. Phys., 2001, 115, 3621–3628. DOI: 10.1063/1.1388201.

[79] J. Troe, “The Struggle for Precise Rate Constants in Gas Phase Reaction Kinetics: The Reaction H + $O_2$ $\rightleftharpoons$ HO + O”, Z. Phys. Chem., 2003, 217, 1303–1317. DOI: 10.1524/zpch.217.10.1303.20494.

[80] B. Ruscic, D. Feller, D. A. Dixon, K. A. Peterson, L. B. Harding, R. L. Asher and A. F. Wagner, “Evidence for a Lower Enthalpy of Formation of Hydroxyl Radical and a Lower Gas-Phase Bond Dissociation Energy of Water”, J. Phys. Chem. A, 2001, 105, 1–4. DOI: 10.1021/jp003711s.

[81] R. B. Stockbridge and R. Wolfenden, “Enhancement of the Rate of Pyrophosphate Hydrolysis by Nonenzymatic Catalysts and by Inorganic Pyrophosphatase”, J. Biol. Chem., 2011, 286, 18538–18546. DOI: 10.1074/jbc.M110.214510.

[82] A. García-Martínez, M. de la Puente, M. Cunin, K. Zinovjev, J. J. Ruiz-Pernía, D. Laage and I. Tuñón, “Protonation and magnesium ions shape the transition state diversity of phosphoanhydride hydrolysis in water”, Nat. Commun., 2026, 17, 6878. DOI: 10.1038/s41467-026-73697-7.

[83] A. A. Baykov, V. A. Anashkin and A. M. Malinen, “Good-Practice Non-Radioactive Assays of Inorganic Pyrophosphatase Activities”, Molecules, 2021, 26, 2356. DOI: 10.3390/molecules26082356.

[84] A. A. Baykov, N. P. Bakuleva and P. A. Rea, “Steady-state kinetics of substrate hydrolysis by vacuolar $H^+$-pyrophosphatase. A simple three-state model”, Eur. J. Biochem., 1993, 217, 755–762. DOI: 10.1111/j.1432-1033.1993.tb18303.x.

[85] Yu. A. Shafranskii, A. A. Baykov, P. F. Andrukovich and S. M. Avaeva, “Comparative kinetic studies of $Mg^{2+}$-activated hydrolysis of tripolyphosphate and pyrophosphate by inorganic pyrophosphatase”, Biokhimiia, 1977, 42, 1244–1251. In Russian, with an English abstract. PMID: 20164. No DOI identified.

[86] R. R. Irani, “Metal Complexing by Phosphorus Compounds. V. Temperature Dependence of Acidity and Magnesium Complexing Constants”, J. Phys. Chem., 1961, 65, 1463–1465. DOI: 10.1021/j100826a516.

[87] C. J. Zea, G. Camci-Unal and N. L. Pohl, “Thermodynamics of binding of divalent magnesium and manganese to uridine phosphates: implications for diabetes-related hypomagnesaemia and carbohydrate biocatalysis”, Chem. Cent. J., 2008, 2, 15. DOI: 10.1186/1752-153X-2-15.

[88] M. D. Toney, “Common Enzymological Experiments Allow Free Energy Profile Determination”, Biochemistry, 2013, 52, 5952–5965. DOI: 10.1021/bi400696j.

[89] I. Barr, “A Bayesian Analysis of Steady–State Enzyme Data leads to Estimates of Rate Constants and Uncertainties in a Multi-Step Reaction, along with Free Energy Profiles”, bioRxiv, version 4, 26 July 2023. DOI: 10.1101/2021.08.04.454956. The calculation in Section S6.4 uses archived code obtained from version 3, posted 19 September 2022.

[90] H. Deng, J. Vedad, R. Z. B. Desamero and R. Callender, “Difference FTIR Studies of Substrate Distribution in Triosephosphate Isomerase”, J. Phys. Chem. B, 2017, 121, 10036–10045. DOI: 10.1021/acs.jpcb.7b08114.

[91] D. Foreman-Mackey, D. W. Hogg, D. Lang and J. Goodman, “emcee: The MCMC Hammer”, Publ. Astron. Soc. Pac., 2013, 125, 306–312. DOI: 10.1086/670067.

[92] T.-Y. Kao, C.-A. Chung and Y.-P. Lee, “Rate Coefficient and Branching Ratio for the Formation of Criegee Intermediate Syn-/Anti-$CH_3CHOO$ from $CH_3CHI$ + $O_2$ and the Self-Reaction of Syn-/Anti-$CH_3CHOO$ Determined with Simultaneous IR/UV Probes”, J. Phys. Chem. A, 2024, 128, 9453–9461. DOI: 10.1021/acs.jpca.4c06588.

[93] Y.-H. Lin and J. J.-M. Lin, “Absolute photodissociation cross section of anti-$CH_3CHOO$”, Phys. Chem. Chem. Phys., 2026, 28, 24422–24430. DOI: 10.1039/D6CP02901A.

[94] H.-P. Trac, P. Raghunath and M.-C. Lin, “Ab Initio Chemical Kinetics for Self- and Cross-Reactions of anti- and syn-$CH_3CHOO$ Conformers”, J. Phys. Chem. A, 2026, 130, 2595–2601. DOI: 10.1021/acs.jpca.6c00127.

[95] Y. Li, Y. Kim, P. V. Bakharev, W. K. Seong, C. Hyun, D. C. Camacho-Mojica, L. Zhang, B. V. Cunning, T. J. Shin, G. Lee and R. S. Ruoff, “Dissolving Diamond: Kinetics of the Dissolution of (100) and (110) Single Crystals in Nickel and Cobalt Films”, Chem. Mater., 2022, 34, 2599–2611. DOI: 10.1021/acs.chemmater.1c03809.

# What Rate Theories Compute

*Supplementary Materials*

Rodney S. Ruoff

Historical source evidence, supplementary derivations, numerical comparisons, and diagnostic procedures

These Supplementary Materials provide the historical sources discussed in the main article (S2), supplementary derivations (S3), numerical comparisons of five modern rate calculations (S4), procedures for testing rate-theory assumptions (S5), and reconstructions, conditional calculations, an enzyme-inference sensitivity analysis, and proposed tests for revisiting rate measurements (S6). Reference numbers follow the main article. Each subsection in S2 identifies the corresponding main-text sections.

## S1 Source conventions and scope

Unless stated otherwise, page numbers refer to the printed source. I prepared the translations from French, German, and Danish. Selected English translations of German passages incorporate wording suggested by Thomas Schultz. Selected English translations of French passages incorporate suggestions from Wolfgang Bacsa and Revathi Bacsa. Peter Bøggild reviewed the two Danish-to-English translations from Christiansen's thesis. Quotations from Eyring and Polanyi use the 2013 English republication and its pagination. Marcelin's doctoral thesis was published in Paris by Gauthier-Villars et Cie in 1914; the digitized copy used here is in the Internet Archive's Leiden University collection (reference [21] of the main text). Marcelin page numbers refer to this thesis unless the 1915 Annales de Physique republication is specified. In Part I, adding 119 to a thesis page number gives the corresponding page in the 1915 republication. Original-language wording is retained except for the bracketed unit conversions in Section S2.6. Numerical energy values are expressed in kJ $mol^{-1}$, with editorial conversions in brackets.

## S2 Historical source record

### S2.1 Van 't Hoff and Arrhenius

Main text Sections 2 and 3. Sources: [14, 15].

Van 't Hoff derived a temperature dependence for rate constants in his 1884 Études de dynamique chimique and wrote it in logarithmic form (pp. 115–116). Arrhenius used the exponential form and credited Van 't Hoff for the chloroacetic-acid case (Arrhenius, pp. 227, 229). This kinetic relation differs from the Van 't Hoff relation for an equilibrium constant. The latter has the modern pressure-standard form $d\ln K^{\circ}/dT = \Delta H^{\circ}/(RT^2)$; its historical fixed-volume form uses the reaction internal-energy change. On p. 115 of the *Études*, Van 't Hoff applies the isochore to the forward and reverse rates to obtain a relation for a rate constant:

"Cette équation n'établit pas la relation cherchée entre la température et la valeur de $k$, mais elle montre que cette relation aura la forme suivante: $d.$ log. $k$ / $d.$ $T = A/T^2 + B$", "This equation does not establish the desired relation between temperature and the value of $k$, but shows that this relation will have the following form: $\frac{d \log k}{d T} = \frac{A}{T^2} + B$" (p. 115).

Van 't Hoff fitted three reactions with two special cases of his expression: $A = 0$ for dibromosuccinic and chloroacetate reactions, and $B = 0$ for chloroacetic acid. He described these fits as "une première approximation", "a first approximation" (p. 116), and explained the restriction in a footnote:

"Les grandeurs $A$ et $B$ pourraient être des fonctions de $T$; en ce qui regarde $A$, cela est même très-probable, parce que $A$ est en rapport avec la chaleur de transformation $q$, qui varie, quoique légèrement, avec la température", "The quantities $A$ and $B$ could be functions of $T$; for $A$, this is particularly likely, since $A$ is related to the heat of transformation $q$, which varies, although only slightly, with temperature" (p. 117, note 1).

Arrhenius proposed a chemically modified form of cane sugar:

"…sondern ein anderer hypothetischer Körper ist, welcher aus Rohrzucker wieder entsteht, sobald er durch die Inversion beseitigt wird. Dieser hypothetische Körper, nennen wir ihn „aktiven Rohrzucker", muss mit zunehmender Temperatur sehr schnell zunehmen (um etwa 12 Prozent pro Grad) auf Kosten des gewöhnlichen „inaktiven" Rohrzuckers", "…but is another hypothetical body, which, if removed by inversion, is immediately reformed out of cane sugar. This hypothetical body, let us call it 'active cane sugar', must increase very rapidly with rising temperature (by about 12 percent per degree) at the expense of the ordinary 'inactive' cane sugar" (p. 233).

Lower on the same page he says how it arises: "entweder durch Umlagerung der Atome oder durch Eintritt von Wasser", "either through rearrangement of the atoms or through entry of water" (p. 233). He thus proposed an isomer or hydrate. Interpreting the active form as an energetically selected fraction of unchanged reactant molecules is a later generalization. He extended his hypothesis to other reactions only by analogy (p. 234).

Arrhenius acknowledged that the temperature dependence alone did not determine a mechanism:

"Ich bin daher der Ansicht, dass man in dieser Weise keine bestimmte Lösung des Problems bekommt, da ja $F(T)$ ganz beliebig sein kann. Man kann daher nicht weiter kommen, ohne eine neue Hypothese einzuführen, welche gewissermassen eine Umschreibung der gefundenen Thatsachen ist", "I am therefore of the opinion that no definite solution of the problem can be obtained in this manner, since $F(T)$ can be entirely arbitrary. One cannot proceed without introducing a new hypothesis, which is to some extent a paraphrase of the established facts" (p. 231).

He compared the 10 to 15 percent increase in rate per degree with the much smaller increase predicted for collision frequency. By kinetic theory, molecular velocity changes "nur um etwa $^{1}/_{6}$ $^{0}/_{0}$ des Wertes bei einer Temperatursteigerung von 1$^{0}$", "only by about ⅙ percent of its value for a temperature rise of 1°"; and the discrepancy is "viel zu gross […], als dass man die Zunahme der Zusammenstösse der reagierenden Moleküle als Grund der Zunahme der Reaktionsgeschwindigkeit mit der Temperatur gut annehmen könnte", "far too great to assume that the increase in the collisions of the reacting molecules causes the increase of the reaction velocity with temperature" (pp. 231–232).

He also allowed the activation parameter to vary with temperature:

"Kleinere Abweichungen von der Formel (1) können dadurch erklärt werden, dass $A$, die halbe Umsetzuugswärme [sic] von der aktiven Modifikation in die inaktive, im Allgemeinen als eine Funktion von der Temperatur anzusehen ist, welche nur in erster Annäherung als konstant [...] zu setzen ist", "Smaller deviations from formula (1) can be explained by the fact that $A$, the half heat of conversion from the active to the inactive modification, should generally be regarded as a function of the temperature, which can be assumed constant [...] only as a first approximation" (p. 234).

Arrhenius did not calculate the active population or its transformation rate from a molecular energy model. Marcelin supplied formal phase integrals in 1914; later electronic-structure and spectroscopic work enabled molecular calculations of absolute rates.

### S2.2 Collision theory and molecular orientation

Main text Section 3. Sources: [16, 17, 18, 19, 20].

Trautz (1916) and Lewis (1918) independently tried to compute absolute rates from collision frequencies and molecular sizes, but they assessed the limitations differently.

Trautz identified a discrepancy between collision frequency and reaction rate: "Ferner ist der Betrag der Reaktionsgeschwindigkeit viel kleiner, als es der Stoßzahl entspräche", "Furthermore the magnitude of the reaction velocity is much smaller than the corresponding collision number" (p. 5). In a separate comparison, he removed the activation exponential from his empirical–thermodynamic expression and kept only the translational part of its heat capacities (p. 5). He then writes: "Die beiden Zahlen sind ihrem Betrag nach um ungefähr 7 Zehnerpotenzen verschieden", "The two numbers differ in magnitude by about 7 powers of ten" (p. 6). The adjusted empirical–thermodynamic value is the larger one. He also names what his gas-theoretical expression omits:

"Denn in der gastheoretischen Formel wurde k e i n e Aktivierungswärme angenommen und die zusammenstoßenden Moleküle sollen keine innere Energie enthalten, die auf Zahl oder Erfolg der Zusammenstöße Einfluß übt", "For no heat of activation was assumed in the gas-theoretical formula, and the colliding molecules are supposed to contain no internal energy that affects the number or the success of the collisions" (p. 5).

He proposed the following correction:

"Um $Z_g$ überhaupt an der Erfahrung prüfen zu können, muß man ihm ein Glied $e^{-\frac{q_0}{RT}}$ zufügen, das dem Umstand Rechnung trägt, daß nicht jeder Stoß von Reaktion gefolgt ist", "Any test of $Z_g$ against experience requires the addition of a term $e^{-\frac{q_0}{RT}}$, which accounts for the circumstance that not every collision is followed by a reaction" (p. 7).

He added a Boltzmann factor and concluded that the calculation required only two substance-specific constants: "der Kenntnis der beiden Stoffkonstanten: Aktivierungswärme und Moleküldurchmesser”, “knowledge of the two substance-specific constants: heat of activation and molecular diameter” (p. 28). His reconciliation also required heat-capacity additivity and a reassignment of temperature factors (pp. 6–8). His final expression contains no separate factor for collisions that meet the energy requirement but fail to react.

Lewis stated in both the body and summary that “at each collision between active molecules, two such molecules react”, and that the chemical efficiency should in theory approach unity at sufficiently high temperature (pp. 473, 475, 491). This assumes unit reaction probability for a collision between active molecules. He lacked an independent collision diameter: “physical theory has not yet succeeded in defining σ with precision, beyond the fact that it is of the order of magnitude of the radius or diameter of the molecule”. He adopted a single mean value, acknowledging that this “will introduce a certain error into the results” and that agreement “is evidently limited mainly by the degree of approximation in the value taken for σ” (pp. 473, 476). He fitted the diameter in two cases, although his proposed calculation required independently determined molecular properties.

Neither the Trautz paper nor the Lewis paper includes a separate orientation or steric factor. Christiansen discusses orientation explicitly in 1921: for the reaction of nitric oxide with oxygen he writes that the orientation of the molecules at the moment of collision can differ greatly, so that “det kun er en vis Brøkdel af NO, NO-Stødkomplekserne, der er i Stand til at reagere”, “it is only a certain fraction of the NO, NO collision complexes that is able to react” (p. 21). [18] Pelzer and Wigner later estimated a steric factor of “~ 1/40” at about 300 K from the energy cost of approaching the saddle point off the line of centers, and noted that experiment gives about five times as much (pp. 463–464). [19]

Tolman described the assumption used in these early collision models in 1920: reaction “is sure to occur” once the colliding molecules have “certain minimum energy contents” and “have come into contact in the sense of the kinetic theory”, and he noted that hypotheses of this kind “have led to useful considerations in the hands of Trautz and W. C. M. Lewis” (p. 2528). He added that a satisfactory calculation required the dependence of molecular energy on coordinates and momenta (p. 2528).

### S2.3 Tolman and the activation-energy interpretation

Main text Sections 2 and 3. Source: [20].

Tolman defined the critical increment as a difference of mean energies. For a photochemical reaction, he described the temperature coefficient as “proportional to the difference between the energy of the molecules which actually react and the average energy of all the molecules of that kind” (p. 2524). In the thermal case, he averaged over molecules and radiation modes because of the activation mechanisms he considered. Omitting radiation modes from his general thermal relation gives the difference between molecular mean energies used in main text Section 3.1. Tolman explicitly allowed this case for polymolecular reactions (pp. 2525–2527).

Tolman also credited Marcelin’s priority: “Marcelin was the first to present any elaborate attempt to apply the methods of statistical mechanics to the problems of reaction velocity” (p. 2508).

### S2.4 Marcelin and the phase-space formulation

Main text Section 3. Sources: [21, 22, 23, 25].

In his 1914 Paris thesis, Marcelin defines a critical surface in the phase space of a molecule, writes the rate as a Gibbsian flux of representative points through that surface, supplies the normalizing phase integral over the reactant region, and defines what he calls the critical affinity through a phase integral over the dividing surface:

"J'appellerai la grandeur A, ainsi définie, l'*affinité critique*. […] D'où la conclusion : *Il existe entre l'affinité et l'énergie critiques une relation semblable à celle qui lie l'affinité et l'énergie,* ces mots étant pris cette fois dans leur sens ordinaire", "I shall call the quantity *A*, defined in this way, the *critical affinity*. […] Hence the conclusion: *There is a relation between critical affinity and critical energy analogous to that between affinity and energy,* with these terms now used in their conventional sense" (pp. 63–64, note 1).

This footnote anticipates the thermodynamic formulation of transition-state theory two decades before Eyring, although it does not supply the universal prefactor.

Marcelin also allowed only a fraction of crossings to produce transformation:

"Parmi toutes les molécules qui traversent une région de la surface critique, il n'y en a qu'une fraction α qui se transforme; cette particularité ne complique en rien la démonstration, car on trouve encore pour la dérivée logarithmique de $\int_{S_1} \alpha\, exp.(-\frac{\mu\varepsilon}{RT})\, d\varpi$, l'énergie critique U",

"Among all the molecules that cross a region of the critical surface, only a fraction $\alpha$ undergoes transformation; this does not complicate the derivation in any way, since the logarithmic derivative of $\int_{S_1} \alpha\, exp.(-\frac{\mu\varepsilon}{RT})\, d\varpi$ still yields the critical energy $U$" (p. 53, note 1).

For the quoted surface integral $I_\alpha$, $RT^2 d \ln I_\alpha / dT$ gives the corresponding flux-weighted molar mean energy when the molecular energies, surface geometry, and transmission weighting are temperature-independent. Here $\mu$ is the number of molecules per mole. Variation of $\alpha$ over the surface changes the weighting; a common temperature-independent multiplier cancels from the logarithmic derivative. Temperature-dependent transmission adds another derivative term. Marcelin separately used a swarm of bees to explain why removing molecules after an outward crossing need not appreciably deplete the equilibrium population supplying the surface:

"A chaque instant, le rapport du nombre d'abeilles actives au nombre total d'*abeilles vivantes* ne sera altéré que d'une manière insignifiante par la suppression des individus qui, ayant traversé la surface critique dans un sens, auraient pu y revenir en sens inverse. Le raisonnement est d'autant plus exact que le rayon R est plus grand, il perdrait toute valeur si la sphère coupait le gros de l'essaim", "At any given instant, removing individuals that have crossed the critical surface in one direction and could have returned across it in the opposite direction will change the ratio of active bees to all *living bees* only slightly. This reasoning becomes more accurate as the radius *R* increases; it would lose its validity if the sphere cut through the main body of the swarm" (p. 47).

The analogy concerns replenishment of the active population after outward crossings. Marcelin excludes inward-moving bees on p. 46 and gives no bound on a dynamical transmission coefficient.

He also defines the difference between the flux-weighted mean energy of molecules crossing the critical surface and the mean energy of the whole population, and names it: “c’est ce que j’appellerai l’énergie critique relative”, “this is what I shall call the relative critical energy” (p. 53). His logarithmic temperature law equates the corresponding molar energy difference with $R\,T^2 d\ln k_1/d\,T$ for a temperature-independent critical surface and temperature-independent transmission weighting. This gives the unimolecular form of Tolman’s 1920 result six years earlier. The equality to a constant Arrhenius activation energy requires the difference of means itself to be temperature-independent.

Marcelin did not evaluate the phase integrals for a molecular reaction or obtain explicit molecular partition-function factors or a universal prefactor. He emphasized the generality of his derivation:

“Ce qui fait l’intérêt de cette remarque, c’est que les raisonnements que nous ferons dans la suite de ce Chapitre seront, précisément indépendants de la nature des molécules, de la forme de la surface S et de l’expression détaillée de *d*λ”, “This observation is of interest because the arguments developed in the rest of this chapter will be independent of the nature of the molecules, the shape of the surface *S*, and the detailed expression for $d\,\lambda$” (p. 38, note 1).

Later work combined potential-energy surfaces built partly from spectroscopic data, as in Eyring and Polanyi’s 1931 surface, [23] with statistical calculations of molecular partition functions, as in Eyring’s 1935 paper. [25] The experimental second half of Marcelin’s thesis compares evaporation velocities into vacuum with a Hertz–Knudsen bound; it does not evaluate a phase integral. Marcelin was killed in action in September 1914 (reference [22], p. 120 of the 1915 republication).

### S2.5 Unimolecular activation and state counting

Main text Section 3. Sources: [18, 31, 32, 33, 34, 35].

Lindemann separated collisional activation from subsequent unimolecular reaction in the 1922 Faraday Society discussion. [31] His proposal followed his criticism of radiation activation: “After all this criticism, I suppose I had better put forward a constructive suggestion,” with “at present of course there is no more evidence in favour of this view than of any other” (p. 599). He assumed that “dissociation can only take place if the centrifugal force due to rotation is above a certain value,” without claiming that criterion as new. The molecule “does not dissociate instantly but on the average only after a time greater than that necessary to re-establish the Maxwell distribution”, allowing the activated population to remain at equilibrium and the kinetics to remain first order. He specified a test: “there must be a limit beyond which the experimental law that reaction velocity is independent of concentration, ceases to hold, namely at concentrations at which the time between collisions is of the same order as the average life of the molecule after attaining its critical velocity”. Low-pressure measurements would be “an interesting check on the theory” (p. 599). He did not formulate a steady-state balance between reaction and collisional deactivation.

Christiansen had described deactivation during the delay in his 1921 thesis. [18] If another collision precedes rearrangement, “saa vil der være en vis Sandsynlighed for, at Molekulet inaktiveres ved Stødet”, “there will be a certain probability that the molecule is inactivated by the collision” (p. 15, note 1; English translations of the two Christiansen passages reviewed by Peter Bøggild). Christiansen used this argument to challenge Perrin’s claim that truly monomolecular reactions must be photochemical, without developing a kinetic scheme. His 1921 thesis and Lindemann’s 1922 proposal both separated collisional activation from the subsequent reaction.

Hinshelwood, Rice and Ramsperger, and Kassel treated energy stored in internal molecular modes as available for reaction. [32, 33, 34] At energy $E$, reaction depends on the probability that sufficient energy reaches the reactive coordinate. Rice and Ramsperger wrote that “it would seem that some kind of collision hypothesis must be used to account for activation in such a case,” while “the old form of the collision hypothesis as developed by Langmuir, Christiansen and Kramers, and Tolman appears entirely inadequate even in the case just cited.” They credited the suggestion “originally made by Lewis and Smith and somewhat extended since by Christiansen and by Hinshelwood and Lindemann, that the internal degrees of freedom may be of some help in causing activation” (p. 1617). They assumed that “the internal energy may be repeatedly redistributed within the molecules between collisions” and that “Hooke’s law of force holds within the molecule”. They assumed deactivation at every collision but qualified the consequence: “if the assumption of deactivation at every collision is not absolutely true, the last sentence does not hold”. They calculated the resulting error (pp. 1618–1621, 1627, 1629) and explicitly wrote the steady-state balance between reaction and collisional deactivation in 1927, citing Hinshelwood’s treatment and noting that Christiansen had obtained an equation of the same general form (p. 1620, notes 12–13).

Hinshelwood credited Lindemann and fitted the effective degrees of freedom: “we will attempt to calculate the value of n which will make this reaction fit in with Lindemann’s theory”. Acetone required more modes, “and this would be true to an even greater extent of the decomposition of nitrogen pentoxide” (pp. 231–232).

Kassel considered Rice and Ramsperger’s discussion of assumptions “so good that it seems useless to consider them further here”. He left explicit rotational and quantum treatments open: rotational energy “might be more explicitly considered, though this would probably be very difficult, or a quantum treatment might be given” (p. 240). The resulting theory is now called RRK.

Marcus developed the energy-resolved formulation in 1952 into what is now called Rice–Ramsperger–Kassel–Marcus (RRKM) theory. [35] He averaged the approximately conserved rotational quantum numbers into the factor $P_1^{\ddagger}/P_1$ (p. 360), which enters his energy-resolved expression (p. 361, Eq. 10). The modern $k(E,J)$ in main text Equation 7 resolves angular momentum explicitly, whereas Marcus averaged over it. Marcus also recovered Eyring’s high-pressure expression (p. 361, note 5).

He wrote that the treatment “differs from previous formulations of the latter in a number of ways, particularly in the expression used for the density of quantum states of the high energy molecules” (p. 359). A classical vibrational state count “generally gives a gross overestimate”; he supplied “a good approximation” in “the following semiclassical expression” valid at high energy (p. 362). He distinguished active modes, “those which can contribute their energy to the breaking bond without restrictions,” from adiabatic modes, “assumed to remain in the same quantum state during the course of decomposition of the molecule,” and inactive modes; angular-momentum restrictions were treated “approximately” (p. 360). He deferred comparison with measurements: “application of the final equations to the available experimental data will be reserved for a later paper” (p. 359). He regarded randomization as unresolved: “vibrational anharmonicity plays an important role in intramolecular energy transfer, and would have to be taken into account in more fundamental approaches to this problem” (p. 364). The collision frequency, “if deactivation does not occur at every collision, should be multiplied by some inefficiency factor” (p. 362).

### S2.6 Potential surfaces and transmission assumptions

Main text Sections 3 and 4. Sources: [19, 23, 24, 25, 27, 28, 56, 57, 58].

Eyring and Polanyi described the limits of their semi-empirical potential-energy surface in 1931. They wrote that knowledge of the required functions “is quite limited, as they are known only for the atom pair H–H, and even in this case only to a fairly rough approximation” (p. 1225), and concluded: “Hence, our result will be an overestimate in many respects” (p. 1225). They identified a failure in the approximation — “This indicates a failure of the underlying approximation” (p. 1229) — and declined to interpret a result for which errors accumulated: “because of the cumulative inaccuracies associated with the approximations involved, it does not seem worth a discussion” (p. 1237). They also left the validity of adiabatic reaction dynamics unresolved: “It is not our aim to come to a decision whether these reactions or chemical reactions in general in reality proceed adiabatically” (p. 1224). They concluded: “It appears that it is justified to identify the height of the saddle of the energy landscape with the heat of activation” (p. 1245). These quotations use the 2013 English republication and its pagination; I have not checked the 1931 German original.

Pelzer and Wigner computed an absolute rate on that surface (Eyring and Polanyi had computed an activation heat) and attributed much of their agreement with experiment to chance:

“Die gute Übereinstimmung zwischen beobachtetem und berechnetem $C$ wird man zweifellos zum grossen Teil einem Zufall zuschreiben, wenn man die vielen Fehlerquellen in Betracht zieht”, “Without doubt, the good agreement between observed and calculated $C$ will be ascribed largely to chance, when accounting for the many sources of error” (p. 468).

They listed three sources of error. The third was the classical treatment of motion. In the following quotation and translation, bracketed molar-energy equivalents replace the source’s parenthetical zero-point-energy expression $1/2\,h\,\nu$ and its numerical value:

“Als dritte Fehlerquelle ist schliesslich in Betracht zu ziehen, dass die Berechnung der Bewegung des Bildpunktes mit der klassischen Theorie ausgeführt worden ist. Das ist wohl im Falle von Wasserstoffatomen nicht berechtigt, wie schon daraus hervorgeht, dass die Nullpunktsenergie von $H_2$, ein Quanteneffekt [etwa 25 kJ mol$^{-1}$] in der Grössenordnung der ganzen Aktivierungswärme ist. Wie gross allerdings der hierdurch verursachte Fehler ist, lässt sich schwer angeben”, “A third source of error to consider is that the calculation of the motion of the image point has been carried out with the classical theory. This is presumably not justified in the case of hydrogen atoms, as already emerges from the fact that the zero-point energy of $H_2$, a quantum effect [approximately 25 kJ $mol^{-1}$], is on the order of magnitude of the whole heat of activation. It is difficult, however, to quantify the size of the resulting error." (p. 469).

They also distinguished the equilibrium crossing calculation from the nonequilibrium reaction — "Was wir in dieser Weise berechnen, ist streng genommen die Anzahl der sekundlichen Gratüberschreitungen im Gleichgewicht, während wir uns sehr weit vom Gleichgewicht befinden", "What we actually calculate in this way is the number of ridge crossings per second at equilibrium, whereas we find ourselves very far from equilibrium" (p. 460) — and assessed recrossing by inspecting the potential surface:

"Wir rechnen aber so, als ob überhaupt keine Gratüberschreitungen in umgekehrter Richtung in das Tal I vorkommen würden. Dies ist […] nicht ganz berechtigt, weil es im allgemeinen möglich ist, dass ein Bildpunkt, der den Grat eben nach Tal II überschritten hat, von dort sofort in das Tal I zurückgeworfen wird. […] Ein Blick auf die Fig. 1 lehrt aber, dass solche Reflexionen des Bildpunktes kaum vorkommen können", "But we calculate as though no ridge crossings in the reverse direction, into Valley I, were to occur at all. This is […] not quite justified, because it is generally possible that an image point that just crossed the ridge into Valley II is immediately thrown back into Valley I. […] A glance at Fig. 1 teaches, however, that such reflections of the image point can hardly occur" (pp. 460–461).

They return to recrossing as a second source of error and give a conditional estimate rather than a trajectory calculation. For the two-ridge potential model examined, approximately equal exits from the intervening region give a factor of ½ at very low temperature; at 1000° they expect a factor already close to one (pp. 468–469).

Eyring included a transmission coefficient in his 1935 expression and expected it to be about unity (p. 110). [25] Hirschfelder and Wigner derived an expression for it in 1939 from the probabilities of reflection on the two sides of the transition state. They assumed that these probabilities were independent of where a system came from and how many times it had crossed, an assumption they stated "is not always justified" (pp. 618–619). [27] The 1941 monograph reproduces their derivation and credits it to them (pp. 147–148). [28] Trajectory calculations and the reactive-flux formalism of the 1970s later enabled calculations of the dynamical correction.

Wigner had also discussed recrossing in his separate 1932 paper, noting that in the multidimensional case "es nicht mehr immer wahr ist, dass die Atome, die den Grat von links überschreiten, von links stammen", "it is no longer always true that the atoms which cross the ridge from the left originate from the left", though "dieser Effekt [spielt] wohl in den meisten Fällen eine nur untergeordnete Rolle", "this effect probably plays only a subordinate role in most cases" (p. 213) — and returned to the problem in 1938.

Wigner's separate 1932 paper calculated a quantum correction to the classical rate, whereas Pelzer and Wigner's paper of that year calculated the saddle-crossing rate. Wigner described his result as "nur eine Korrektion für die klassisch berechnete Reaktionsgeschwindigkeit", "only a correction to the classically calculated reaction rate" (p. 204). He identified a divergence and a convergence problem: "Es ist aber sehr störend, zu sehen, dass es über alle Grenzen steigt, wenn $V_2$ zu Null geht", "It is however very disturbing to see that it grows beyond all bounds when $V_2$ goes to zero", and "unsere Rechnung [ist] wegen Konvergenzschwierigkeiten mathematisch keineswegs streng", "our calculation is, on account of convergence difficulties, by no means mathematically rigorous" (pp. 211–212). For para-hydrogen conversion, the correction reduced agreement with experiment: "Wir sehen, dass durch die Quantenkorrektur die Übereinstimmung […] etwas verschlechtert wurde", "We see that the quantum correction reduced the agreement […] to some degree" (p. 216). He also corrected an error in his earlier paper with Pelzer: "Der theoretische Wert war von Pelzer und Wigner wegen eines Fehlers in ihrer Formel (22) zweifach zu hoch angegeben worden", "The theoretical value had been given twice too high by Pelzer and Wigner, owing to an error in their formula (22)" (p. 216, note 2).

Eckart introduced his 1930 barrier to obtain an analytic solution. Previous treatments were based "either on a potential function which has discontinuous derivatives, or else on approximate treatments involving asymptotic (i.e., divergent) series", whereas his potential allowed an exact solution: "there is an analytic function which represents some of the types of potential barriers which have been discussed and whose associated Schrödinger equation is soluble" (p. 1303). His title is "The Penetration of a Potential Barrier by Electrons". He considered electrons; Wigner in 1932 and Bell in 1933 subsequently applied his result to chemical reactions and cited his paper.

Wigner's 1938 account separates the molecular energy surface and dynamics, the statistical calculation of elementary rates, and the cooperation of elementary reactions that produces the observed reaction (pp. 29–30). [24] He also notes that restoration of the thermal distribution need not be fast.

### S2.7 Eyring and the activated-complex assumptions

Main text Sections 2 and 3. Sources: [25, 26, 28].

In Eyring's 1935 paper, equation (2) is a ratio of partition functions multiplied by $p / m^*$, which he describes as "the average velocity of activated complex along the normal coordinate corresponding to passing over the barrier," with $m^*$ the reduced mass and $p$ the average momentum along that coordinate (p. 110). The factor $k_B T / h$ results from combining the reaction-coordinate partition function with the mean forward velocity; it is not assumed in equation (2). On p. 109, he justifies the translational treatment: "this mode because of the small curvature can be treated statistically as a translational degree of freedom". He neglects tunneling because "the barriers are so flat near the top that tunneling may be neglected without appreciable error" and describes the surfaces as "very approximate" (p. 109).

In their 1941 monograph, Glasstone, Laidler, and Eyring state the quasi-equilibrium assumption: "it will be supposed that the initial reactants are always in equilibrium with the activated complexes", with a footnote calling it "one of the fundamental postulates of the theory developed here" and stating the condition under which it holds (p. 185). They also discuss a case in which it fails (p. 289).

They devote a section to the transmission coefficient (pp. 146–150) and retain $\kappa$ in the master equation (p. 190). In the instances I examined, they state its omission — "the transmission coefficient being omitted for the sake of simplicity" (p. 195). Eyring himself retained a transmission coefficient in the general theory and set it to unity in specific applications, giving his reason; in the 1935 *Chemical Reviews* paper, “the transmission coefficient is set equal to unity, as it can hardly differ materially from this value in the reactions to be considered” (p. 74). [26]

Glasstone, Laidler, and Eyring qualify the Arrhenius parameters on the first page of their monograph: “provided that the temperature range is not large, the quantities *A* and *E* can be taken as constant” (p. 1), and give a temperature-dependent experimental activation energy on pp. 194–195.

On p. 12, Glasstone, Laidler, and Eyring call the rate of crossing “a universal frequency, dependent only on temperature and independent of the nature of the reactants and the type of reaction,” after factoring the reaction-coordinate contribution out of the concentration of activated complexes over a short length along the reaction coordinate. In their later derivation, they state the statistical and dynamical conditions explicitly (pp. 185–190). Pages 13–14 are missing from the copy I used, preventing assessment of any immediate qualification.

### S2.8 Kramers, Marcus, Vineyard, and Bell

Main text Section 4. Sources: [10, 38, 40, 41, 43, 49, 59].

Kramers restricted his 1940 calculation to one dimension: “for the sake of simplicity only a one-dimensional model is studied” (p. 284). He also assumed position-independent friction, writing “($\eta$ is taken independent of $q$)” beside the diffusion equation (p. 288). He could not solve the intermediate-friction regime:

“I have not been able to solve exactly the problem of this quasi-stationary diffusion in $p$, $q$-space for arbitrary values of the viscosity $\eta$.” (p. 292)

“I have not been able to find a trustworthy method for extending (28) to $\eta$-values which are not small compared to $4\pi\omega$.” (p. 298)

In §6, “Relation of our model to actual reaction processes”, he questioned whether a single reaction coordinate remained adequate:

“It may of course be that in the activated state the distinction of one particular reactive vibration, which is subject to perturbations from the other vibrations, is no longer sound and that consequently a one-dimensional model of the transition state is largely at fault.” (p. 303)

Marcus restricted his 1956 treatment to a mechanism “in which there is very little spatial overlap of the electronic orbitals of the two reacting molecules in the activated complex” (p. 966). He calls this the “slight-overlap” assumption on the same page and uses “slight overlap” again in the body (p. 967). He also describes the orbitals as “weakly interacting” (p. 967). The words “continuum” and “parabolic” do not occur in the paper, although the approximations they name are present: the solvent outside a rigid, dielectrically saturated sphere about each reactant is treated as a dielectric continuum, and the sphere radii are “taken to be essentially unchanged by the reaction” (pp. 970, 974). His later accounts present those approximations in the two-parabola form now taught. The effect of nuclear tunneling on the inverted-region falloff was calculated by Siders and Marcus in 1981, building on earlier quantum treatments of electron transfer (pp. 748–749). [41]

Vineyard derived the many-body harmonic prefactor within absolute-rate theory, criticizing applications limited to one-body models (p. 121). He made two simplifying assumptions: “at all stages quantum effects will be neglected” and “motions near the saddle point will be treated by the theory of small oscillations — i.e., the vibrations will be assumed to be nearly simple harmonic” (p. 121). He did not discuss or correct for recrossing. His dividing surface “separates the region around A from that around B, and any representative point reaching S with finite velocity crosses over into the region around B”; the crossing rate “is assumed equal to the desired average jump rate” (p. 122). Crossing this local surface does not prevent an anharmonic trajectory from returning. Vineyard’s result is the whole-lattice frequency ratio within the stated classical harmonic treatment (p. 123).

Bell gives the validity condition with his tunneling correction and both series expansions, describing this regime as “the case which is most frequently of interest in chemical problems” (p. 3). He also treats the other two regimes and states that the large-tunneling case “has not yet been encountered in chemical kinetics” (p. 2). He writes that “no exact solution of the Schrödinger equation has been obtained for a parabolic barrier” and builds his expression to satisfy stated conditions that an exact one must meet (p. 2). He does not discuss the divergence of the compact correction at the crossover temperature. Later reviews do: Truhlar, Garrett, and Klippenstein describe the one-dimensional parabolic barrier as one “which diverges at low temperature and that fails to capture the physics of multidimensional tunneling at other temperatures” (pp. 12786–12787). Bao and Truhlar give it as an equation, stating that the transmission coefficient “blows up when $h|\nu‡|/(2k_B T) = \pi$” (p. 7557), which is Bell’s condition in their variables.

Bell also required the reduced barrier height to be below his dimensionless barrier-thickness parameter, and the exponential of their difference to be small enough to neglect the second term in his fuller expression. The fuller expression itself used an approximate permeability (pp. 2–3). [59] The temperature restriction in main text Section 4.4 applies together with these additional conditions.

### S2.9 Variational and dynamical rate calculations

Main text Sections 4, 5, 7, and 8. Sources: [10, 44, 45, 46, 47, 48, 49, 50, 51, 52, 53, 54, 60, 63, 64, 69].

Truhlar, Garrett, and Klippenstein give the upper-bound condition explicitly: “In classical mechanics, TST provides an upper bound to the rate constant if reactant equilibration replenishes reactant states fast enough” (p. 12772). [10] For equality, forward-crossing trajectories must have “originated at reactants and will not reach the surface again before being thermalized or captured in a product state” (p. 12771). They attribute these assumptions to Wigner. A separate obstruction arises in quantum mechanics: “The difficulty of incorporating quantum effects into transition-state theory essentially results from the classical nature of the concept of one-way flux through a surface” (p. 12773). They state the coordinate dependence of reaction-path frequencies (p. 12777). On ab initio barrier heights, “the challenge survives today, largely unmet”; on ion–molecule substitution, “gas-phase SN2 reactions illustrate some of the major unknowns in the current status of TST” (p. 12778). Bao and Truhlar note that, after non-rigorous quantization, the CVT rate constant “no longer necessarily provides an upper bound to the exact quantum rate constant” (p. 7555). [49]

Wigner stated the upper bound in his 1937 abstract and the variational principle in the text (pp. 722–723): the true rate is the minimum one-way flux over admissible phase-space surfaces. [44] He did not implement the variation; his working formula “is only an upper limit … and one could probably somewhat lower it by choosing $H_0$ more appropriately” (p. 723). For iodine-atom recombination, his calculated rates lay below measurements for all five bath gases (p. 725). His formula applied only to atomic third bodies, making the discrepancy for $H_2$, $N_2$, and $O_2$ “perhaps, not so serious,” whereas for helium and argon the calculated rates “should be indeed materially higher than the observed ones.” For these two gases, the result conflicts with the bound if the molecular model is accurate. Wigner considers collision radii, potential-curve shape, and additional attractive electronic states, leaving the question open (p. 725).

Horiuti varied the surface in 1938, naming the minimizing surface the critical surface. [45] He balanced the normal potential gradient divided by $k_B T$ against the sum of inverse principal radii of curvature. For a model two-atom association with a potential maximum, he found that the conventional transition-state rate exceeded his minimized rate by a factor of 1.1–1.5 at 25 °C (pp. 213–216).

Keck applied the principle to multidimensional three-body recombination in 1960; [46] Pechukas and McLafferty derived conditions for exactness of classical transition-state theory for collinear atom–diatom collisions (pp. 1622–1623). [47] Garrett and Truhlar applied canonical variational theory to 37 atom–diatom reactions (p. 5207). [48] Variation of the dividing surface is particularly useful for loose transition states with weakly bound, distant fragments and no sharp saddle, including barrierless association, ion–molecule reactions, and three-body recombination.

Miller’s 1974 reactive-flux expression starts from quantum scattering (pp. 1823–1824). [60] Miller, Schwartz, and Tromp explicitly credit Yamamoto’s earlier flux–flux correlation expression (p. 4889). [50, 52] They show that their correlation function and Yamamoto’s differ but have the same rate integral (p. 4891, Eqs. 2.11–2.13). [52] Dellago and co-workers later formulate the calculation of reactive-flux correlations as an equivalent reversible-work calculation using transition paths (p. 1964). [63] The representations differ, but they yield the same rate for the specified system.

Later authors also stated restrictions on their methods. Chandler states in the abstract of his 1978 paper that the validity of the reactive-flux expression “hinges on a separation of time scales and the plateau value problem” (p. 2959). Craig and Manolopoulos describe their ring-polymer method’s deep-tunneling performance as “satisfactory (although not perfect)”. Bolhuis, Chandler, Dellago, and Geissler write, “Often, a significant challenge in transition path sampling work is the characterization of the stable states” (p. 302), and “Characterizing stable states and generating mechanistic interpretations, however, remain subjective endeavors” (p. 315). Referring to earlier path-finding methods and excepting transition path sampling, E, Ren, and Vanden-Eijnden write that “most of these methods seem to require that the energy landscape be relatively smooth” (p. 052301-1). They present the string method for systems with such landscapes. Yang and co-workers focus their review on the training-data problem, noting that machine-learned models “perform poorly when used in out-of-distribution scenarios” (p. 388).

### S2.10 Interpretation of the historical record

Main text Sections 2–4 and 10. Sources: [1, 28, 49].

The sources document limitations that their authors recognized: Kramers could not solve the intermediate-friction regime; Pelzer and Wigner suspected cancellation of errors; Kassel left a quantum state count open; and Lewis lacked an independently determined collision diameter. Other limitations were not discussed. Vineyard did not address recrossing, and Bell did not discuss the divergence of his compact correction at its domain boundary.

Later reviews and monographs retained many of these qualifications. Hänggi, Talkner, and Borkovec reproduce Kramers’ one-dimensionality and his constant damping rate and include a subsection on the regime of validity of Kramers’ theory (§IV.G, pp. 278–279); they also credit the transmission coefficient to Eyring, who introduced it, they write, “as an ad hoc fudge factor,” and they call the $\kappa = 1$ result the rate “commonly known as” the transition-state-theory rate (p. 255). Bao and Truhlar state Bell’s low-temperature limit as an equation. Glasstone, Laidler, and Eyring attach the temperature-range condition to the Arrhenius parameters on the first page of text.

These conditions affect the interpretation of fitted activation parameters. An activation enthalpy obtained over a narrow temperature range depends on the linearity assumption, the treatment of the transmission coefficient, the standard state, and the assignment of the rate-limiting step. Eyring included $\kappa$ in the general expression and gave a reason for setting it to unity in his applications. Omitting $\kappa$ without stating that assumption leaves the fitted parameters ambiguous.

The authors’ statements distinguish problems they could estimate from those they could only identify. Rice and Ramsperger estimated the error caused by their collisional-deactivation assumption. Kramers identified an intermediate-friction problem he could not solve, and Marcelin wrote phase integrals he did not evaluate. Arrhenius’s 1889 account preceded molecular potential-energy surfaces.

Lewis asserted unit efficiency and fitted the collision diameter in two examples although his stated program sought quantities determined independently of measured reaction rates. Trautz added an activation exponential and assumptions about heat-capacity additivity and temperature factors (pp. 5–8), then proposed that essentially two substance constants should suffice (p. 28). I regard those statements as stronger than the evidence supplied.

Several limitations identified in the original papers became problems addressed by later theories. Those theories changed the representations and intermediate quantities used to calculate rates, and several also corrected earlier formulations.

## S3 Supplementary derivations and conventions

### S3.1 The population above an energy threshold

For a classical canonical distribution with $m$ independent positive quadratic energy terms, let $E$ denote their summed energy expressed on a molar scale. The density is

$$p(E)=\frac{E^{m/2-1}\exp[-E/(RT)]}{\Gamma(m/2)(RT)^{m/2}}, E\geq 0. \tag{S1}$$

The fraction above a fixed threshold $E_0$ is therefore

$$P(E\geq E_0)=\frac{\Gamma(m/2, E_0/(RT))}{\Gamma(m/2)}. \tag{S2}$$

Here $\Gamma(a,x)$ is the upper incomplete gamma function. Only $m=2$ gives $\exp[-E_0/(RT)]$. For positive even $m>2$, a polynomial in $E_0/(RT)$ multiplies the exponential; for $m=1$, the result is a complementary error function. The distribution follows by convolving the canonical distributions of the independent quadratic terms. This population distribution does not specify which molecules react or how often they cross a dividing surface. An activation energy inferred from the temperature dependence of a rate therefore need not equal $E_0$; Tolman's result weights the reacting states (Section S2.3). [20]

### S3.2 Concentration and pressure standard states

For an elementary step of molecularity $n$, write the concentration-based rate constant with a fixed standard concentration $c^\circ$ as

$$k=\kappa(T)\frac{k_\mathrm{B}T}{h}(c^\circ)^{1-n}\exp\left[-\frac{\Delta G_c^\ddagger(T)}{RT}\right]. \tag{S3}$$

The subscript $c$ denotes this fixed-concentration convention. Along the specified temperature path, define $\mathscr{H}_c^\ddagger=-T^2\,\mathrm{d}(\Delta G_c^\ddagger/T)/\mathrm{d}T$ and $\mathcal{S}_c^\ddagger=-\mathrm{d}\Delta G_c^\ddagger/\mathrm{d}T$. They satisfy $\Delta G_c^\ddagger=\mathscr{H}_c^\ddagger-T\mathcal{S}_c^\ddagger$. The calligraphic symbols identify the slope and intercept parameters defined by these derivatives, which need not equal physical standard activation enthalpy and entropy. The logarithm used for an Eyring plot is dimensionless:

$$\ln\left[\frac{hk(c^\circ)^{n-1}}{k_\mathrm{B}T}\right]=\ln\kappa+\frac{\mathcal{S}_c^\ddagger}{R}-\frac{\mathscr{H}_c^\ddagger}{RT}. \tag{S4}$$

For an ideal gas, changing from pressure standard $p^\circ$ to fixed concentration standard $c^\circ$ changes the chemical potential of each species by $RT\ln(c^\circ RT/p^\circ)$. Formation of one activated complex from $n$ reactant molecules then gives

$$\begin{aligned}\Delta G_c^\ddagger&=\Delta G_p^\ddagger+(1-n)RT\ln\left(\frac{c^\circ RT}{p^\circ}\right),\\ \mathscr{H}_c^\ddagger&=\Delta H_p^\ddagger+(n-1)RT.\end{aligned} \tag{S5}$$

The second relation follows by differentiating the first, including the temperature inside the logarithm. For ideal gases, formation of one activated complex from $n$ reactants gives $\Delta(pV)^{\ddagger}=(1-n)RT$, so $\mathscr{H}_c^{\ddagger}=\Delta U^{\ddagger}$. The physical ideal-gas enthalpy remains independent of pressure. At fixed composition, the Gibbs differential is $\mathrm{d}(G/T)=-H\,\mathrm{d}T/T^2+V\,\mathrm{d}p/T$; along the standard-state path $p=c^{\circ}RT$, its second term must be retained. The derivative then yields $H-pV=U$.

Entropy requires the same distinction. Writing $r=c^{\circ}RT/p^{\circ}$, the physical ideal-gas activation entropy at the concentration standard is $\Delta S_{c,\mathrm{phys}}^{\ddagger}=\Delta S_p^{\ddagger}+(n-1)R\ln r$, whereas $\mathcal{S}_c^{\ddagger}=\Delta S_{c,\mathrm{phys}}^{\ddagger}+(n-1)R$. For reactions that change the number of gas molecules, these derivative parameters differ from the corresponding fixed-pressure thermodynamic quantities. Differentiating Equation S3 gives

$$\begin{aligned} E_a &= RT^2\frac{\mathrm{d}\ln k}{\mathrm{d}T}=\mathscr{H}_c^{\ddagger}+RT+RT^2\frac{\mathrm{d}\ln\kappa}{\mathrm{d}T} \\ &= \Delta H_p^{\ddagger}+nRT+RT^2\frac{\mathrm{d}\ln\kappa}{\mathrm{d}T}. \end{aligned} \tag{S6}$$

The pressure-standard equality and the identification $\mathscr{H}_c^{\ddagger}=\Delta U^{\ddagger}$ here apply to ideal gases. For a solution or interface, the result depends on the specified temperature path and standard chemical potentials. A concentration-based coefficient in a nonideal system may also contain an activity-coefficient ratio, whose temperature derivative contributes to the slope. A standard-state conversion does not account for solvation or nonideal interactions. [25, 28, 29, 30]

For bimolecular association, changing from a 1 bar ideal-gas standard to a 1 mol L$^{-1}$ concentration standard lowers $\Delta G^{\ddagger}$ by $RT\ln(c^{\circ}RT/p^{\circ})$. The decreases are 7.95 kJ mol$^{-1}$ at 298 K and 45.92 kJ mol$^{-1}$ at 1200 K, using consistent volume and pressure units. The 1200 K value illustrates the standard-state conversion; it does not presume a liquid at that temperature. Solvation free energies contribute separately. A consistent conversion leaves the physical rate unchanged.

If $k_c(T)=k_{\mathrm{intr}}(T)\Phi(T)$, with the remaining experimental conditions fixed, the activity factor contributes an additional $RT^2\,\mathrm{d}\ln\Phi/\mathrm{d}T$ to the apparent activation energy. For an elementary bimolecular activated-complex description, $\Phi=\gamma_A\gamma_B/\gamma^{\ddagger}$ when the intrinsic coefficient and transmission are defined separately. Changes in composition or speciation require additional terms.

### S3.3 Activation parameters when transmission is omitted

Suppose the true rate has the form in Equation S3 but an apparent activation free energy is inferred by setting $\kappa=1$. At the same fixed standard-state convention and temperature path,

$$\Delta G_{\mathrm{app}}^{\ddagger}=\Delta G_{\mathrm{TST}}^{\ddagger}-RT\ln\kappa. \tag{S7}$$

Differentiation gives the apparent local Eyring parameters. These describe the fitted rate; they do not imply a change in physical equilibrium enthalpy or entropy:

$$\mathcal{H}^{\ddagger}_{\mathrm{app}} = \mathcal{H}^{\ddagger}_{\mathrm{TST}} + RT^2 \frac{\mathrm{d}\ln\kappa}{\mathrm{d}T}, \tag{S8}$$

$$\mathcal{S}^{\ddagger}_{\mathrm{app}} = \mathcal{S}^{\ddagger}_{\mathrm{TST}} + R\ln\kappa + RT \frac{\mathrm{d}\ln\kappa}{\mathrm{d}T}. \tag{S9}$$

A constant $\kappa$ therefore changes the apparent intercept parameter by $R\ln\kappa$ without changing the slope parameter. Temperature-dependent transmission affects both. These are local derivative identities; parameters obtained by a straight-line fit over a finite temperature interval average over any unresolved curvature. For a specified classical model with equilibrated reactants and recrossing as the only correction, $0 < \kappa \le 1$. A factor that also includes nuclear tunneling is not subject to that bound. [10, 49, 51] The interpretation therefore depends on the stated definition of $\kappa$.

### S3.4 Correlated errors in activation enthalpy and entropy

Activation enthalpy and entropy fitted to the same temperature series have correlated uncertainties, especially over a narrow interval. Krug, Hunter, and Grieger showed that when these errors dominate the true differences among reactions, fitted pairs lie near a line whose slope is the harmonic mean experimental temperature, $T_{\mathrm{hm}}$ (p. 2336). [11] Such a correlation does not by itself establish a chemical compensation relationship. Their second paper regresses activation enthalpy against activation free energy evaluated at $T_{\mathrm{hm}}$, where the estimation errors are uncorrelated under their regression assumptions. Measurements near an estimated compensation temperature can then test whether the rates actually coincide there (pp. 2341, 2343, 2345).

Exner illustrated the statistical problem with a hypothetical pair of temperature measurements: independent logarithms of rate constants at temperatures with $T_2/T_1 = 0.9$ give a correlation coefficient of 0.998 between the derived Arrhenius parameters (p. 489). [12] A common intersection of rate curves remains possible. In a later comparison of rates, Exner found an approximately valid relation in 78 of 100 reaction and equilibrium series and left 14 undecided; he also concluded that many reported isokinetic temperatures near the experimental temperature resulted from incorrect statistical treatment (p. 2762). [13] A proposed compensation relation should therefore be tested against the rates and their uncertainty, not inferred from a plot of two correlated fitted parameters alone.

## S4 Numerical evidence for five modern rate comparisons

The five studies below compare calculated rates with experiments, electronic-structure calculations using a fixed rate protocol, or unbiased dynamics under the same Hamiltonian. Table S1 lists the conditions, comparisons, and limitations. Because the references differ, these results do not provide a common accuracy ranking. The following subsections give the source pages and identify fitted quantities.

| System and conditions | Reference | Numerical comparison | Meaning and limits |
|---|---|---|---|
| OH + $CH_4$ → $H_2O$ + $CH_3$; gas phase, 200–1500 K. [72, 73] | Experiment | Computed/experimental rate: 0.79–1.26 | Potential and rate-method errors remain combined (S4.1). |
| 2H* → $H_2$ on Pt(111); zero coverage, second-order rate, 653–953 K. [68] | Experiment | Original: 3.9–6.2; adjusted: 0.95–2.12 | Adjusted barrier fitted to experiment; original error sources remain combined (S4.2). |
| $CH_3$ + $CH_3$ and other radical recombinations; 50–3500 K for $CH_3$ + $CH_3$. [74] | CASPT2/cc-pVDZ with the same VRC-TST protocol | Maximum relative rate error: 10.0% for $CH_3$ + $CH_3$; up to 20% for larger fragments | Tests learned interaction energies within the rate protocol; not experimental accuracy (S4.3). |
| Chignolin (CLN025); CHARMM22*, 340 K. [75] | Same-Hamiltonian equilibrium dynamics, 120 μs | Three independent runs: 2.49, 13.4, 10.1 $\mu s^{-1}$; reference: 2.5 ± 0.5 $\mu s^{-1}$ | 5.4-fold spread tests sampling and reweighting; no convergence bound is established (S4.4). |
| Alanine dipeptide, $C_7$eq → $C_7$ax; vacuum, AMBER99SB, 300 K. [76] | Same-Hamiltonian unbiased trajectories | Short-time rate error within about a factor of 10; conventional error with poor variable can exceed $10^3$ | Tests time inference; depends on coordinate, acceleration, and sample size (S4.5). |

Table S1. Conditions, references, and limitations of five published rate comparisons. Comparisons using the same Hamiltonian, state definitions, ensemble, and target dynamics test the accelerated calculation independently of the physical model's agreement with experiment.

### S4.1 OH + $CH_4$ and sensitivity to the potential

Gui, Fan, Sun, and Li computed rate coefficients for gas-phase OH + $CH_4$ → $H_2O$ + $CH_3$ over 200–1500 K using ring-polymer molecular dynamics (main text Section 4). [72] Li and Guo's permutation-invariant polynomial neural-network surface was fitted to about 135,000 UCCSD(T)-F12a/AVTZ energies, with a root-mean-square error of 0.38 kJ $mol^{-1}$ and no kinetic data. [73] With the conventional propagator, computed/experimental ratios at the tabulated temperatures range from 0.79 to 1.26: 1.26 at 200 K, 1.09 at 300 K, 0.79 near 1000 K, and 0.875 at 1500 K (reference [72], Table 1, p. 5207). Measurements near 500, 700, and 1000 K were at 498, 698, and 1005 K. The authors state errors of "less than 20%"; their 200 K table entry gives 26%. The Cayley propagator gives a ratio of 1.33 at 200 K, and Cayley bead-count tests changed rates by 13% at 200 K and 5% at 300 and 500 K (Supplementary Materials of reference [72], Fig. S8).

Earlier ring-polymer calculations overestimate the 200 K rate by an order of magnitude on PES-2000 (reference [72], p. 5207) and by a factor of 3.9 on PES-2014 (reference [72], Table 1, p. 5207). The authors attribute the discrepancy mainly to shallower van der Waals wells on those surfaces (p. 5206). These calculations show sensitivity to the potential, although numerical differences among the studies prevent a complete separation of potential and rate-method errors. Below the 326 K crossover temperature, ring-polymer dynamics can overestimate rates for this asymmetric barrier (p. 5207), so agreement may include error cancellation.

### S4.2 Hydrogen desorption from Pt(111)

Zhang, Nitz, Borodin, Wodtke, and Guo computed zero-coverage second-order rates for 2H* → $H_2$ desorption from Pt(111) with ring-polymer dynamics, comparing with velocity-resolved kinetics over 653–953 K. [68] Their neural-network surface was fitted to 1457 density-functional (GGA) energies with rigid metal atoms. On the original surface, the transmission coefficient $\kappa$ is 0.36–0.37 over this range. With identical symmetry and electronic-spin factors applied to both rates, recrossing lowers the centroid transition-state estimate by a factor of 2.7–2.8 (reference [68], Table 1). Computed rates exceed measurements by factors of 3.9–6.2 (reference [68], Table 2, p. 322). The authors report a computed Arrhenius activation energy of 63.7 kJ mol$^{-1}$ and compare it with an experimentally estimated zero-coverage barrier of 72.4 kJ mol$^{-1}$ (p. 323). They empirically increased the desorption barrier by 8.7 kJ mol$^{-1}$, using that energy difference. The adjusted computed/experimental ratios are 0.95–2.12, and the relative isotopic branching trends for $H_2$, HD, and $D_2$ are consistent with measurements (reference [68], Fig. 5).

The authors identified four possible sources of the original discrepancy (p. 323): rigid metal atoms, freely translating adsorbed hydrogen without a diffusion barrier, the density functional, and ring-polymer dynamics. They interpreted the barrier adjustment as evidence for a largely enthalpic surface error. For this temperature range, they identified reactant zero-point energy as the main nuclear quantum contribution and tunneling as relatively unimportant.

To compare rate magnitude and temperature dependence separately, I calculated $r(T)=k_{calc}(T)/k_{exp}(T)$ for the four temperatures common to the original calculation, adjusted calculation, and experiment. Table S2 reproduces the source rate coefficients, converted from cm$^2$ s$^{-1}$ to m$^2$ s$^{-1}$, and gives the resulting ratios. The calculations retain the same symmetry and electronic-spin conventions.

| Temperature (K) | Original rate (m$^2$ s$^{-1}$) | Adjusted rate (m$^2$ s$^{-1}$) | Experimental rate (m$^2$ s$^{-1}$) | Original ratio | Adjusted ratio |
|---|---|---|---|---|---|
| 653 | $4.11 \times 10^{-13}$ | $8.14 \times 10^{-14}$ | $6.61 \times 10^{-14}$ | 6.218 | 1.231 |
| 753 | $2.07 \times 10^{-12}$ | $4.99 \times 10^{-13}$ | $5.27 \times 10^{-13}$ | 3.928 | 0.947 |
| 853 | $7.21 \times 10^{-12}$ | $2.11 \times 10^{-12}$ | $1.69 \times 10^{-12}$ | 4.266 | 1.249 |
| 953 | $1.88 \times 10^{-11}$ | $6.38 \times 10^{-12}$ | $3.01 \times 10^{-12}$ | 6.246 | 2.120 |

Table S2. Published Pt(111) rate coefficients and calculated-to-experimental ratios. Rates are from Table 2 of reference [68], p. 322; ratios are calculated here from the printed values. No uncertainty intervals are supplied in the source table.

The ratio $\max_T r(T)/\min_T r(T)$ is 1.590 before adjustment and 2.239 afterward. Any temperature-independent multiplier of the calculated rates cancels from this ratio. The barrier adjustment therefore improves all four absolute rate comparisons but widens the remaining variation of their ratios with temperature. This is a descriptive result for four tabulated central values. It does not establish statistical significance, determine the temperature dependence outside this interval, or show that the original potential is physically more accurate.

A second calculation checks the response to the imposed barrier change. For $\delta E = 8.68368$ kJ mol$^{-1}$, the ratio $k_{adj} / [k_{orig} \exp(-\delta E / RT)]$ is 0.9804, 0.9649, 0.9956, and 1.0153 at 653, 753, 853, and 953 K. The recalculated rates therefore follow the simple barrier-rescaling response within 3.6%. This confirms how the imposed correction changes the rates; it does not independently identify the source of the discrepancy with experiment.

The reported activation energy also requires a qualification. An unweighted least-squares fit of $\ln(k / k_{ref})$ against $1/T$, with fixed $k_{ref} = 1$ m$^2$ s$^{-1}$, to all seven printed original-surface rates in Table 1 gives 66.21 kJ mol$^{-1}$, whereas the text reports 63.7 kJ mol$^{-1}$. The temperatures are 300, 400, 500, 653, 753, 853, and 953 K, and the corresponding rates are $2.37 \times 10^{-19}$, $1.88 \times 10^{-16}$, $9.81 \times 10^{-15}$, $4.11 \times 10^{-13}$, $2.07 \times 10^{-12}$, $7.21 \times 10^{-12}$, and $1.88 \times 10^{-11}$ m$^2$ s$^{-1}$. The activation energy is obtained by multiplying the fitted slope by $-R$. The published information does not resolve this difference. The 8.7 kJ mol$^{-1}$ adjustment quoted here is the correction explicitly imposed by the authors, not an energy difference reconstructed from matched experimental and calculated Arrhenius fits.

### S4.3 Radical recombination with a learned interaction surface

Vari, de Falco, and Cavallotti used a neural network for radical-fragment interaction energies in high-pressure-limit variable-reaction-coordinate transition-state calculations (VRC-TST; main text Section 4). [74] They compared the network with CASPT2/cc-pVDZ energies under the same rate protocol, using system-specific active spaces. For $CH_3$ + $CH_3$, with a (2e,2o) active space, the maximum rate error is 10.0% over 50–3500 K (reference [74], Table 1, p. 10379). Errors remain below 10% for their systems with up to two heavy atoms and reach 20% for larger fragments (p. 10380). They report a nearly fourfold reduction in the number of electronic-structure calculations (p. 10381).

Expanding the dividing-surface and pivot-displacement search to 328,565 network evaluations lowered the $C_2H_5$ + $CH_3$ rate by about 10%, improving agreement with earlier theoretical values (reference [74], Fig. 10, p. 10381). The earlier theoretical values are not exact rates, so this comparison establishes improved agreement rather than absolute accuracy. The sampled interaction surface uses frozen fragments, but separate corrections account for geometry relaxation and higher-level energies; final rates also include recrossing factors of 0.85, or 0.9 for $CH_3$ + H (pp. 10375–10376). The strict classical TST upper-bound argument therefore cannot be applied directly to these corrected rates. The authors do not claim experimental accuracy (p. 10381).

### S4.4 Variation among independent chignolin folding calculations

Lazzeri, Jung, Bolhuis, and Covino trained a neural-network committor during transition path sampling (main text Sections 4 and 5) and reweighted paths to recover free energies and rates. [75] For CLN025 folding with CHARMM22* at 340 K, folded and unfolded states have ≥99% and ≤1% native contacts, respectively (p. 9067). The reference from 120 μs of equilibrium simulation is 2.5 ± 0.5 μs$^{-1}$. After 50 sampling steps containing 20 transition paths, run 1 gave 2.49 μs$^{-1}$ with 1.10 μs of path sampling plus 0.90 μs of basin simulations for ensemble extension; the other runs gave 13.4 and 10.1 μs$^{-1}$ (p. 9069). All three lie within an order of magnitude of the reference.

The authors reported the three estimates separately; their 5.4-fold spread indicates substantial variation among independent runs. For run 1, folding-rate estimates read from Fig. S4d are approximately 0.9–1.3 times the reference over committor values 0.40–0.55, about 1.6–1.8 times at 0.60–0.66, and about 3.5 times near 0.35 (Supplementary Materials of reference [75], p. S3). These ratios are approximate readings from the plot, not uncertainty bounds. All calculations use the same force field, so the comparison tests sampling rather than experimental accuracy.

### S4.5 Recovering time from biased alanine dipeptide trajectories

Infrequent metadynamics rescales biased trajectory times to estimate an unbiased mean first-passage time, whose inverse is the rate for Poisson escape. Blumer, Reuveni, and Hirshberg tested this inference for the $C_{7eq} \rightarrow C_{7ax}$ transition of alanine dipeptide in vacuum with AMBER99SB at 300 K. [76] They started trajectories from a fixed $C_{7eq}$ structure and stopped them when $0.5 < \varphi < 1.5$ rad (p. 3490). About 7% of unbiased runs did not reach the product; an exponential survival fit gave a reference time of about 3.5 μs (p. 3487; Supplementary Materials of reference [76], pp. S2–S3). Biasing this dihedral angle gave matching conventional and short-time estimates at a bias-deposition rate of 50 $\text{ns}^{-1}$. With $\psi$, the conventional error exceeded three orders of magnitude at high acceleration. Short-time fitting stayed within about an order of magnitude across tested accelerations and gave 3.4 μs at the same bias-deposition rate (reference [76], Fig. 4, pp. 3487–3488). The plotted distributions use bootstrap sets of 200 trajectories. The reported errors therefore apply to that sample size and the specified collective variables (main text Section 5). [65]

## S5 Diagnostic procedures and their limits

### S5.1 Separation of relaxation and escape timescales

Waiting-time measurements require defined reactant and product basins, a criterion for commitment to product, and an initial ensemble. For a first-passage experiment or simulation, the survival probability $S(t)$ is the probability that product has not yet been reached. Its hazard is $-\mathrm{d}\ln S(t)/\mathrm{d}t$. A constant hazard over a resolved interval supports exponential escape on that interval, although hidden heterogeneity or more than one mechanism may produce the same behavior. Right-censored trajectories, which end before reaching product, must remain in the analysis. [76] In a two-state Markov model $A \rightleftharpoons B$, a small population perturbation relaxes with rate $k_{AB}+k_{BA}$, whereas exponential first-passage escape from $A$ to $B$ has rate $k_{AB}=1/\langle \tau_{AB} \rangle$. [51, 63]

Within a candidate reactant basin, compare relaxation of relevant coordinates with the escape timescale. Repeat the calculation from distinct initial configurations and determine whether their populations relax to the same distribution before most escapes occur. A reactive-flux plateau after microscopic recrossings and before substantial population relaxation is a complementary test. [51, 54, 63] Without this interval, a single phenomenological rate may be inadequate. Barrier height alone does not establish timescale separation, and the tested coordinate autocorrelations may miss other slow variables.

### S5.2 A proposed reaction coordinate

Sample configurations near the proposed dividing surface. Launch independent trajectories with momenta or stochastic noise drawn from the stated physical distribution, and estimate the fraction reaching product before reactant. The distribution of these commitment probabilities tests whether configurations assigned the same coordinate value have comparable probabilities of reaching product. [54, 55, 65] A broad distribution at the proposed transition surface indicates that the coordinate omits information needed to predict commitment. Sufficient trajectories are needed to distinguish variation among configurations from binomial uncertainty in each estimated probability.

A distribution concentrated near one-half supports the proposed configuration-space transition region. It does not determine recrossing or establish a minimum transition-state rate; dynamical transmission requires a separate calculation. [66] In inertial dynamics, commitment generally depends on momentum as well as position. Changing the initial momentum distribution changes the configuration-averaged committor being tested.

### S5.3 Errors in potentials, sampling, and rate calculations

To test accelerated sampling or recovery of physical times, compare the method with adequately converged dynamics using the same Hamiltonian, states, ensemble, and target dynamics. To test a learned potential, compare its energies and forces for sampled reactive configurations with the electronic-structure reference. Comparison with experiment tests the physical model and rate treatment together. [68, 69, 72, 73, 74, 75, 76]

Increase the amount of sampling, repeat independent runs, and report the variation among them. Vary numerical settings that can change a rate, including bead count for ring-polymer calculations or bias-deposition rate for infrequent metadynamics. If a barrier has been adjusted to measured kinetics, test the resulting model against conditions or observables not used in fitting. [68] Agreement after calibration does not separate the errors in the potential, prefactor, sampling, and dynamics.

### S5.4 The modeled event and the measured observable

The comparison with experiment must distinguish an elementary-step rate constant, an escape rate, a turnover frequency, and a conversion affected by transport. Changes in composition, pressure, surface coverage, or sample geometry can affect activities, competing reaction steps, or transport. Distinguishing these contributions requires independent constraints on the model used to interpret the measured response. [9, 29, 30] For an elementary activation analysis, specify the rate law and standard-state conventions (Section S3).

## S6 Reinterpreting older rate measurements

### S6.1 Hydrogen–oxygen chain branching: thermochemistry and dynamics

Define the forward direction as $\mathrm{H} + \mathrm{O}_2 \rightarrow \mathrm{OH} + \mathrm{O}$. Harding and co-workers' statistical treatment and Troe and Ushakov's subsequent trajectory calculation used different equilibrium-constant expressions to infer the reverse rate from measured forward rates. [77,78] The revised expression incorporated the lower OH formation enthalpy supported by independent photoionization measurements and electronic-structure calculations by Ruscic and co-workers. [80] Troe subsequently reviewed the reconciliation. [79] Holding the combined experimental forward-rate fit fixed isolates the thermochemical contribution. The following equilibrium-constant expressions apply over 1000–5000 K:

$$\begin{aligned} k_f &= 1.62\times10^{-16}\exp[-7474\,\mathrm{K}/T]\,\mathrm{m}^3\ \mathrm{molecule}^{-1}\ \mathrm{s}^{-1}, \\ K_{old} &= 17.0(T/1000\,\mathrm{K})^{-0.213}\exp[-8404\,\mathrm{K}/T], \\ K_{revised} &= 17.0(T/1000\,\mathrm{K})^{-0.213}\exp[-8144\,\mathrm{K}/T]. \end{aligned} \tag{S10}$$

The forward-rate fit and old equilibrium-constant expression are Equations 6.1 and 6.4 of reference [77]; the revised equilibrium constant is Equation 1.5 of reference [78]. The forward fit combines selected experimental data as reported by Du and Hessler and reproduced in these theory papers. Both reaction directions are bimolecular, and the dimensionless equilibrium constants use the same ideal-gas convention. Consequently,

$$k_r = \frac{k_f}{K},\ \frac{k_{r,revised}}{k_{r,old}} = \exp[-260\,\mathrm{K}/T]. \tag{S11}$$

The exponential shift corresponds to $R(260\,\mathrm{K}) = 2.162\ \mathrm{kJ\ mol^{-1}}$. It changes the reverse rate inferred from experiment, without changing the measured forward rate.

Table S3. Controlled evaluation of the published fits. Reverse rates are in $10^{-17}$ $\mathrm{m^3\ molecule^{-1}\ s^{-1}}$; the percentage change is $100(k_{r,revised}/k_{r,old}-1)$. These are evaluations of analytic fits, not individual measurements.

| $T$ (K) | Reverse rate, old $K$ | Reverse rate, revised $K$ | Change (%) | Trajectory rate |
|---|---|---|---|---|
| 1000 | 2.415 | 1.862 | −22.9 | 1.946 |
| 1500 | 1.931 | 1.624 | −15.9 | 1.640 |
| 2000 | 1.758 | 1.544 | −12.2 | 1.468 |
| 3000 | 1.642 | 1.506 | −8.3 | 1.289 |

The last column evaluates Equation 3.1 of reference [78], valid over approximately 300–5000 K and converted to SI volume units:

$$k_{r,CT} = 10^{-17}\left[0.026(T/1000\,\mathrm{K})^{1.47} + 1.92(1000\,\mathrm{K}/T)^{0.46}\right]\mathrm{m}^3\ \mathrm{molecule}^{-1}\ \mathrm{s}^{-1}. \tag{S12}$$

The thermochemical conversion changes the reverse rate inferred from the fixed forward fit. The trajectory treatment changes calculated reactive branching: in this model, additional nonstatistical return of energized $HO_2$ to $O+OH$ lowers the reverse rate. The trajectory fit lies 4.9% below the revised inferred rate at 2000 K and 14.4% below it at 3000 K. These deterministic evaluations introduce no new trajectories or fitted rate parameters. The covariance of the experimental fit parameters has not been obtained, so the table is not an uncertainty analysis. The reconstruction reproduces the published 2000–2001 reconciliation; it is not a present-day recommended rate expression. The original authors also cautioned that agreement of thermal rates did not uniquely establish the dynamics. [78,79]

### S6.2 Pyrophosphate: a conditional population-and-rate calculation

Hydrolysis cleaves the bridging P–O–P linkage of pyrophosphate to produce two orthophosphate ions. García-Martínez and co-workers calculated potential-of-mean-force barriers for $MgP_2O_7^{2-}$ and its protonated form, $MgHP_2O_7^{-}$. [82] For this conditional calculation, treat these heights as molar Eyring activation free energies, use the common prefactor $k_B T/h$, and set both transmission coefficients to unity:

$$k_j = \frac{k_B T}{h} \exp[-\Delta G_j^{\ddagger}/(RT)]. \tag{S13}$$

At 298.15 K, the converted source values 138.072 and 118.8256 kJ mol$^{-1}$ give $k_0 = 4.0188 \times 10^{-12}$ s$^{-1}$ and $k_H = 9.4606 \times 10^{-9}$ s$^{-1}$, a ratio of 2354. The conversion digits preserve the inputs; they do not indicate corresponding physical precision.

For rapid interconversion within a mixture restricted to these two complexes, let $f$ be the protonated fraction. The apparent rate and the protonated fraction of reactive flux, $\phi_H$, are

$$k_{app} = (1-f)k_0 + f k_H, f = \frac{k_{app} - k_0}{k_H - k_0}, \phi_H = \frac{f k_H}{k_{app}}. \tag{S14}$$

Inserting Stockbridge and Wolfenden's extrapolated $k_{app} = 2.8 \times 10^{-10}$ s$^{-1}$ gives $f = 0.02918$ and $\phi_H = 0.9861$. [81] This fraction is inferred from the rate, rather than determined independently.

Assigning this fraction to pH 8 solely as an illustrative reference gives $\mathrm{p}K_{a,ill} = 6.478$. For a fixed reference rate $k_{ref}$ with the same units as $k_{app}$, and at fixed temperature, ionic strength, and species rates,

$$f = \frac{1}{1 + 10^{\mathrm{pH} - \mathrm{p}K_{a,ill}}}, \frac{d \log_{10}(k_{app}/k_{ref})}{d\,\mathrm{pH}} = f - \phi_H. \tag{S15}$$

Table S4. Conditional pH dependence of the two-complex model at 298.15 K. The pH-8 rate is the assigned reference.

| pH | Protonated fraction (%) | $k_{app}$ (s$^{-1}$) | Rate relative to pH 8 |
|---|---|---|---|
| 7.5 | 8.681 | $8.249 \times 10^{-10}$ | 2.946 |
| 8.0 | 2.918 | $2.800 \times 10^{-10}$ | 1.000 |
| 8.7 | 0.596 | $6.040 \times 10^{-11}$ | 0.216 |

The predicted decrease is 13.66-fold across this interval; the local logarithmic slope at pH 8 is −0.957. This curve follows from assigning the rate-fitted fraction to pH 8 and fitting an illustrative $\mathrm{p}K_{a,ill}$; it is not an independent prediction of the historical pH dependence. The historical room-temperature rate was extrapolated from 61 pooled measurements at higher temperatures, with $R^2 = 0.88$. Its abstract specifies pH 8.5, whereas the three experimental recipes used nominal pH 7.5, 8.0, and 8.7. [81] Table S4 therefore does not test directly measured 298.15 K rates or reproduce the full composition of those recipes. The independently sourced equilibrium constants in Section S6.3 are not derived from this fitted $pK_{a,ill}$.

Table S5. Conditions pooled in the historical magnesium-containing series. Concentrations are analytical totals.

| Nominal pH | Mg (mM) | Pyrophosphate (mM) | Buffer; assay |
|---|---|---|---|
| 7.5 | 10 | 2 | 10 mM ethyl phosphonate; $^{31}$P NMR |
| 8.0 | 10 | 2 | 2 mM ethyl phosphonate; acid-molybdate |
| 8.7 | 0.2 | 0.02 | 2 mM sodium borate; acid-molybdate |

### S6.3 Independent speciation and missing rate inputs

The six $H^+$ and $Mg^{2+}$ constants compiled by Baykov and co-workers [83, 84] were measured by Shafranskii and co-workers using potentiometric titration. [85] Their solutions contained 1 mM pyrophosphate and 0–10 mM $MgCl_2$ in 0.1 M tetramethylammonium chloride at 25 °C.

Retain $P_2O_7^{4-}$, $HP_2O_7^{3-}$, $H_2P_2O_7^{2-}$, their three singly magnesium-bound forms, and neutral $Mg_2P_2O_7$. Use the reported 25 °C values $pK_{a3}=6.38$ and $pK_{a4}=9.11$. The association constants for binding magnesium to these three free pyrophosphate forms are $b_0=10^{5.89}$, $b_H=10^{3.60}$, and $b_{H2}=10^{2.10}$ L mol$^{-1}$. The second-magnesium association constant for $MgP_2O_7^{2-}$ is $b_2=10^{2.70}$ L mol$^{-1}$.

Table 1 of the original study reports errors of ±0.02 for both acid-dissociation $pK$ values. The errors in the base-10 logarithms of the numerical values of $b_0$, $b_H$, $b_{H2}$, and $b_2$ in L mol$^{-1}$ are ±0.07, ±0.10, ±0.10, and ±0.07, respectively. [85] The source does not define these uncertainties statistically or report parameter covariances. The population calculation uses central values without assigning confidence intervals.

For free magnesium concentration $m$, define

$$\begin{gathered} u=10^{9.11-\mathrm{pH}}, v=u\,10^{6.38-\mathrm{pH}}, \\ q_0=b_0 m, q_H=b_H m u, q_{H2}=b_{H2} m v, d=b_0 b_2 m^2, \\ D=1+u+v+q_0+q_H+q_{H2}+d. \end{gathered} \tag{S16}$$

These seven weights, divided by $D$, are the species fractions relative to total dissolved pyrophosphate, $P_T$. Its free fully deprotonated concentration is $P_T/D$. With all concentrations in mol L$^{-1}$, solve the magnesium balance

$$M_T=m+\frac{P_T}{D}(q_0+q_H+q_{H2}+2d) \tag{S17}$$

for $0\le m\le M_T$. At pH 8, $M_T=0.010$ and $P_T=0.002$ mol L$^{-1}$ give $m=0.006496$ mol L$^{-1}$ and Table S6.

Table S6. Illustrative 25 °C populations for the pH-8 recipe. Every percentage uses total dissolved pyrophosphate as its denominator.

| Species | Fraction (%) |
|---|---|
| $P_2O_7^{4-}$ | 0.00459 |
| $HP_2O_7^{3-}$ | 0.05908 |
| $H_2P_2O_7^{2-}$ | 0.00142 |
| $MgP_2O_7^{2-}$ | 23.124 |
| $MgHP_2O_7^{-}$ | 1.528 |
| $MgH_2P_2O_7$ | 0.00116 |
| $Mg_2P_2O_7$ | 75.282 |

The thermodynamic cycle gives $pK_{a,Mg} = 9.11 + 3.60 - 5.89 = 6.82$. The protonated fraction within only $MgP_2O_7^{2-} + MgHP_2O_7^{-}$ is therefore 6.20% at pH 8, distinct from its 1.53% fraction of total pyrophosphate in Table S6.

This illustration uses the tabulated $pK_a$ values directly with pH, without an additional activity correction. It omits alkali-metal complexation, buffer binding, changes in activity coefficients between media, and precipitation. The original titration methods mention converting hydrogen-ion activity to concentration in acidic solution but do not specify how that correction enters each tabulated $pK_a$. [85] Applying the constants to the historical solutions requires consistent activity conventions, each solution's composition, and temperature-dependent equilibrium data. Some heated samples contained precipitate, and reference [81] does not specify pH corrections at incubation temperature.

Temperature-dependent equilibrium data are available for part of this model. Irani measured the two weakest pyrophosphate acid dissociations over 0–65 °C in specified ionic media. [86] However, precipitation prevented measurement of the $MgP_2O_7^{2-}$ and $Mg_2P_2O_7$ binding constants at 65 °C. The magnesium-complex measurements used an ionic strength of 1.0 M, adjusted with tetramethylammonium bromide. Their temperature dependence cannot be transferred directly to the historical hydrolysis solutions or combined with the nominal 0.1 M constants above without accounting for the different ionic media.

Zea and co-workers measured apparent magnesium–pyrophosphate binding by isothermal titration calorimetry in 100 mM HEPES at pH 7.5 and 37 °C. [87] Their Table S1 reports $K = (4.10 \pm 0.20) \times 10^4$ L mol$^{-1}$ and $\Delta H_{bind} = -7.41 \pm 0.04$ kJ mol$^{-1}$ from a model with one binding site. Figure S6 gives the pyrophosphate-specific titration conditions: 4.5 mM $MgCl_2$ added to 0.5 mM pyrophosphate. No correction for buffer protonation heat is reported. These conditional parameters do not separately determine the enthalpies of the protonation and magnesium-binding steps in the seven-species model.

Reference [81] cites that calorimetry when quoting a magnesium–pyrophosphate dissociation constant at 25 °C and an enthalpy of +15.5 kJ mol$^{-1}$. The cited pyrophosphate binding enthalpy instead corresponds to +7.41 kJ mol$^{-1}$ for the reverse process at 37 °C. Reference [81] gives no conversion that reconciles the two values or measurement conditions. The cited enthalpy therefore needs clarification before it is used to extrapolate species populations. This discrepancy alone does not establish an error in the measured hydrolysis rates.

A total initial rate of pyrophosphate consumption requires $k_{app} = \sum_i f_i k_i$, assuming species exchange is rapid relative to hydrolysis. Each $f_i$ must refer to total dissolved pyrophosphate. Reference [82] supplies no hydrolysis barrier for $Mg_2P_2O_7$; additional populated species require rates or justified bounds. The observed phosphate-appearance rate is twice the pyrophosphate-consumption rate for this stoichiometry. At appreciable conversion, phosphate accumulation and magnesium redistribution can change the species fractions. Comparison with complete reaction time courses therefore requires evolving mass balances, and any precipitation, rather than a constant initial $k_{app}$.

Absolute species rates require consistent reactant-basin normalization, channel multiplicities, and transmission coefficients. The authors' response accompanying reference [82] estimates absolute functional errors of about 4.2–8.4 kJ mol$^{-1}$, corresponding to rate factors of approximately 5–30 at 298.15 K. These systematic errors are separate from the reported sampling uncertainty; errors shared across species also affect the rate mixture differently from errors in their barrier differences. The 298.15 K PMFs and 25 °C equilibrium constants do not determine rates at the original elevated incubation temperatures. Those predictions require temperature-dependent equilibria and species rates, which may need additional molecular calculations or independent measurements. The individual historical rate records and solution conditions are also needed. The calculations above neither validate nor refute the protonation interpretation; they specify the additional inputs required for a quantitative test.

### S6.4 Triosephosphate isomerase and prior sensitivity

#### S6.4.1 Reaction model and source version

Triosephosphate isomerase (TIM) interconverts dihydroxyacetone phosphate, S, and glyceraldehyde 3-phosphate, P. The kinetic scheme is $E + S \rightleftharpoons ES \rightleftharpoons EZ \rightleftharpoons EP \rightleftharpoons E + P$, where EZ denotes the enzyme-bound enediol(ate) intermediate. Number the successive forward steps $k_1$, $k_2$, $k_3$, and $k_4$, and the reverse steps $k_{-1}$, $k_{-2}$, $k_{-3}$, and $k_{-4}$. Association rates $k_1$ and $k_{-4}$ have units mM$^{-1}$ s$^{-1}$; the other rates have units s$^{-1}$. Define $K_1 = k_1 / k_{-1}$ in mM$^{-1}$ and $K_4 = k_4 / k_{-4}$ in mM.

Toney combined turnover, isotope-effect, viscosity, and intermediate-partition observations with NMR limits on bound species to infer microscopic rates and free-energy profiles. He explicitly identified the weak constraint on the intermediate energy. [88] Barr subsequently applied Bayesian inference to this problem. [89] I independently implemented the likelihood and prior specified in the archived simulate.stan and enrg.R files obtained from the September 2022 version of Barr's preprint. The calculation uses those files' inputs and conventions. It is not a reproduction of the posterior reported in the July 2023 revision. The executable implementation, input values, sampled distributions, profile fits, and figure-generation code are provided in Supplementary Data 1 (https://doi.org/10.5281/zenodo.23557458).

The molar intermediate energy difference, equilibrium intermediate fraction $f_{EZ}=[EZ]/([ES]+[EZ]+[EP])$, and equilibrium ratio of bound product to bound substrate are

$$\Delta G_{EZ-ES}=-RT\ln\left(\frac{k_2}{k_{-2}}\right) \quad (S18)$$

$$f_{EZ}=\left(\frac{k_{-2}}{k_2}+\frac{k_3}{k_{-3}}+1\right)^{-1},q=\frac{[EP]}{[ES]}=\frac{k_2k_3}{k_{-2}k_{-3}} \quad (S19)$$

The energy convention is $T=298$ K, as in the archived implementation, with $R=0.00831446261815324$ kJ mol$^{-1}$ K$^{-1}$. The reference state is ES; the calculated difference is neither an activation barrier nor the energy relative to free enzyme and substrate.

### S6.4.2 Inputs and likelihood

Table S7 lists the twelve input values and assigned standard deviations. The model predicts forward and reverse turnover rates, Michaelis constants, kinetic isotope effects, the viscosity dependence of the specificity constant, an isotope-partition quantity, the intermediate fraction, and the equilibrium constant. The archived implementation treats all twelve residuals as independent Gaussian variables. With prediction $F_j(\theta)$, input $y_j$, and assigned standard deviation $\sigma_j$, $\chi^2=\sum_{j=1}^{12}[(F_j-y_j)/\sigma_j]^2$ and the likelihood is proportional to $\exp(-\chi^2/2)$. This assumes independence and uses the supplied uncertainties; it does not estimate covariance from the original measurements. Barr derived the archived equilibrium input from literature Haldane estimates, rather than an independent direct measurement of the unhydrated-species equilibrium. Its covariance with the kinetic inputs has not been reconstructed; independence is part of this sensitivity calculation’s assumptions.

Table S7. Inputs retained from the 2022 implementation. The isotope effects are dimensionless light/heavy ratios for deuterium substitution. The specificity isotope effects refer to $k_{cat}/K_M$, not to $K_M$ alone.

| Quantity | Input | Assigned SD | Unit |
|---|---|---|---|
| Forward turnover $k_f$ | 750 | 50 | s$^{-1}$ |
| Reverse turnover $k_r$ | 8350 | 350 | s$^{-1}$ |
| Forward $K_{M,f}$ | 1.35 | 0.15 | mM |
| Reverse $K_{M,r}$ | 0.05 | 0.01 | mM |
| Forward turnover isotope effect | 3.4 | 0.1 | — |
| Reverse specificity isotope effect | 1.6 | 0.1 | — |
| Forward specificity isotope effect | 3.4 | 0.1 | — |
| Reverse turnover isotope effect | 1.6 | 0.1 | — |
| Viscosity slope | 0.8 | 0.1 | — |
| Isotope-partition quantity $\theta$ | 3 | 1 | — |
| Intermediate fraction $f_{EZ}$ | 0.025 | 0.025 | — |

| Quantity | Input | Assigned SD | Unit |
|---|---|---|---|
| Equilibrium constant $K_{eq}$ | 0.0035 | 0.0001 | — |

The intermediate input requires a qualification. It encodes a reported upper limit of approximately 0.05 as $0.025 \pm 0.025$; it is not a positive measurement of a 2.5% intermediate population. This encoding gives different likelihoods to populations that all satisfy the upper limit. I retained it in both prior comparisons to isolate the effect of changing the rate prior. A model for censored observations would require information about the detection limit and measurement errors.

The forward and reverse specificity isotope effects are calculated from the same microscopic scheme as the turnover rates. The viscosity input is the slope of $(k_{cat}/K_M)_{\eta_0}/(k_{cat}/K_M)_{\eta}$ against $\eta/\eta_0$, where $\eta_0$ is the viscosity without added viscosogen. The isotope-partition quantity $\theta$ is the ratio of the net rates of DHAP formation and GAP formation from EZ. [88] The model expression for the viscosity slope is $(k_{-1}k_{-2}k_{-3}+k_2k_3k_4)/(k_{-1}k_{-2}k_{-3}+k_2k_3k_4+k_{-1}k_3k_4+k_{-1}k_{-2}k_4)$, and the isotope-partition quantity is $\theta = k_{-2}(1+k_{-3}/k_4)/[k_3(1+k_2/k_{-1})]$. The equilibrium constant is $K_{eq} = K_1k_2k_3K_4/(k_{-2}k_{-3})$. Supplementary Data 1 provides the complete forward equations for all twelve predictions. The isotope-effect equations retain the source's rapid isotope-washout assumption, with one isotopically sensitive chemical step for each labeled reaction direction. The calculation does not test this assumption or resolve additional protonation or conformational substates.

### S6.4.3 Priors and computation

The six directly parameterized microscopic rates are $k_2$, $k_3$, $k_4$, $k_{-1}$, $k_{-2}$, and $k_{-3}$. The first three have lower bounds of 750 $s^{-1}$ and the last three have lower bounds of 8350 $s^{-1}$. All six have an upper bound of $10^{12}$ $s^{-1}$. The bounds on $K_1$ and $K_4$ are 0–1000 in their stated units; those on the intrinsic isotope effects for steps 2 and −3 are 0–500. For numerical optimization only, a lower floor of $10^{-14}$ in the corresponding numerical units was applied to parameters whose source lower bound is zero. Posterior sampling used logarithmic coordinates and the original positive support.

In the archived model, the six microscopic rates share an exponential rate parameter with a Gamma(1,1) hyperprior. To state its integrated form without adding dimensional quantities, let $u_i = k_i/(1\,s^{-1})$. Integrating the shared parameter in the literal archived target gives

$$p(u) \propto \left(1+\sum_{i=1}^{6} u_i\right)^{-7} \qquad \text{(S20)}$$

within the specified bounds. The archived target does not include parameter-dependent normalizing factors for truncated exponential distributions; none were added here. The alternative replaces only this six-rate prior by independent bounded log-uniform priors, $p(u) \propto \prod_i u_i^{-1}$. The likelihood, bounds, and priors on the other parameters remain the same. In particular, the two identical exponential prior statements for $K_4$ in the archived code were both retained; together they give $\exp[-0.05K_4/(1\,\mathrm{mM})]$. The prior on $K_1$ is flat within its bounds, and both intrinsic isotope effects have log-normal priors with logarithmic mean 1 and standard deviation 0.5. The Jacobian for all ten logarithmically transformed parameters was included.

I sampled the two targets with the affine-invariant ensemble sampler in emcee 3.1.6. [91] Each target used two independently initialized ensembles, with seeds 71 and 97 and 48 interacting walkers per ensemble. After 4000 warm-up steps, the original-prior runs each retained 16,000 steps per walker; the alternative-prior runs each retained 32,000. The retained length exceeded 60 estimated integrated autocorrelation times for each sampled coordinate. The two ensembles and the first and second halves gave consistent energy quantiles. Every twentieth step was saved for the reported summaries. The walkers within an ensemble are not independent chains.

Profile fits used bounded nonlinear least squares with multiple starting points. At each prescribed intermediate energy, the relation $\ln k_{-2} = \ln k_2 + \Delta G_{\mathrm{EZ-ES}}/(RT)$ removed one free parameter, and the remaining parameters were optimized without a prior. The relative profile likelihood is $\exp[-(\chi^2_{\mathrm{profile}} - \chi^2_{\mathrm{min}})/2]$.

### S6.4.4 Results and interpretation

Table S8. Sensitivity of posterior summaries to the prior on six microscopic rates. Intervals contain the central 90% of the sampled posterior under the stated model. They are conditional on the likelihood, input uncertainties, bounds, and remaining priors.

| Quantity | Archived shared-scale prior | Bounded log-uniform prior |
|---|---|---|
| Median $\Delta G_{\mathrm{EZ-ES}}$ (kJ mol$^{-1}$) | 7.3 | 17.2 |
| Central 90% interval (kJ mol$^{-1}$) | 6.0–9.1 | 7.8–38.2 |
| Median $q = [\mathrm{EP}]/[\mathrm{ES}]$ | 0.083 | 0.085 |
| Central 90% interval for $q$ | 0.054–0.125 | 0.054–0.130 |

The minimum residual sum of squares was $\chi^2 = 0.2064$. Constraining the intermediate energy to 10 or 25 kJ mol$^{-1}$ gave $\chi^2 = 0.2536$ or 1.0413, respectively; every prediction remained within one assigned standard deviation of its input. At 40 kJ mol$^{-1}$, $\chi^2 = 1.0443$, corresponding to a relative profile likelihood of 0.658 (Figure S1A).

The paired change $k_{-2} \to c\,k_{-2}$ and $k_3 \to c\,k_3$ helps explain the weak constraint at high intermediate energy. It increases $\Delta G_{\mathrm{EZ-ES}}$ by $RT \ln c$, while leaving $q$, $K_{\mathrm{eq}}$, the viscosity slope, $\theta$, and the two specificity isotope effects unchanged. The turnover rates and associated Michaelis constants approach finite limits as $c$ grows, while $f_{\mathrm{EZ}}$ decreases. The resulting weak sensitivity over a range of parameters does not establish exact nonidentifiability throughout the model.

The alternative posterior has substantial probability near the imposed rate ceiling: 25.4% of saved draws place at least one of the six rates above $10^{11}$ s$^{-1}$, and 8.0% place one above $5 \times 10^{11}$ s$^{-1}$. I assessed lower ceilings by retaining only saved draws in which all six rates lie below the selected ceiling. This conditions the sampled posterior on the restricted support; it is not a fresh MCMC calculation. Ceilings of $10^9$, $10^{10}$, and $10^{11}$ $s^{-1}$ retain 34.9%, 52.8%, and 74.6% of the alternative-prior draws. Their energy medians are 12.5, 14.0, and 15.5 kJ $mol^{-1}$, with central 90% intervals of 7.4–24.5, 7.5–29.1, and 7.7–33.7 kJ $mol^{-1}$, respectively. All saved shared-scale-prior draws remain below $10^9$ $s^{-1}$. The prior sensitivity therefore persists under these restrictions, while the alternative median and upper tail depend materially on the rate ceiling. Neither the alternative prior nor any of these ceilings is established as physically preferred; the upper endpoints are not experimentally determined energy limits. The diagnostic script, retained counts, and quantiles are provided in Supplementary Data 1.

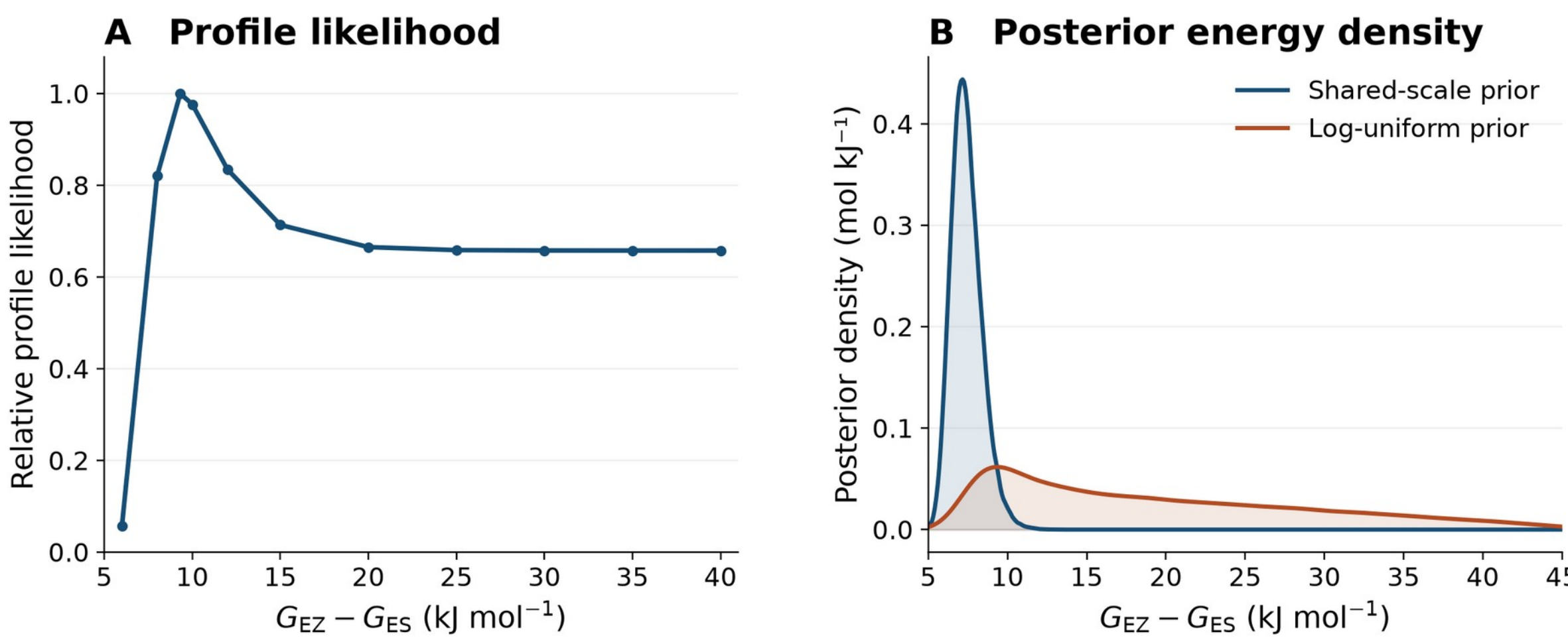

Figure S1. Inference of the TIM intermediate energy using the archived 2022 model. (A) Relative profile likelihood after optimizing the other parameters without a prior. Points are calculated fits, with connecting lines as guides. (B) Posterior energy densities using the archived shared-scale prior or independent bounded log-uniform priors on the same six microscopic rates. Curves are kernel-density estimates from the saved posterior draws. Both calculations use the same twelve Gaussian inputs, other priors, and parameter bounds; energies use 298 K. The alternative calculation samples rates near the imposed $10^{12}$ $s^{-1}$ ceiling, on which its energy interval is conditional. Table S8 reports quantiles directly from the draws rather than from the density curves.

Deng and co-workers used the absence of an isotope-shifted GAP carbonyl signal in equilibrium yeast-TIM/substrate mixtures to estimate the aldehydic GAP population as less than 5% of the keto-DHAP population. [90] Hydrated GAP and DHAP were not quantified by these measurements. Comparing this approximate spectroscopic upper bound with the modeled $q$ therefore requires accounting for the detected chemical forms, free and enzyme-bound contributions, and differences in solvent isotope composition, pH, and temperature. The infrared result does not directly measure $f_{\mathrm{EZ}}$ or $\Delta G_{\mathrm{EZ-ES}}$. A useful additional constraint would be a measurement sensitive to the equilibrium EZ population, or a time-resolved signal that constrains its decay, under the same enzyme, solvent, pH, and temperature conditions as the kinetic data. Temporal resolution and signal sensitivity would need to be established before selecting an experiment. The analysis would need calibrated signal responses and a likelihood for any non-detection. A changed aldehydic-GAP bound would constrain a different population ratio and would not directly determine the EZ energy.

### S6.5 Proposed joint analysis of Criegee optical records

A reanalysis of the $CH_3CHOO$ self- and cross-reactions should fit the measured signals rather than treat previously inferred concentrations as independent data. [92] If $S$ and $A$ denote the *syn* and *anti* number concentrations, the natural-log ultraviolet optical depth is $L(\sigma_s S+\sigma_a A)$, before adding background contributions; $L$ is the optical path length. The selective infrared signal is proportional to $S$. Its proportionality factor was calibrated through the ultraviolet record, so the absolute scales are correlated. The precursor measurement constrains production, subject to uncertainty in the total Criegee yield and conformer branching.

For the bimolecular loss terms alone,

$$\frac{\mathrm{d}S}{\mathrm{d}t}=-2k_{ss}S^2-k_{as}SA,\ \frac{\mathrm{d}A}{\mathrm{d}t}=-2k_{aa}A^2-k_{as}SA. \quad \text{(S21)}$$

Formation, other reactions, and physical loss terms would be included as required by the experimental conditions. The factor of two counts the two molecules consumed in a self-reaction event; the convention must remain consistent when comparing coefficients.

Interpolation of Lin and Lin's Table S5 gives $\sigma_a$ at 335 nm of approximately $1.06\times10^{-21}$ m$^2$, compared with the earlier $0.85\times10^{-21}$ m$^2$. [92, 93] Nearby spectral averaging changes this comparison slightly. The reported 12.5% absolute normalization uncertainty is shared across the spectrum and cannot be reduced by treating different wavelengths as independent measurements. The *syn* calibration also requires consistent treatment. Lin and Lin's inferred conformer ratio depends on the new calibration and applies to different generation conditions; it is not an independent extra constraint on Kao's experiments.

The proposed comparison would first reproduce the original analysis, then replace the calibration information while holding the reaction model and fitting procedure fixed. Joint uncertainty estimates for $k_{aa}/k_{ss}$ and $k_{as}$ would reveal which combinations are constrained. Changes to competing-reaction models should be assessed separately. Original detector records, calibration information, and experimental conditions are needed; fitted summaries alone are insufficient.

Trac and co-workers' predictions should be compared at matched temperature, pressure, and bath gas. [94] The appropriate theoretical quantity is total Criegee loss through both fragmentation and stabilized-dimer formation. The predictions should remain a separate comparison, rather than being imposed during fitting and subsequently counted as validation.

At 298 K in 5 Torr He, Trac and co-workers' Tables S4–S6 give approximately equal stabilization and acetaldehyde-channel coefficients for each conformer pair. Their main text instead states that stabilization is about four times the acetaldehyde-channel coefficient for the cross-reaction and more than 2.5 times for the *syn* self-reaction. The reported total coefficients agree across the main text and SI, but the branching cannot be resolved from these documents. A quantitative comparison of the channels would therefore require clarification of their rate coefficients and statistical conventions. [94]

### S6.6 Diamond dissolution and thermal-history tests

Li and co-workers measured diamond recession beneath solid nickel and cobalt films in the presence of water vapor and interpreted the higher-water conditions as predominantly controlled by carbon removal at the buried interface. [95] The measured observable is a recession velocity, not a first-order molecular rate constant. The original study already included rapid heating and cooling in a Joule-heated cold-wall system. The proposed work concerns systematic control of the full sample-temperature history and its effect on local recession.

For a fixed interface and film state, the local depth removed is the time integral of the recession velocity along the measured temperature history. Changing ramp times can therefore change the final depth without changing the mechanism. A persistent change in the rate after return to the same temperature would require an additional explanation, such as changed interface structure, film continuity, oxidation, or carbon inventory. Use matched specimens stopped at successive stages to estimate interval-average rates, and compare the excursion sequence with a sequence held at the reference temperature for the same elapsed duration. Include replicate variation and uncertainty in temperature, timing, height, and spatial registration.

Before/after maps require reference regions whose height remains unchanged and checks that metal removal does not change the diamond surface. Laser microscopy and possibly AFM can resolve spatial variation within their measured resolution; neither instrument alone determines the operating atomistic pathway. Matched specimens with the metal retained would allow examination of the buried interface and film. A quenched state may differ from the state during reaction, so a structural assignment requires attention to changes during cooling.

Across the extended temperature range, assess sensitivity to water supply and film thickness while characterizing continuity and oxidation. These comparisons can bound external-removal and transport contributions; a water-supply plateau alone does not establish exclusive interface control. The group’s deposited films adopt an epitaxial arrangement. A separate possible comparison would bond pre-existing Ni(111) crystals at selected in-plane rotations, conditional on the orientations and adequate interfacial contact surviving the thermal treatment. No result from these proposed experiments is reported here.

## References cited in the Supplementary Materials

Numbering follows the main article.

[1] P. Hänggi, P. Talkner and M. Borkovec, “Reaction-rate theory: fifty years after Kramers”, Rev. Mod. Phys., 1990, 62, 251–341. DOI: 10.1103/RevModPhys.62.251.

[9] M. Bello, W. Yang and A. Heyden, “First-Principles Modeling in Heterogenous Catalysis: Where Are We?”, ACS Catal., 2026, 16, 9618–9644. DOI: 10.1021/acscatal.6c01630.

[10] D. G. Truhlar, B. C. Garrett and S. J. Klippenstein, “Current Status of Transition-State Theory”, J. Phys. Chem., 1996, 100, 12771–12800. DOI: 10.1021/jp953748q.

[11] R. R. Krug, W. G. Hunter and R. A. Grieger, “Enthalpy-Entropy Compensation. 1. Some Fundamental Statistical Problems Associated with the Analysis of van’t Hoff and Arrhenius Data”, J. Phys. Chem., 1976, 80, 2335–2341. DOI: 10.1021/j100562a006; “Enthalpy-Entropy Compensation. 2. Separation of the Chemical from the Statistical Effect”, J. Phys. Chem., 1976, 80, 2341–2351. DOI: 10.1021/j100562a007.

[12] O. Exner, “Concerning the Isokinetic Relationship”, Nature, 1964, 201, 488–490. DOI: 10.1038/201488b0.

[13] O. Exner, “The Enthalpy-Entropy Relationship in Organic Reactions”, Collect. Czech. Chem. Commun., 1975, 40, 2762–2780. DOI: 10.1135/cccc19752762.

[14] J. H. van ’t Hoff, Études de dynamique chimique, Frederik Muller & Co., Amsterdam, 1884.

[15] S. Arrhenius, “Über die Reaktionsgeschwindigkeit bei der Inversion von Rohrzucker durch Säuren”, Z. Phys. Chem., 1889, 4, 226–248. DOI: 10.1515/zpch-1889-0416.

[16] M. Trautz, “Das Gesetz der Reaktionsgeschwindigkeit und der Gleichgewichte in Gasen. Bestätigung der Additivität von $C_v$ – 3/2 R. Neue Bestimmung der Integrationskonstanten und der Moleküldurchmesser”, Z. Anorg. Allg. Chem., 1916, 96, 1–28. DOI: 10.1002/zaac.19160960102.

[17] W. C. McC. Lewis, “Studies in Catalysis. Part IX. The Calculation in Absolute Measure of Velocity Constants and Equilibrium Constants in Gaseous Systems”, J. Chem. Soc., Trans., 1918, 113, 471–492. DOI: 10.1039/CT9181300471.

[18] J. A. Christiansen, Reaktionskinetiske Studier: Problemer fra de kemiske Reaktioners Kinetik med en eksperimentel Undersøgelse af Fosgenets termiske Spaltning, doctoral (dr. phil.) thesis, University of Copenhagen, Vilhelm Priors Kgl. Hofboghandel, Copenhagen, 1921.

[19] H. Pelzer and E. Wigner, “Über die Geschwindigkeitskonstante von Austauschreaktionen”, Z. Phys. Chem. B, 1932, 15, 445–471. DOI: 10.1515/zpch-1932-1539.

[20] R. C. Tolman, “Statistical Mechanics Applied to Chemical Kinetics”, J. Am. Chem. Soc., 1920, 42, 2506–2528. DOI: 10.1021/ja01457a008.

[21] R. Marcelin, Contribution à l’étude de la cinétique physico-chimique, doctoral thesis, Faculté des sciences de Paris, Gauthier-Villars et Cie, Paris, 1914. Digitized copy, Internet Archive, Leiden University collection: https://archive.org/details/IA41551612_0077.

[22] R. Marcelin, “Contribution à l’étude de la cinétique physico-chimique”, Ann. Phys. (Paris), 1915, série 9, 3, 120–231. DOI: 10.1051/anphys/191509030120.

[23] H. Eyring and M. Polanyi, “Über einfache Gasreaktionen”, Z. Phys. Chem. B, 1931, 12, 279–311; English translation, “On Simple Gas Reactions”, Z. Phys. Chem., 2013, 227, 1221–1245. DOI: 10.1524/zpch.2013.9023.

[24] E. Wigner, “The Transition State Method”, Trans. Faraday Soc., 1938, 34, 29–41. DOI: 10.1039/tf9383400029.

[25] H. Eyring, “The Activated Complex in Chemical Reactions”, J. Chem. Phys., 1935, 3, 107–115. DOI: 10.1063/1.1749604.

[26] H. Eyring, “The Activated Complex and the Absolute Rate of Chemical Reactions”, Chem. Rev., 1935, 17, 65–77. DOI: 10.1021/cr60056a006.

[27] J. O. Hirschfelder and E. Wigner, “Some Quantum-Mechanical Considerations in the Theory of Reactions Involving an Activation Energy”, J. Chem. Phys., 1939, 7, 616–628. DOI: 10.1063/1.1750500.

[28] S. Glasstone, K. J. Laidler and H. Eyring, The Theory of Rate Processes: The Kinetics of Chemical Reactions, Viscosity, Diffusion and Electrochemical Phenomena, McGraw-Hill, New York, 1941.

[29] M. G. Evans and M. Polanyi, “Some applications of the transition state method to the calculation of reaction velocities, especially in solution”, Trans. Faraday Soc., 1935, 31, 875–894. DOI: 10.1039/tf9353100875.

[30] W. F. K. Wynne-Jones and H. Eyring, “The Absolute Rate of Reactions in Condensed Phases”, J. Chem. Phys., 1935, 3, 492–502. DOI: 10.1063/1.1749713; erratum, “Errata: The Absolute Rate of Reactions in Condensed Phases”, J. Chem. Phys., 1936, 4, 740. DOI: 10.1063/1.1749779.

[31] F. A. Lindemann, S. Arrhenius, I. Langmuir, N. R. Dhar, J. Perrin and W. C. McC. Lewis, "Discussion on 'The Radiation Theory of Chemical Action'", Trans. Faraday Soc., 1922, 17, 598–606. DOI: 10.1039/TF9221700598.

[32] C. N. Hinshelwood, "On the Theory of Unimolecular Reactions", Proc. R. Soc. London, Ser. A, 1926, 113, 230–233. DOI: 10.1098/rspa.1926.0149.

[33] O. K. Rice and H. C. Ramsperger, "Theories of Unimolecular Gas Reactions at Low Pressures", J. Am. Chem. Soc., 1927, 49, 1617–1629. DOI: 10.1021/ja01406a001.

[34] L. S. Kassel, "Studies in Homogeneous Gas Reactions I", J. Phys. Chem., 1928, 32, 225–242. DOI: 10.1021/j150284a007.

[35] R. A. Marcus, "Unimolecular Dissociations and Free Radical Recombination Reactions", J. Chem. Phys., 1952, 20, 359–364. DOI: 10.1063/1.1700424.

[38] H. A. Kramers, "Brownian motion in a field of force and the diffusion model of chemical reactions", Physica, 1940, 7, 284–304. DOI: 10.1016/S0031-8914(40)90098-2.

[40] R. A. Marcus, "On the Theory of Oxidation-Reduction Reactions Involving Electron Transfer. I", J. Chem. Phys., 1956, 24, 966–978. DOI: 10.1063/1.1742723.

[41] P. Siders and R. A. Marcus, "Quantum Effects for Electron-Transfer Reactions in the 'Inverted Region'", J. Am. Chem. Soc., 1981, 103, 748–752. DOI: 10.1021/ja00394a004.

[43] G. H. Vineyard, "Frequency factors and isotope effects in solid state rate processes", J. Phys. Chem. Solids, 1957, 3, 121–127. DOI: 10.1016/0022-3697(57)90059-8.

[44] E. Wigner, "Calculation of the Rate of Elementary Association Reactions", J. Chem. Phys., 1937, 5, 720–725. DOI: 10.1063/1.1750107.

[45] J. Horiuti, "On the Statistical Mechanical Treatment of the Absolute Rate of Chemical Reaction", Bull. Chem. Soc. Jpn., 1938, 13, 210–216. DOI: 10.1246/bcsj.13.210.

[46] J. C. Keck, "Variational Theory of Chemical Reaction Rates Applied to Three-Body Recombinations", J. Chem. Phys., 1960, 32, 1035–1050. DOI: 10.1063/1.1730846.

[47] P. Pechukas and F. J. McLafferty, "On transition-state theory and the classical mechanics of collinear collisions", J. Chem. Phys., 1973, 58, 1622–1625. DOI: 10.1063/1.1679404.

[48] B. C. Garrett and D. G. Truhlar, "Generalized Transition State Theory. Canonical Variational Calculations Using the Bond Energy-Bond Order Method for Bimolecular Reactions of Combustion Products", J. Am. Chem. Soc., 1979, 101, 5207–5217. DOI: 10.1021/ja00512a016.

[49] J. L. Bao and D. G. Truhlar, "Variational transition state theory: theoretical framework and recent developments", Chem. Soc. Rev., 2017, 46, 7548–7596. DOI: 10.1039/c7cs00602k.

[50] T. Yamamoto, "Quantum Statistical Mechanical Theory of the Rate of Exchange Chemical Reactions in the Gas Phase", J. Chem. Phys., 1960, 33, 281–289. DOI: 10.1063/1.1731099.

[51] D. Chandler, "Statistical mechanics of isomerization dynamics in liquids and the transition state approximation", J. Chem. Phys., 1978, 68, 2959–2970. DOI: 10.1063/1.436049.

[52] W. H. Miller, S. D. Schwartz and J. W. Tromp, "Quantum mechanical rate constants for bimolecular reactions", J. Chem. Phys., 1983, 79, 4889–4898. DOI: 10.1063/1.445581.

[53] I. R. Craig and D. E. Manolopoulos, "Chemical reaction rates from ring polymer molecular dynamics", J. Chem. Phys., 2005, 122, 084106. DOI: 10.1063/1.1850093.

[54] P. G. Bolhuis, D. Chandler, C. Dellago and P. L. Geissler, "Transition Path Sampling: Throwing Ropes Over Rough Mountain Passes, in the Dark", Annu. Rev. Phys. Chem., 2002, 53, 291–318. DOI: 10.1146/annurev.physchem.53.082301.113146.

[55] W. E and E. Vanden-Eijnden, “Towards a Theory of Transition Paths”, J. Stat. Phys., 2006, 123, 503–523. DOI: 10.1007/s10955-005-9003-9.

[56] E. Wigner, “Über das Überschreiten von Potentialschwellen bei chemischen Reaktionen”, Z. Phys. Chem. B, 1932, 19, 203–216. DOI: 10.1515/zpch-1932-1920.

[57] C. Eckart, “The Penetration of a Potential Barrier by Electrons”, Phys. Rev., 1930, 35, 1303–1309. DOI: 10.1103/PhysRev.35.1303.

[58] R. P. Bell, “The Application of Quantum Mechanics to Chemical Kinetics”, Proc. R. Soc. London, Ser. A, 1933, 139, 466–474. DOI: 10.1098/rspa.1933.0031.

[59] R. P. Bell, “The tunnel effect correction for parabolic potential barriers”, Trans. Faraday Soc., 1959, 55, 1–4. DOI: 10.1039/tf9595500001.

[60] W. H. Miller, “Quantum mechanical transition state theory and a new semiclassical model for reaction rate constants”, J. Chem. Phys., 1974, 61, 1823–1834. DOI: 10.1063/1.1682181.

[63] C. Dellago, P. G. Bolhuis, F. S. Csajka and D. Chandler, “Transition path sampling and the calculation of rate constants”, J. Chem. Phys., 1998, 108, 1964–1977. DOI: 10.1063/1.475562.

[64] W. E, W. Ren and E. Vanden-Eijnden, “String method for the study of rare events”, Phys. Rev. B, 2002, 66, 052301. DOI: 10.1103/PhysRevB.66.052301.

[65] L. Maragliano, A. Fischer, E. Vanden-Eijnden and G. Ciccotti, “String method in collective variables: Minimum free energy paths and isocommittor surfaces”, J. Chem. Phys., 2006, 125, 024106. DOI: 10.1063/1.2212942.

[66] R. G. Mullen, J.-E. Shea and B. Peters, “Transmission Coefficients, Committors, and Solvent Coordinates in Ion-Pair Dissociation”, J. Chem. Theory Comput., 2014, 10, 659–667. DOI: 10.1021/ct4009798.

[68] L. Zhang, F. Nitz, D. Borodin, A. M. Wodtke and H. Guo, “Ring Polymer Molecular Dynamics Rates for Hydrogen Recombinative Desorption on Pt(111)”, Precis. Chem., 2025, 3, 319–325. DOI: 10.1021/prechem.4c00099.

[69] Y. Yang, S. Zhang, K. D. Ranasinghe, O. Isayev and A. E. Roitberg, “Machine Learning of Reactive Potentials”, Annu. Rev. Phys. Chem., 2024, 75, 371–395. DOI: 10.1146/annurev-physchem-062123-024417.

[72] X. Gui, W. Fan, J. Sun and Y. Li, “New Stable and Fast Ring-Polymer Molecular Dynamics for Calculating Bimolecular Rate Coefficients with an Example of OH + $CH_4$”, J. Chem. Theory Comput., 2022, 18, 5203–5212. DOI: 10.1021/acs.jctc.2c00522.

[73] J. Li and H. Guo, “Communication: An accurate full 15 dimensional permutationally invariant potential energy surface for the OH + $CH_4$ → $H_2O$ + $CH_3$ reaction”, J. Chem. Phys., 2015, 143, 221103. DOI: 10.1063/1.4937570.

[74] S. Vari, C. de Falco and C. Cavallotti, “NN-VRCTST: Neural Network Potentials Meet Variable Reaction Coordinate Transition State Theory for the Rate Constant Determination of Barrierless Reactions”, J. Chem. Theory Comput., 2025, 21, 10373–10384. DOI: 10.1021/acs.jctc.5c01288.

[75] G. Lazzeri, H. Jung, P. G. Bolhuis and R. Covino, “Molecular Free Energies, Rates, and Mechanisms from Data-Efficient Path Sampling Simulations”, J. Chem. Theory Comput., 2023, 19, 9060–9076. DOI: 10.1021/acs.jctc.3c00821.

[76] O. Blumer, S. Reuveni and B. Hirshberg, “Short-Time Infrequent Metadynamics for Improved Kinetics Inference”, J. Chem. Theory Comput., 2024, 20, 3484–3491. DOI: 10.1021/acs.jctc.4c00170.

[77] L. B. Harding, A. I. Maergoiz, J. Troe and V. G. Ushakov, “Statistical rate theory for the HO + O $\rightleftharpoons$ $HO_2$ $\rightleftharpoons$ H + $O_2$ reaction system: SACM/CT calculations between 0 and 5000 K”, J. Chem. Phys., 2000, 113, 11019–11034. DOI: 10.1063/1.1314374.

[78] J. Troe and V. G. Ushakov, “Theoretical studies of the HO + O $\rightleftharpoons$ $HO_2$ $\rightleftharpoons$ H + $O_2$ reaction. II. Classical trajectory calculations on an ab initio potential for temperatures between 300 and 5000 K”, J. Chem. Phys., 2001, 115, 3621–3628. DOI: 10.1063/1.1388201.

[79] J. Troe, “The Struggle for Precise Rate Constants in Gas Phase Reaction Kinetics: The Reaction H + $O_2$ $\rightleftharpoons$ HO + O”, Z. Phys. Chem., 2003, 217, 1303–1317. DOI: 10.1524/zpch.217.10.1303.20494.

[80] B. Ruscic, D. Feller, D. A. Dixon, K. A. Peterson, L. B. Harding, R. L. Asher and A. F. Wagner, “Evidence for a Lower Enthalpy of Formation of Hydroxyl Radical and a Lower Gas-Phase Bond Dissociation Energy of Water”, J. Phys. Chem. A, 2001, 105, 1–4. DOI: 10.1021/jp003711s.

[81] R. B. Stockbridge and R. Wolfenden, “Enhancement of the Rate of Pyrophosphate Hydrolysis by Nonenzymatic Catalysts and by Inorganic Pyrophosphatase”, J. Biol. Chem., 2011, 286, 18538–18546. DOI: 10.1074/jbc.M110.214510.

[82] A. García-Martínez, M. de la Puente, M. Cunin, K. Zinovjev, J. J. Ruiz-Pernía, D. Laage and I. Tuñón, “Protonation and magnesium ions shape the transition state diversity of phosphoanhydride hydrolysis in water”, Nat. Commun., 2026, 17, 6878. DOI: 10.1038/s41467-026-73697-7.

[83] A. A. Baykov, V. A. Anashkin and A. M. Malinen, “Good-Practice Non-Radioactive Assays of Inorganic Pyrophosphatase Activities”, Molecules, 2021, 26, 2356. DOI: 10.3390/molecules26082356.

[84] A. A. Baykov, N. P. Bakuleva and P. A. Rea, “Steady-state kinetics of substrate hydrolysis by vacuolar $H^+$-pyrophosphatase. A simple three-state model”, Eur. J. Biochem., 1993, 217, 755–762. DOI: 10.1111/j.1432-1033.1993.tb18303.x.

[85] Yu. A. Shafranskii, A. A. Baykov, P. F. Andrukovich and S. M. Avaeva, “Comparative kinetic studies of $Mg^{2+}$-activated hydrolysis of tripolyphosphate and pyrophosphate by inorganic pyrophosphatase”, Biokhimiia, 1977, 42, 1244–1251. In Russian, with an English abstract. PMID: 20164. No DOI identified.

[86] R. R. Irani, “Metal Complexing by Phosphorus Compounds. V. Temperature Dependence of Acidity and Magnesium Complexing Constants”, J. Phys. Chem., 1961, 65, 1463–1465. DOI: 10.1021/j100826a516.

[87] C. J. Zea, G. Camci-Unal and N. L. Pohl, “Thermodynamics of binding of divalent magnesium and manganese to uridine phosphates: implications for diabetes-related hypomagnesaemia and carbohydrate biocatalysis”, Chem. Cent. J., 2008, 2, 15. DOI: 10.1186/1752-153X-2-15.

[88] M. D. Toney, “Common Enzymological Experiments Allow Free Energy Profile Determination”, Biochemistry, 2013, 52, 5952–5965. DOI: 10.1021/bi400696j.

[89] I. Barr, “A Bayesian Analysis of Steady–State Enzyme Data leads to Estimates of Rate Constants and Uncertainties in a Multi-Step Reaction, along with Free Energy Profiles”, bioRxiv, version 4, 26 July 2023. DOI: 10.1101/2021.08.04.454956. The calculation in Section S6.4 uses archived code obtained from version 3, posted 19 September 2022.

[90] H. Deng, J. Vedad, R. Z. B. Desamero and R. Callender, “Difference FTIR Studies of Substrate Distribution in Triosephosphate Isomerase”, J. Phys. Chem. B, 2017, 121, 10036–10045. DOI: 10.1021/acs.jpcb.7b08114.

[91] D. Foreman-Mackey, D. W. Hogg, D. Lang and J. Goodman, “emcee: The MCMC Hammer”, Publ. Astron. Soc. Pac., 2013, 125, 306–312. DOI: 10.1086/670067.

[92] T.-Y. Kao, C.-A. Chung and Y.-P. Lee, “Rate Coefficient and Branching Ratio for the Formation of Criegee Intermediate Syn-/Anti-$CH_3CHOO$ from $CH_3CHI$ + $O_2$ and the Self-Reaction of Syn-/Anti-$CH_3CHOO$ Determined with Simultaneous IR/UV Probes”, J. Phys. Chem. A, 2024, 128, 9453–9461. DOI: 10.1021/acs.jpca.4c06588.

[93] Y.-H. Lin and J. J.-M. Lin, “Absolute photodissociation cross section of anti-$CH_3CHOO$”, Phys. Chem. Chem. Phys., 2026, 28, 24422–24430. DOI: 10.1039/D6CP02901A.

[94] H.-P. Trac, P. Raghunath and M.-C. Lin, “Ab Initio Chemical Kinetics for Self- and Cross-Reactions of anti- and syn-$CH_3CHOO$ Conformers”, J. Phys. Chem. A, 2026, 130, 2595–2601. DOI: 10.1021/acs.jpca.6c00127.

[95] Y. Li, Y. Kim, P. V. Bakharev, W. K. Seong, C. Hyun, D. C. Camacho-Mojica, L. Zhang, B. V. Cunning, T. J. Shin, G. Lee and R. S. Ruoff, “Dissolving Diamond: Kinetics of the Dissolution of (100) and (110) Single Crystals in Nickel and Cobalt Films”, Chem. Mater., 2022, 34, 2599–2611. DOI: 10.1021/acs.chemmater.1c03809.